\newcommand{\beten}{\mbox{${}^{10}{\rm Be}$}} 
\newcommand{\altwosix}{\mbox{${}^{26}{\rm Al}$}}
\newcommand{\clthirtysix}{\mbox{${}^{36}{\rm Cl}$}}
\newcommand{\cafortyone}{\mbox{${}^{41}{\rm Ca}$}} 
\newcommand{\mnfivethree}{\mbox{${}^{53}{\rm Mn}$}}
\newcommand{\fesixty}{\mbox{${}^{60}{\rm Fe}$}} 
\newcommand{\pdoneohseven}{\mbox{${}^{107}{\rm Pd}$}} 
\newcommand{\ionetwonine}{\mbox{${}^{129}{\rm I}$}} 
\newcommand{\hfoneeighttwo}{\mbox{${}^{182}{\rm Hf}$}}
\newcommand{\beratio}{\mbox{${}^{10}{\rm Be}/{}^{9}{\rm Be}$}}
\newcommand{\alratio}{\mbox{${}^{26}{\rm Al}/{}^{27}{\rm Al} $}}

\newcommand{\feratio}{\mbox{${}^{60}{\rm Fe}/{}^{56}{\rm Fe}$}}
\newcommand{\mnratio}{\mbox{${}^{53}{\rm Mn}/{}^{55}{\rm Mn}$}}
\newcommand{\hfratio}{\mbox{${}^{182}{\rm Hf}/{}^{180}{\rm Hf}$}}

\documentclass[preprint2]{pp7}

\usepackage{xcolor}
\usepackage[switch]{lineno}
\usepackage{makeidx}
\makeindex

\input{pp7.h}

\begin{document}

\title{\textbf{\LARGE Short-Lived Radionuclides in Meteorites \\ and the Sun's Birth Environment}}

\author {\textbf{\large Steven J. Desch}}
\affil{\small\it Arizona State University}
\author {\textbf{\large Edward D. Young}}
\affil{\small\it University of California, Los Angeles}
\author {\textbf{\large Emilie T. Dunham}}
\affil{\small\it University of California, Los Angeles}
\author {\textbf{\large Yusuke Fujimoto}}
\affil{\small\it University of Aizu / Carnegie Institution for Science}
\author {\textbf{\large Daniel R. Dunlap}}
\affil{\small\it Oak Ridge National Laboratory}

%\linenumbers

\begin{abstract}
\baselineskip = 11pt
\leftskip = 1.5cm 
\rightskip = 1.5cm
\parindent=1pc
{\small 
The solar nebula contained a number of short-lived radionuclides (SLRs) with half-lives of tens of Myr or less, comparable to the timescales for formation of protostars and protoplanetary disks.  Therefore, determining the origins of SLRs would provide insights into star formation and the Sun's astrophysical birth environment. In this chapter, we review how isotopic studies of meteorites reveal the existence and abundances of these now-extinct radionuclides; and the evidence that the SLR $\beten$, which uniquely among the SLRs is not produced during typical stellar nucleosynthesis, was distributed homogeneously in the solar nebula.
We review the evidence that the SLRs $\altwosix$, $\mnfivethree$, and $\hfoneeighttwo$, and other radionuclides, were also homogeneously distributed and can be used to date events during the Solar System's planet-forming epoch.
The homogeneity of the SLRs, especially $\beten$, strongly suggests they were all inherited from the Sun's molecular cloud, and that production by irradiation within the solar nebula was very limited, except for  $\clthirtysix$.
We review astrophysical models for the origin of $\beten$, showing that it requires that the Sun formed in a spiral arm of the Galaxy with higher star formation rate than the Galaxy-wide average.
Likewise, we review the astrophysical models for the origins of the other SLRs and show that they likely arose from contamination of the Sun's molecular cloud by massive stars over tens of Myr, most likely dominated by ejecta from Wolf-Rayet stars. 
% This probably needs to be justified:
The other SLRs also demand formation of the Sun in a spiral arm of the Galaxy with a star formation rate as high as demanded by the Solar System initial $\beten$ abundance.
We discuss the astrophysical implications, and suggest further tests of these models and future directions for the field. 
 \\~\\~\\~}
 %leave this in to get the correct vertical space after the abstract
\end{abstract}  

%\maketitle

%\section{\textbf{LEVEL ONE HEAD (all caps, bold)}}
%\subsection{\textbf{Level Two Head (upper and lower case, flush left, bold)}}
%\subsubsection{\textbf{Level three head (italic)}}
%{\bf Footnotes} are not allowed. Information that you would include in a footnote should be worked into the main text.

%--------------------------------------------------------------------
%
% SECTION I. INTRODUCTION 
%
\section{\textbf{INTRODUCTION}}

\bigskip

\subsection{\textbf{Discovery of the Early Solar System's SLRs}}

The elements heavier than He in the Sun and Earth, for the most part, were forged within stars that have lived and died throughout the evolution of the Galaxy.
When those stars ejected outflows after evolving into red giant or Wolf-Rayet\index{Wolf-Rayet stars} (WR) stars, and/or exploded as supernovae, these atoms were released back into the interstellar medium (ISM)\index{Interstellar medium}, and eventually into molecular clouds from which new stars form. 
As stars have turned hydrogen atoms into heavier species over Galactic history, the abundances of stable nuclei and long-lived (half-lives of billions of years) radionuclides like ${}^{238}{\rm U}$ have steadily built up in subsequent generations of stars.

Evolved stars also eject short-lived radionuclides (SLRs) with half-lives of tens of Myr or less. 
These also build up in the ISM and in molecular clouds, but the balance between production and decay means they are present at much lower levels.
Unlike stable isotopes and long-lived isotopes like ${}^{238}{\rm U}$, those SLRs inherited by the Solar System at its birth 4,569 Myr ago have by now completely decayed.
With few exceptions---such as live ${}^{60}{\rm Fe}$ at\index{Radionuclides!Iron-60} the seafloor from a nearby supernova\index{supernova} $\sim$ 3 Myr ago \citep{KnieEtal2004,WallnerEtal2020}, or radionuclides created by Galactic cosmic rays (GCRs\index{Galactic cosmic rays}, which are ions from the ISM, probably liberated from dust grains, that are accelerated to high energies by supernova-driven shocks)---these SLRs are now absent in the Solar System.
However, substantial evidence from meteorites\index{meteorites} shows these SLRs were abundant in the solar nebula\index{Solar nebula} as the Solar System was born.

It was long ago suspected that the Solar System may have formed quickly enough to incorporate SLRs produced in stars that had recently died and contaminated the Sun's molecular cloud.
\citet{Urey1955} compellingly argued that the melting of asteroids and protoplanets demands radioactive heating by the specific SLR\index{Radionuclides!Aluminum-26}  ${}^{26}{\rm Al}$ ($t_{1/2} \!\!= \!\!0.72$ Myr; \citealt{AuerEtal2009}), which must have been in the solar nebula, synthesized in stars shortly before the Sun's birth.
Remarkably, Urey asserted this just months after the true half-life of ${}^{26}{\rm Al}$ had been determined [see the account by \citet{Kohman1997}], and two decades before the existence of ${}^{26}{\rm Al}$ in the early Solar System was established.
Moreover, Urey argued the Solar System must have formed as rapidly as protostars in the Perseus cluster, whose age \cite{Blaauw1952} had recently determined to be 1.5 Myr.
Only if the Solar System formed within $\sim 10^7$ years after stellar nucleosynthesis could asteroids be melted by the radioactive decay of ${}^{26}{\rm Al}$.
By extension, other SLRs, while not required to melt planetary materials, should also have been incorporated in the solar nebula.
Decades later, this brilliant, pioneering synthesis of the studies of {\it protostars} and {\it planets} remains essentially correct.

The first support for this hypothesis came soon after, from the discovery by \citet{Reynolds1960} that meteorites formed at the birth of the Solar System incorporated the SLR\index{Radionuclides!Iodine-129} ${}^{129}{\rm I}$ ($t_{1/2} \!\!= \!\!16.1$ Myr; \citealt{GarciaToranoEtal2018}).
The evidence came from excesses of the daughter product, ${}^{129}{\rm Xe}$, that correlated with the ${\rm I}$ abundance in different minerals.
Much later the existence of ${}^{26}{\rm Al}$ in the solar nebula was established, by isotopic studies of Ca-rich, Al-rich Inclusions (CAIs\index{Calcium-rich, aluminum-rich inclusions (CAIs)}) in chondritic meteorites, which had only recently been described and understood to be the first solids formed in the Solar System \citep{ChristopheMichelLevy1968,MarvinEtal1970,Grossman1972}.

The SLR ${}^{26}{\rm Al}$ is\index{Radionuclides!Aluminum-26} extinct but still detectable via excesses of its daughter product, ${}^{26}{\rm Mg}$.
Both Al and Mg are abundant in CAIs, and the abundances of their isotopes (${}^{27}{\rm Al}$, ${}^{24}{\rm Mg}$, and ${}^{26}{\rm Mg}$) can be measured precisely using mass spectrometry on different minerals within the same inclusion.
Different Mg-bearing minerals within an igneous CAI should have formed with the same isotopic molar ratio ${}^{26}{\rm Mg} / {}^{24}{\rm Mg}$; but, if they also contain Al, they must have originally incorporated some live ${}^{26}{\rm Al}$ that has now decayed to ${}^{26}{\rm Mg}$, raising the ${}^{26}{\rm Mg} / {}^{24}{\rm Mg}$ ratio by an amount proportional to the ${}^{27}{\rm Al}/{}^{24}{\rm Mg}$ ratio.
If (and {\it only} if) such a linear proportionality can be established in a CAI, the former existence of ${}^{26}{\rm Al}$ can be inferred in the starting materials of the sample, and its abundance quantified.
The linear relationship is called an internal isochron\index{isochron}, and the slope of the isochron is the initial ratio $({}^{26}{\rm Al} / {}^{27}{\rm Al})_0$  in the CAI at the time of its formation.
\citet{GrayCompston1974} found excesses of ${}^{26}{\rm Mg}$ in CAIs, and \citet{LeeEtal1976} determined from isochrons an initial value $({}^{26}{\rm Al} / {}^{27}{\rm Al})_0 \approx 5 \times 10^{-5}$ in CAIs.
Today, a preponderance of CAIs appear consistent with a refined initial value $({}^{26}{\rm Al} / {}^{27}{\rm Al})_0 = 5.23 \times 10^{-5}$ \citep{JacobsenEtal2008}, and this is interpreted as the isotopic composition of Al in the solar nebula at its birth: 52 out of every million Al atoms were the radioactive isotope ${}^{26}{\rm Al}$.

In a similar manner, the existence in the solar nebula of a dozen short-lived (half-lives up to tens of Myr) radionuclides has been established from isotopic studies of meteorites\index{Meteorites}.
In {\bf Table 1} we\index{Radionuclides!Calcium-41}\index{Radionuclides!Chlorine-36}\index{Radionuclides!Aluminum-26}\index{Radionuclides!Beryllium-10}\index{Radionuclides!Iron-60}\index{Radionuclides!Manganese-53}\index{Radionuclides!Palladium-107}\index{Radionuclides!Hafnium-182}\index{Radionuclides!Curium-247}\index{Radionuclides!Iodine-129}\index{Radionuclides!Lead-205}\index{Radionuclides!Niobium-92}\index{Radionuclides!Samarium-146}\index{Radionuclides!Plutonium-244}\index{Radionuclides!Cesium-135} list the ``Solar System" abundance of each SLR, at a time $t\!\!=\!\!0$ defined to be the time when CAIs formed (or, more precisely, when they recorded ${}^{26}{\rm Al} / {}^{27}{\rm Al} = 5.23 \times 10^{-5}$).
The abundances of SLRs are listed relative to the abundance of a common, stable isotope of the parent element, e.g., ${}^{26}{\rm Al} / {}^{27}{\rm Al}$, or ${}^{10}{\rm Be} / {}^{9}{\rm Be}$.
This is because it is the ratios of isotopes that are measured with sufficient precision to derive the slopes of isochrons.  
Isochrons\index{isochron} are determined using plots of ${}^{26}{\rm Mg} / {}^{24}{\rm Mg}$ vs.\ ${}^{27}{\rm Al} / {}^{24}{\rm Mg}$, or ${}^{10}{\rm B} / {}^{11}{\rm B}$ vs.\ ${}^{9}{\rm Be} / {}^{11}{\rm B}$, for example.
After ${}^{129}{\rm I}$ \citep{Reynolds1960} and ${}^{26}{\rm Al}$ \citep{LeeEtal1976},
the existence of ${}^{107}{\rm Pd}$ in the solar nebula was discovered
\citep{KelleyWasserburg1978}, followed by reports of 
${}^{53}{\rm Mn}$ \citep{BirckAllegre1985},
${}^{205}{\rm Pb}$ \citep{ChenWasserburg1987},
${}^{244}{\rm Pu}$ \citep{HudsonEtal1989},
${}^{41}{\rm Ca}$ \citep{SrinivasanEtal1994}, 
${}^{182}{\rm Hf}$ \citep{LeeHalliday1995},
${}^{92}{\rm Nb}$ \citep{Harper1996b},
${}^{10}{\rm Be}$ \citep{McKeeganEtal2000}, 
${}^{135}{\rm Cs}$ \citep{HidakaEtal2001}, 
${}^{60}{\rm Fe}$ \citep{TachibanaHuss2003}, 
${}^{36}{\rm Cl}$ \citep{LinEtal2005}, and 
${}^{247}{\rm Cm}$ \citep{BrenneckaEtal2010}. 

It must be noted that initial discoveries of evidence for the presence of an SLR in the early Solar System often take time to become well established, and the detections of some of these SLRs---especially ${}^{135}{\rm Cs}$, and to some extent ${}^{205}{\rm Pb}$---are not yet universally accepted.\index{Radionuclides!Cesium-135}\index{Radionuclides!Lead-205}
The SLRs ${}^{126}{\rm Sn}$ ($t_{1/2} = 0.23$ Myr) and ${}^{97}{\rm Tc}$ ($t_{1/2} = 4.2$ Myr) and ${}^{98}{\rm Tc}$ ($t_{1/2} = 4.2$ Myr) have been sought but not found \citep{BrenneckaEtal2017,BerminghamMeyer2019}.
As well, the initial abundances of all SLRs remain controversial at some level. 
This is particularly true for ${}^{60}{\rm Fe}$, for\index{Radionuclides!Iron-60} which it has been debated whether ${}^{60}{\rm Fe}/{}^{56}{\rm Fe}$ in the early Solar System was closer to $10^{-8}$ or $10^{-6}$. 
It is also a major point of contention whether or not many SLRs were distributed equally throughout the Sun's protoplanetary disk. 
It has been suggested that $({}^{53}{\rm Mn}/{}^{55}{\rm Mn})_0$ followed a heliocentric gradient \citep{ShukolyukovLugmair2000}, and that $(\beratio)_0$ ratios vary widely among CAIs \citep{FukudaEtal2019,FukudaEtal2021}.   
Even for ${}^{26}{\rm Al}$ it is debated whether $\altwosix$ was distributed uniformly, or if CAIs record a different ${}^{26}{\rm Al} / {}^{27}{\rm Al}$ ratio from the rest of the Solar System \citep[e.g.,][]{LarsenEtal2011,BollardEtal2019}.

The last two decades have seen tremendous growth of data concerning SLRs in the solar nebula, and the use of radioisotopes in meteorites as chronometers in general. 
The existence of ${}^{247}{\rm Cm}$ in the solar nebula has been inferred, and its effects on the U-Pb chronometer have been quantified. 
Uranium-corrected Pb dating of objects is becoming routine, allowing precise dating, even as more and more precise Al-Mg, Mn-Cr, and Hf-W dates are obtained. 
The number of CAIs for which $(\beratio)_0$ has been measured has exploded from a few to over sixty. 
There are now dozens of analyses yielding $({}^{60}{\rm Fe}/{}^{56}{\rm Fe})_0$, and a half dozen CAIs for which $({}^{36}{\rm Cl}/{}^{35}{\rm Cl})_0$ has been derived. 
Precisely because of this explosion of new measurements, however, the necessary synthesis of the data has lagged. 
Improved precision in Pb-Pb dating and the proliferation of Al-Mg, Mn-Cr and Hf-W measurements have made it difficult to reconcile dates obtained using these systems. 
This has led to suggestions of widespread heterogeneities in $\altwosix$. 
The sheer number of $(\beratio)_0$ measurements gives the impression that inclusions have a wide range of values, and unanimity on the initial $({}^{60}{\rm Fe}/{}^{56}{\rm Fe})$ and $({}^{36}{\rm Cl}/{}^{35}{\rm Cl})$ values of the Solar System is lacking.

It is vital to understand the issues that go into obtaining and interpreting data concerning SLRs. 
Only then can questions about initial abundances or heterogeneities in the SLRs be answered.
The answers to these questions, in turn, are key to resolving long-standing issues about the astrophysical origins of the SLRs. 

%---------------------------------
% TABLE 1. Meteoritic Data 

\begin{deluxetable}{c|c|cr|cr|cr}
\rotate
\renewcommand{\thefootnote}{\alph{footnote}}
\tablenum{1}
\tablecaption{Short-lived radionuclides in the early Solar System \label{table:one} }
\tablewidth{0pt}
\tablehead{
\colhead{SLR} & 
\colhead{Daughter} & 
\colhead{Half-life (Myr)} &
\colhead{Ref.} &
\colhead{Solar System Abundance at $t\!\!=\!\!0$} &
\colhead{Ref.} &
\colhead{ISM Production Ratios} &
\colhead{Ref.}
}
\startdata
${}^{41}{\rm Ca}$ & ${}^{41}{\rm K}$ & $0.0994 \pm 0.0015$ & [1] & ${}^{41}{\rm Ca}/{}^{40}{\rm Ca} \approx 4.2 \times 10^{-9}$ & [2] & ${}^{41}{\rm Ca}/{}^{40}{\rm Ca} = 2.3 \times 10^{-3}$ & [3] \\
${}^{36}{\rm Cl}$ & ${}^{36}{\rm Ar}$, ${}^{36}{\rm S}$ & $0.301 \pm 0.002^{\dagger}$ & & ${}^{36}{\rm Cl}/{}^{35}{\rm Cl} \approx (1.7 - 3) \times 10^{-5}$ & [4] & ${}^{36}{\rm Cl}/{}^{35}{\rm Cl} = 2.63 \times 10^{-2}$ & [5]\\
${}^{26}{\rm Al}$ & ${}^{26}{\rm Mg}$ & $0.717 \pm 0.017$ & [6] & ${}^{26}{\rm Al}/{}^{27}{\rm Al} = 5.23 \times 10^{-5}$ & [7] & ${}^{26}{\rm Al}/{}^{27}{\rm Al} = 1.667 \times 10^{-2}$ & [8]  \\
${}^{10}{\rm Be}$ & ${}^{10}{\rm B}$ & $1.387 \pm 0.012$ & [9] & ${}^{10}{\rm Be}/{}^{9}{\rm Be} = (7.1 \pm 0.2) \times 10^{-4}$ & [10]\\
${}^{60}{\rm Fe}$ & ${}^{60}{\rm Fe}$ & $2.62 \pm 0.04$ & [11] & ${}^{60}{\rm Fe}/{}^{56}{\rm Fe} = (0.9 \pm 0.1) \times 10^{-8}$ & [12] &  ${}^{60}{\rm Fe}/{}^{56}{\rm Fe} = 1.23 \times 10^{-4}$ & [13] \\
${}^{53}{\rm Mn}$ & ${}^{53}{\rm Cr}$ & $3.98 \pm 0.11$ & [14] & ${}^{53}{\rm Mn}/{}^{55}{\rm Mn} = (7.8 \pm 0.4) \times 10^{-6}$ & [15] & ${}^{53}{\rm Mn}/{}^{55}{\rm Mn} = 7.52 \times 10^{-1}$ & [16] \\
${}^{107}{\rm Pd}$ & ${}^{107}{\rm Ag}$ & $6.5 \pm 0.3$ & [17] & ${}^{107}{\rm Pd}/{}^{108}{\rm Pd} \approx (7.7 \pm 0.5) \times 10^{-5}$ & [18] & ${}^{107}{\rm Pd}/{}^{108}{\rm Pd} = 6.5 \times 10^{-1}$ & [19] \\
${}^{182}{\rm Hf}$ & ${}^{182}{\rm W}$ & $8.896 \pm 0.089$ & [20] & ${}^{182}{\rm Hf}/{}^{180}{\rm Hf} = (1.04 \pm 0.1) \times 10^{-4}$ & [21] & ${}^{182}{\rm Hf}/{}^{180}{\rm Hf} = 2.9 \times 10^{-1}$ & [22]  \\
${}^{247}{\rm Cm}$ & ${}^{235}{\rm U}$ & $15.6 \pm 0.5^{\dagger}$ & & ${}^{247}{\rm Cm}/{}^{235}{\rm U} = (5.6 \pm 0.3) \times 10^{-5}$ & [23] & ${}^{247}{\rm Cm}/{}^{232}{\rm Th} = 1.01 \times 10^{-1}$ & [24] \\
${}^{129}{\rm I}$ & ${}^{129}{\rm Xe}$ & $16.14 \pm 0.12$ & [25] & ${}^{129}{\rm I}/{}^{127}{\rm I} = (1.74 \pm 0.02) \times 10^{-4}$ & [26] & ${}^{129}{\rm I}/{}^{127}{\rm I} = 1.25 \times 10^{0}$ & [27] \\
${}^{205}{\rm Pb}$ & ${}^{205}{\rm Tl}$ & $17.3 \pm 0.7^{\dagger}$ & & ${}^{205}{\rm Pb}/{}^{204}{\rm Pb} = (1.4 \pm 0.3) \times 10^{-4}$ & [28] & $\,$ & $\,$  \\
${}^{92}{\rm Nb}$ & ${}^{92}{\rm Zr}$ & $34.7 \pm 0.7^{\dagger}$ & & ${}^{92}{\rm Nb}/{}^{93}{\rm Nb} = (1.7 \pm 0.6) \times 10^{-5}$ & [29] & ${}^{92}{\rm Nb}/{}^{93}{\rm Nb} = 5.65 \times 10^{-3}$ & [30] \\
${}^{146}{\rm Sm}$ & ${}^{142}{\rm Nd}$ &
$103 \pm 5$
% $68 \pm 7$ 
& [31] & ${}^{146}{\rm Sm}/{}^{144}{\rm Sm} = (8.28 \pm 0.44) \times 10^{-3}$ 
% (9.4 \pm 0.5) \times 10^{-3}$ 
& [32] & ${}^{146}{\rm Sm}/{}^{144}{\rm Sm} = 9.5 \times 10^{-1}$ & [33] \\
${}^{244}{\rm Pu}$ & ${}^{236}{\rm U}$, ${}^{232}{\rm Th}$ & $80.0 \pm 0.9^{\dagger}$ & & ${}^{244}{\rm Pu}/{}^{238}{\rm U} = (7 \pm 1) \times 10^{-3}$ & [34] & ${}^{244}{\rm Pu}/{}^{232}{\rm Th} = 6.67 \times 10^{-1}$ & [35] \\
${}^{135}{\rm Cs}$ & ${}^{135}{\rm Ba}$ & $2.3 \pm 0.3^{a}$ & & ${}^{135}{\rm Cs}/{}^{133}{\rm Cs} = (4.8 \pm 0.8) \times 10^{-4}$ & [36] & ${}^{135}{\rm Cs}/{}^{133}{\rm Cs} = 8.0 \times 10^{-1}$ & [37]  \\
$\,$ & $\,$ & $\,$ & $\,$ & ${}^{135}{\rm Cs}/{}^{133}{\rm Cs} < 2.8 \times 10^{-6}$ & $\,$ 
\enddata
\tablecomments{
$t\!\!=\!\!0$ defined to be when $(\alratio) = 5.23 \times 10^{-5}$. % in the solar nebula.
Half-life uncertainties are $1\sigma$, abundance uncertainties are $2\sigma$.
${}^{\dagger}$ National Nuclear Data  Center, Brookhaven National Laboratory, \url{https://www.nndc.bnl.gov/ }
References: ${}^{41}{\rm Ca}$: [1] \citet{JorgEtal2012}; [2] \citet{SrinivasanEtal1994}, \citet{LiuEtal2012}; [3] \citet{HussEtal2009}.
${}^{36}{\rm Cl}$: [4] \citet{LinEtal2005}, \citet{TangEtal2017}; [5] \citet{HussEtal2009}. 
${}^{26}{\rm Al}$: [6] \citet{AuerEtal2009}; [7] \citet{LeeEtal1976}, \citet{JacobsenEtal2008}; [8] \citet{HussEtal2009}.
${}^{10}{\rm Be}$: [9] \citet{ChmeleffEtal2010}, \citet{KorschinekEtal2010}; [10] \citet{McKeeganEtal2000}, \citet{DunhamEtal2022}. 
${}^{60}{\rm Fe}$: [11] \citet{RugelEtal2009}; [12] \citet{TachibanaHuss2003}, \citet{DeschEtal2022b}; [13] \citet{HussEtal2009}.  
${}^{53}{\rm Mn}$: [14] \citet{DeschEtal2022b}, \citet{SanbornEtal2019}, \citet{HondaImamura1971}; [15] \citet{BirckAllegre1985}, \citet{TissotEtal2017}, \citet{DeschEtal2022b}; [16] \citet{HussEtal2009}.
${}^{107}{\rm Pd}$: [17] \citet{FlynnGlendenin1969}; [18] \citet{KelleyWasserburg1978}, 
\citet{SchonbachlerEtal2008}, \citet{DeschEtal2022b};
[19] \citet{LugaroEtal2014}. 
${}^{182}{\rm Hf}$: [20] \citet{VockenhuberEtal2004}; [21] \citet{LeeHalliday1995}, \citet{KruijerEtal2014}, \citet{DeschEtal2022b}; [22] \citet{LugaroEtal2014}.  
${}^{247}{\rm Cm}$: [23] \citet{BrenneckaEtal2010}, \citet{TissotEtal2016}, \citet{TangEtal2017}; [24] \citet{GorielyArnould2001}.
${}^{129}{\rm I}$: [25] \citet{GarciaToranoEtal2018}; [26] \citet{Reynolds1960}, \citet{PravdivtsevaEtal2017}; [27] \citet{LugaroEtal2014}. 
${}^{205}{\rm Pb}$: [28] \citet{ChenWasserburg1987}, \citet{NielsenEtal2006}, \citet{PalkEtal2018}.  
${}^{92}{\rm Nb}$: [29] \citet{Harper1996b}, \citet{HabaEtal2021}; [30] \citet{SchonbachlerEtal2002}.
${}^{146}{\rm Sm}$: [31] \citet{MeissnerEtal1987};
%\citet{KinoshitaEtal2017}; 
[32] %\citet{KinoshitaEtal2017}, \citet{MarksEtal2014}; [33] \citet{Jacobsen2005}, \citet{JacobsenWasserburg1984}, \citet{LugmairGaler1992}.
${}^{244}{\rm Pu}$: [34] \citet{HudsonEtal1989}; [35] \citet{HussEtal2009}. 
${}^{135}{\rm Cs}$: [36] \citet{HidakaEtal2001}, \citet{BrenneckaKleine2017}; [37] \citet{Harper1996a}.
}
\end{deluxetable}
%---------------------------------

\subsection{\textbf{Interpretation of Meteoritic Evidence}} 

\subsubsection{Caveats with Isochrons}

As described above, the abundance of an SLR in the early Solar System is not directly measured but must be inferred from a correlation.
If an SLR was present, then a linear relationship between isotopic ratios and elemental abundances is expected. 
However, several physical processes---heating, aqueous alteration, irradiation---can alter the isotopic abundances and/or the relevant elemental ratios in a mineral, thereby disturbing or even obliterating the original isochron\index{isochron}. 
To interpret an isochron as an initial abundance of an SLR, it is vital to establish that there is a linear relationship and that these alterations did not occur. 
Departure from a linear relationship does not mean the SLR was not present, but it does mean some other physical process occurred, and unless that process can be modeled, the initial abundance of the SLR cannot be determined. 

For inclusions with valid isochrons, the abundance of the SLR at the time of the formation of the inclusion can be found, but the meaning of `formation' must be defined on a case-by-case basis.  
Isochrons essentially record conditions at the time when isotopes were no longer mobile, and the isotopic ratios were established, a state known as `isotopic closure.' 
Isotopic closure\index{isotopic closure} in a mineral is usually achieved by cooling below a critical temperature, but many physical processes can prevent isotopic closure, so different events are recorded by the different isotopic systems in various inclusions. 
Within an inclusion, elements and isotopes are able to diffuse and move from their original lattice positions and out of their minerals at specific rates.
Usually the diffusion coefficient\index{diffusion} of atoms in a mineral obeys an Arrhenius relationship:
\begin{equation}
D(T) = D_0 \, \exp \left( -E / R T \right),
\end{equation}
where $D_0$ is the diffusion coefficient pre-factor and $E$ the activation energy per mole, both specific to the element and mineral, and $R$ is the gas constant.
If mineral grains have a typical radius $a$ and spend a time $t > a^2 / D(T)$ at temperatures $T$, isotopes are effectively exchanged between grains, and isotopic differences between minerals are erased faster than the time spent at that temperature, preventing development of an isochron.
Conversely, for a given cooling rate, there is a mineral-specific critical temperature above which isochrons will not form.

\begin{figure*}[t!]
    \centering
    \includegraphics[width=1\textwidth]{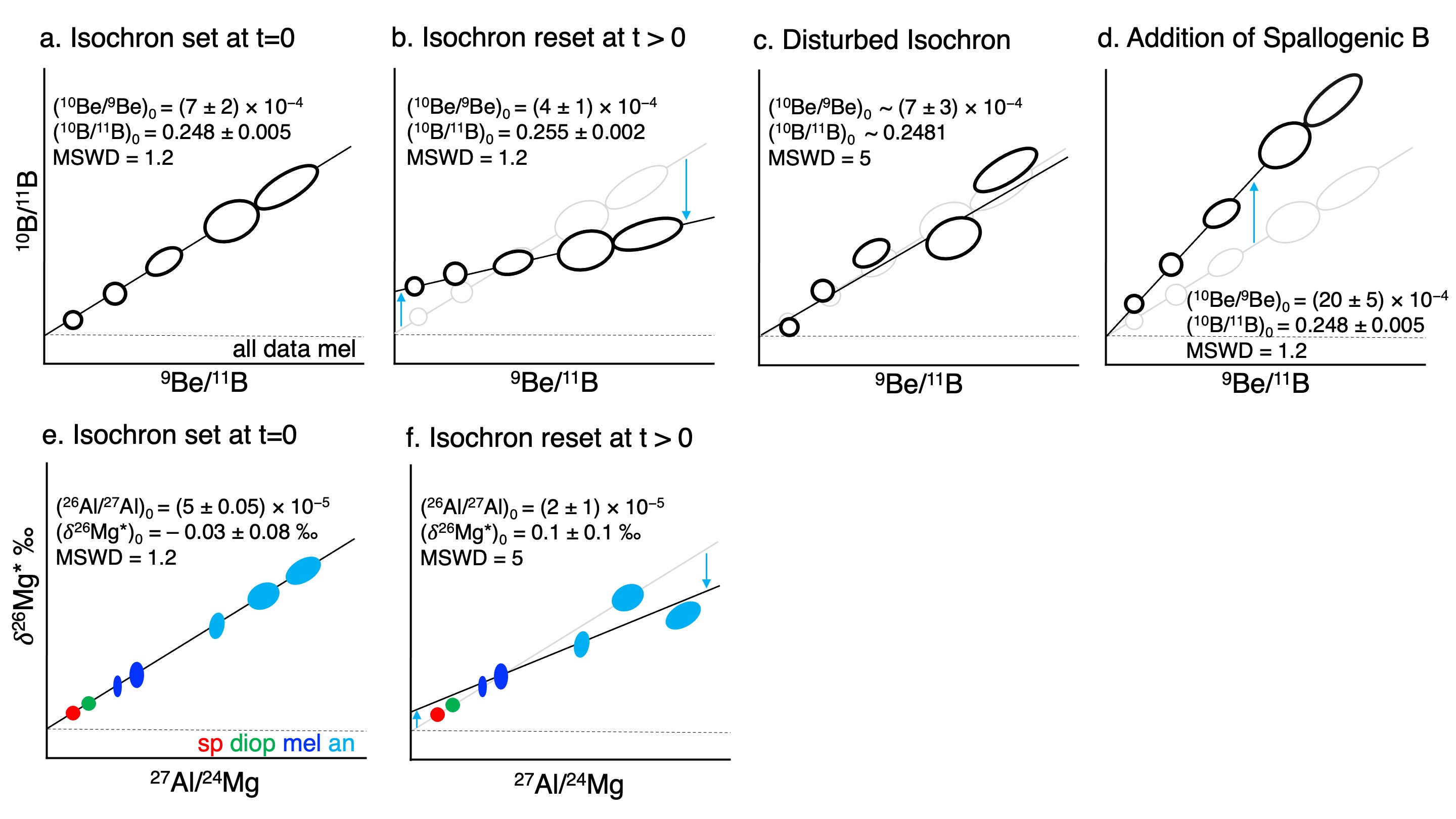}
    \caption{Illustration of isochrons\index{isochron} and how various processes affect them. We use the examples of ${}^{10}{\rm Be}$-${}^{10}{\rm B}$ and ${}^{26}{\rm Al}$-${}^{26}{\rm Mg}$ isotope systematics. Panel a shows a ${}^{10}{\rm Be}$-${}^{10}{\rm B}$ isochron, set at $t\!\!=\!\!0$ and unaffected by later processing; often only melilite is measured to build such isochrons (this mineral can have a range of Be/B). This isochron is repeated in panels b-d in light gray. Panel e shows an ${}^{26}{\rm Al}$-${}^{26}{\rm Mg}$ isochron, set at $t\!\!=\!\!0$ and unaffected by later processing; often spinel (red), Ti-Al-diopside (green), melilite (dark blue), and anorthite (light blue) are measured (if present) to produce a ${}^{26}{\rm Al}$-${}^{26}{\rm Mg}$ isochron (allowing a spread in Al/Mg). Panel b shows a ${}^{10}{\rm Be}$-${}^{10}{\rm B}$ isochron that has been reset at a time $t > 0$ by partial or complete melting due to heating in the nebula or on the parent body. All melilite will respond similarly to partial or complete melting, so if the resetting happened before the complete decay of ${}^{10}{\rm Be}$, the ${}^{10}{\rm Be}$-${}^{10}{\rm B}$ regression first flattens, pivoting around the bulk Be/B ratio, then $^{10}$Be will continue to decay, resulting in an elevated intercept and lowered slope (blue arrows). Panel f shows a ${}^{26}{\rm Al}$-${}^{26}{\rm Mg}$ isochron that has been reset by heating without melting. Mg has distinct diffusion coefficients for each mineral, so depending on the extent of heating and the grain sizes, some minerals (e.g., anorthite) can be reset while others remain undisturbed. The regression is characterized by ${\rm MSWD} \gg 1$ but the disturbance is easily modeled. Panel c shows a disturbed ${}^{10}{\rm Be}$–${}^{10}{\rm B}$ isochron with high MSWD. The disturbance could be due to aqueous alteration, or a SIMS measurement on multiple minerals or a crack (cracks tend to have terrestrial contamination; when measured, this drives isotope ratios toward terrestrial values), or other unknown reasons. High MSWD indicates that the excesses of daughter isotope (e.g., ${}^{10}{\rm B}$ or ${}^{26}{\rm Mg}$) were not successfully modeled, and therefore the initial abundance of the SLR cannot be quantified. ${\rm MSWD} \approx 1$ does not always guarantee an undisturbed isochron, however; Panel d shows how the slope of a ${}^{10}{\rm Be}$–${}^{10}{\rm B}$ regression is raised but the MSWD unchanged when extra spallogenic boron is added to an inclusion due to irradiation on the parent body. }
    \label{fig:regressions}
\end{figure*}

An illustrative example is the Al-Mg system in CAIs. 
A Type B CAI may typically contain the minerals spinel, melilite, Al-Ti-rich diopside, and anorthite, listed here in increasing order of their Al/Mg ratios. 
Over several Myr, these minerals may develop different ${}^{26}{\rm Mg}/{}^{24}{\rm Mg}$ ratios as $\altwosix$ decays.
By virtue of their very different Al/Mg ratios, these minerals can produce a robust isochron in a plot of $\rm ^{26}Mg/^{24}Mg$ vs. ${}^{27}{\rm Al}/{}^{24}{\rm Mg}$ with a slope corresponding to the initial $(\alratio)_0$, as depicted in {\bf Figure~\ref{fig:regressions}e}. 
Mg has distinct diffusion coefficients for each mineral, so depending on the extent of heating and the grain sizes, some minerals (e.g., anorthite) can be reset while others remain undisturbed.
Mg in anorthite has a high diffusion coefficient, so if the CAI is heated above the critical closure temperature, Mg atoms will diffuse in and out of the anorthite.
For the form of the diffusion coefficient above (Equation 1), the mineral-specific closure temperature\index{isotopic closure} is given by
\begin{equation}
T_{\rm c} = \frac{E}{R} \,
\left[ \ln \left( \frac{ A \, T_{\rm c}^2 \, D_0 }{ (E / R) \, \left| dT/dt \right| \, a^2 } \right) \right]^{-1}
\label{eqn:closure}
\end{equation}
\citep{Dodson1973}, where $A\!\!=\!\!55$, $a$ is the radius of the mineral grain, $dT/dt$ is the cooling rate, and it is assumed that the cooling starts from a peak temperature far above $T_{\rm c}$.
Assuming heating above the closure temperature for anorthite, but not above the closure temperature of most or all of the other minerals, the points for some subset of spinel, melilite, and diopside (in a plot of ${}^{26}{\rm Mg}/{}^{24}{\rm Mg}$ vs.\ ${}^{27}{\rm Al}/{}^{24}{\rm Mg}$) may fall on the original isochron, while the point(s) for anorthite would fall far below the isochron (reflecting more of a CAI-averaged ${}^{26}{\rm Mg}/{}^{24}{\rm Mg}$), as depicted in {\bf Figure~\ref{fig:regressions}f}. 
Anorthite also could exchange with an aqueous fluid on the host planetesimal, a sceond reason why the disturbance might only be seen in the anorthite. 
In these examples, the cause of the disturbance would be obvious, and with knowledge of the diffusion coefficients, the disturbance readily could be modeled. 
One could confidently build the isochron with the other points, excluding the anorthite. 
The isochron would record the time the CAI was initially assembled, but one could also infer the peak temperature and cooling rate that might have led to disturbance of anorthite but not the other minerals. 

Usually, though, a thermal event heats a CAI above the closure temperature of all its minerals (or even completely melts it). 
The ${}^{26}{\rm Mg}/{}^{24}{\rm Mg}$ ratios would be homogenized across the CAI after this event, but then would continue to increase again, differently in each mineral, as any remaining ${}^{26}{\rm Al}$ decayed.
In that case, the isochron would record the time of the heating event, not the initial formation, and it is said that the Al-Mg system was `reset'., as in {\bf Figure~\ref{fig:regressions}f}.
The example of a Be-B isochron set at $t\!\!=\!\!0$ is given in {\bf Figure~\ref{fig:regressions}a}, and of a Be-B isochron reset ac time $t>0$ in {\bf Figure~\ref{fig:regressions}b}.
% Likewise, an Al-Mg isochron set at $t\!\!=\!\!0$ is depicted in {\bf Figure~\ref{fig:regressions}e}, and a an Al-Mg isochron reset at $t>0$ shown in {\bf Figure~\ref{fig:regressions}f}.
The primary effect of isotopic homogenization is to raise up spots with ${}^{10}{\rm B}/{}^{11}{\rm B}$ %or ${}^{26}{\rm Mg}/{}^{24}{\rm Mg}$
below the bulk value, and draw down spots with ratios greater than the bulk value, tending to reduce the slope of the isochron---possibly to zero right after the event. 
Comparison of the slope $({}^{10}{\rm Be}/{}^{9}{\rm Be})_0$ of this latter isochron with one set at $t\!\!=\!\!0$ would allow an estimate of how much $\beten$ had decayed between those two events, and the time difference between them.

For more complicated thermal histories, it may not be possible to model the thermal alteration of the isochron, and one must simply accept that it has been disturbed.
% This would usually manifest itself as a non-linear relationship. 
One physical mechanism that can disturb an isochron is aqueous alteration. 
For example, the mineral nepheline (NaAlSiO$_{4}$) can replace the mineral melilite (solid solution between Ca$_{2}$Al$_{2}$SiO$_{7}$ and Ca$_{2}$MgSi$_{2}$O$_{7}$) when Na is introduced into melilite by water and then the water carries Ca away, altering Al/Mg and potentially even ${}^{26}{\rm Mg}/{}^{24}{\rm Mg}$ ratios. 
Unlike thermal alteration, the effect of aqueous alteration on an isochron is not predictable, because of uncertainties in what elements are contained in the water and what temperatures were reached. 
However, the presence of alteration minerals such as nepheline, sodalite, grossular and ilmenite is a good indicator of the occurrence of aqueous alteration in a CAI, signalling that the isochron may have been disturbed. 
Disturbance will probably manifest itself again as a non-linear relationship, as depicted in {\bf Figure~\ref{fig:regressions}c}.
It is likely that the slope of the regression would change by a small but unpredictable amount, and the fit to a line much worsened.

Another mechanism, especially important to the Be-B system, is irradiation by high-energy ($>$ tens of MeV per nucleon) particles, which can induce nuclear reactions, changing the stable isotope ratios in meteoritic inclusions. 
These particles, typically protons or alpha particles, may be solar energetic particles (SEPs\index{Solar energetic particles}) accelerated in solar flares, or GCRs\index{Galactic cosmic rays} accelerated in supernova\index{supernova} shocks throughout the Galaxy. 
Both sorts of irradiation can directly produce cosmogenic nuclides such as ${}^{10}{\rm Be}$, ${}^{26}{\rm Al}$, and ${}^{53}{\rm Mn}$, directly in a meteoroid while on the surface of its parent body, or during its journey from the asteroid belt to the Earth; the abundances of live SLRs can be used to constrain the cosmic-ray exposure time \citep{Eugster2003}.
Separate from this, irradiation can alter isochrons in subtler ways, especially the ${}^{10}{\rm Be}$ isochron. 
Energetic particles hitting abundant O nuclei within an inclusion can spall them, producing B isotopes with a typical ratio ${}^{10}{\rm B}/{}^{11}{\rm B} \approx 0.44$, higher than the $\approx 0.25$ more common in chondrites \citep{LiuEtal2010}.
Addition of this spallogenic B to the pre-existing B in a mineral (both primordial and radiogenic due to decay of ${}^{10}{\rm Be}$) can alter the ${}^{10}{\rm B}/{}^{11}{\rm B}$ ratio in the minerals, disturbing the isochrons.
Because the increase in ${}^{10}{\rm B}/{}^{11}{\rm B}$ correlates with $1/{}^{11}{\rm B}$, this has the effect of increasing the slope without even worsening the fit to a line \citep{DunhamEtal2020}.  
This could lead to interpretations that the CAI had a much higher $(\beratio)_0$ than it did, as depicted in {\bf Figure~\ref{fig:regressions}d}.

%---------------------------
% FIGURE 1
%------------------------------------

Whether the isotopic ratios in different minerals in an inclusion array along a line or not is determined by linear regression. 
In the example of the isochron for ${}^{10}{\rm Be}$, the data points $x_i = {}^{9}{\rm Be}/{}^{11}{\rm B}$ and $y_i = {}^{10}{\rm B}/{}^{11}{\rm B}$ in each mineral or analysis spot $i$ each have measurement uncertainties $\sigma_{x,i}$ and $\sigma_{y,i}$ (which are usually correlated with correlation coefficient $r_i$, because the ${}^{11}{\rm B}$ abundance appears in both), and are usually analyzed using the York regression \citep{YorkEtal2004}.
This yields the slope $b = ({}^{10}{\rm Be}/{}^{9}{\rm Be})_0$ (the initial ratio at isotopic closure) and intercept $a = ({}^{10}{\rm B}/{}^{11}{\rm B})_0$ (initial B isotopic composition), uncertainties in $a$ and $b$, as well as the goodness-of-fit parameter `mean squared weighted deviation' or MSWD: 
\begin{equation}
   {\rm MSWD} = \dfrac{1}{(N-2)} \sum_{i=1}^{N} \dfrac {(y_{i}-a-bx_{i})^2} {\sigma^{2}_{y{i}}+b^{2} \sigma^{2}_{x{i}}},
\end{equation}
where $N$ is the number of data points.\index{isochron}
MSWD is used to assess the degree of over- or under-dispersion of the data.
If the scatter in the data is entirely attributable to measurement error, MSWD will be $\approx\!\!1$, and there is a 95\% probability that the MSWD will lie between roughly $1 - 2 \sigma_{\rm MSWD}$ and $1 + 2 \sigma_{\rm MSWD}$, where $\sigma_{\rm MSWD} = \left[ 2 / (N-2) \right]^{1/2}$
\citep{WendtCarl1991}. 
It is standard in meteoritics and geochronology to report uncertainties using the 95\% confidence interval (roughly 2 sigma). 
If ${\rm MSWD} > 1 + 2 \sigma_{\rm MSWD}$, then there is a $< 5\%$ chance that the data would scatter so badly from a line due to measurement error alone, and the data are judged to not conform to a line and to not describe an isochron.
For example, for a $^{10}$Be–$^{10}$B isotope dataset with 20 $x$-$y$ pairs, the MSWD should not exceed 1.67, nor should it be too small (roughly 0.3). 
If the MSWD of a linear regression for a CAI is out of this range, this does not preclude the one-time existence of the SLR in the CAI, but it does mean the data conform to a line so poorly that an additional process affected the data, and unless that process can be modeled, the initial abundance of the SLR cannot be quantified. 

% OK, Ed, hope this works for you. You're right, though, this was needed.
Provided a sample has a valid ${}^{26}{\rm Al}-{}^{26}{\rm Mg}$ isochron, the initial $(\alratio)_0$ within it at the time of its formation can be derived. 
For example, a chondrule might exhibit an isochron with slope $(\alratio)_0 = 1.3 \times 10^{-5}$, 1/4 the value that CAIs recorded. 
If the chondrules and CAIs formed from the same reservoir, the chondrules must have formed two half-lives of $\altwosix$, or 1.44 Myr, after the CAIs.
The Al-Mg system allows a relative time of formation between two objects to be derived. 
Besides the Al-Mg system, the Mn-Cr and Hf-W systems are important relative chronometers.

\subsubsection{Pb-Pb Chronometry} 

Similar but additional issues arise when constructing isochrons to determine the Pb-Pb age of inclusions. 
Unlike the so-called `relative' chronometers described above, the Pb-Pb system\index{Pb-Pb ages} provides, in theory, the ability to derive an `absolute' age of a sample, as reviewed recently by \citet{ConnellyEtal2017}. 
The two isotopes of uranium, ${}^{235}{\rm U}$ and ${}^{238}{\rm U}$, ultimately\index{Radionuclides!Uranium-235}\index{Radionuclides!Uranium-238} decay respectively to ${}^{207}{\rm Pb}$ and ${}^{206}{\rm Pb}$, with half-lives of $703.81 \pm 0.96 (1\sigma)$ Myr and $4468.3 \pm 4.8 (1\sigma)$ Myr \citep{JaffeyEtal1971,VillaEtal2016}, adding to the primordial abundances of ${}^{207}{\rm Pb}$ and ${}^{206}{\rm Pb}$, whereas no natural radionuclide contributes to the abundance of primordial ${}^{204}{\rm Pb}$. 
This leads to the relationship 
\begin{equation}
\frac{ \exp(+t/\tau_{235}) - 1 }{ \exp(+t/\tau_{238}) - 1 } = 
\left( \frac{ {}^{238}{\rm U} }{ {}^{235}{\rm U} } \right) \, \left( \frac{ {}^{207}{\rm Pb} }{ {}^{206}{\rm Pb} } \right)_{\rm r},
\label{eq:pbpb}
\end{equation}
where $t$ is the time since isotopic closure of the Pb in the mineral (e.g., pyroxene), 
$({}^{238}{\rm U}/{}^{235}{\rm U})$ is the U isotopic ratio measured in the sample today, and the $({}^{207}{\rm Pb}/{}^{206}{\rm Pb})_{\rm r}$ is the isotopic ratio of just the radiogenic Pb attributable to U decay.
If the latter quantities can be found, the time $t$ since formation---the age---of the sample can be solved for.

The radiogenic Pb ratio, $({}^{207}{\rm Pb}/{}^{206}{\rm Pb})_{\rm r}$, is found by treating a sample with various acids, creating multiple washes and leachates with different amounts of primordial (or ``common") Pb and radiogenic Pb, then deriving $({}^{207}{\rm Pb}/{}^{206}{\rm Pb})_{\rm r}$ from an isochron.
Traditionally, this quantity was obtained as the slope of a line found by plotting measured ${}^{207}{\rm Pb}/{}^{204}{\rm Pb}$ vs.\ ${}^{206}{\rm Pb}/{}^{204}{\rm Pb}$, but this requires knowledge of the isotopic composition of primordial Pb.
Greater precision has been obtained by plotting the ``inverse isochron" of ${}^{207}{\rm Pb}/{}^{206}{\rm Pb}$ vs. ${}^{204}{\rm Pb}/{}^{206}{\rm Pb}$ and using the intercept that involves only the radiogenic species to determine age  \citep{TeraWasserburg1974,AmelinEtal2002}. 
This approach allows determination of the intercept using the most radiogenic fractions.
% This also minimizes uncertainties, as $\rm ^{206}Pb / ^{204}Pb \gg 1$ in these ancient rocks. 
In order to standardize results across laboratories, before 2010 it was assumed that all Solar System samples were characterized by $({}^{238}{\rm U}/{}^{235}{\rm U}) = 137.88$.
\citet{BrenneckaEtal2010} demonstrated that there were variations in $({}^{238}{\rm U}/{}^{235}{\rm U})$ ratios, possibly due to decay of ${}^{247}{\rm Cm}$ to\index{Radionuclides!Curium-247} ${}^{235}{\rm U}$ in\index{Radionuclides!Uranium-235} the solar nebula, sufficient to cause $\pm 1$ Myr imprecision in the age. 
It is now standard to only report ``U-corrected" Pb-Pb ages that have been found using the measured U isotopic composition. 

An important caveat with Pb-Pb absolute ages\index{Pb-Pb ages} is that they are actually uncertain by $\pm 9$ Myr, due to the uncertainties in the ${}^{238}{\rm U}$ and especially ${}^{235}{\rm U}$ half-lives\index{Radionuclides!Uranium-235}\index{Radionuclides!Uranium-238} \citep{TissotEtal2017}. 
However, as long as the same half-lives are used, differences in ages between two samples
%, even samples between different laboratories,
can be found to a precision determined only by the measurement uncertainties, which introduce an uncertainty in the intercept of the inverse isochron. 

Another caveat unique to the Pb-Pb isochron is how to deal with the inevitability of contamination by terrestrial Pb in samples. % ({\bf ??}). 
Some of the washes and/or leachates obtained during acid dissolution of a sample must be discarded, but without objective, physical criteria for determining which points should be excluded from a regression, inaccurate results may be obtained. 
If points are excluded solely on the basis that they do not fit a prescribed line, then the fit to a line of the remaining data points will be excellent, and the uncertainty in the intercept and the age will appear small, perhaps only $\pm 0.2$ Myr; but this approach is vulnerable to confirmation of the initial guess for the linear fit, whether or not it is correct.  
Unless the exclusion of specific points from the regression can be physically justified, a more typical uncertainty in Pb-Pb ages is $\pm 0.5$ Myr. 

Somewhat paradoxically, the uncertainties in uranium half-lives make the ``absolute" Pb-Pb chronometer work far better as a relative chronometer, allowing a useful determination of the differences in times between two events. 
It is not as precise as other relative chronometers (e.g., Al-Mg chronometry typically has a precision of $\pm 0.1$ Myr), but it provides an important, independent method for dating samples.
This allows an independent assessment of the homogeneity of SLRs through their ability to predict the times of formation of inclusions as recorded by their Pb-Pb ages. 

\subsubsection{Hf-W Chronometry} 

In particular cases, it is possible to date events without constructing an internal isochron, in particular for the Hf-W system\index{Radionuclides!Hafnium-182}, as reviewed by \citet{KleineWalker2017}.
The SLR ${}^{182}{\rm Hf}$ decays to ${}^{182}{\rm W}$ (via ${}^{182}{\rm Ta}$, with a half-life of 115 days) with a half-life of 8.9 Myr, leading to an excess in the ${}^{182}{\rm W} / {}^{180}{\rm W}$ ratio.
This excess is expressed as a fractional deviation from a standard (often bulk silicate Earth, `BSE'),
\begin{equation}
\epsilon^{182}{\rm W} = 
\left[\frac{ ({}^{182}{\rm W}/{}^{180}{\rm W}) }{ ({}^{182}{\rm W}/{}^{180}{\rm W})_{\rm std} } - 1\right] \times 10,000,
\end{equation}
and is given in epsilon units (parts per ten thousand). 
The deviations in this ratio must be corrected for isotopic fractionation, using the abundances of the other ${\rm W}$ isotopes (${}^{180}{\rm W}$, ${}^{183}{\rm W}$, ${}^{184}{\rm W}$ and ${}^{186}{\rm W}$). 
They must also be corrected for  creation of W isotopes by cosmic ray irradiation on the parent body, which can be modeled by using dosimeters such as Pt isotopic ratios \citep{KruijerEtal2013}.
At $t\!\!=\!\!0$, before ${}^{182}{\rm Hf}$ had decayed, the bulk Solar System material must have started with a deficit of ${}^{182}{\rm W}$, and $\epsilon^{182}{\rm W} \approx -3.5$ relative to the bulk silicate Earth today. 
Similarly, due to complete decay of $\hfoneeighttwo$ within them, chondrites (sometimes referred to as the CHondritic Uniform Reservoir, or CHUR) have $\epsilon^{182}{\rm W} \approx -1.9$, greater than the initial value of $-3.5$. 
(The CHUR value is lower than bulk silicate Earth because the silicate Earth has a higher concentration of Hf than chondrites.)  
Measurements of $\epsilon^{182}{\rm W}$ in a bulk sample can provide a proxy for the amount of ${}^{182}{\rm Hf}$ that has decayed, and therefore the time since $t\!\!=\!\!0$, when an internal isochron cannot be constructed. 

One of the most important applications of this technique is to date the time of core formation in planets. 
After a planet or differentiated planetesimal (represented today by achondritic meteorites\index{Meteorites!achondrites}) forms, its mantle contains primordial ${\rm W}$, extra ${}^{182}{\rm W}$ from prior decay of ${}^{182}{\rm Hf}$, plus live ${}^{182}{\rm Hf}$. 
Upon formation of a metal core by differentiation, the very lithophile element Hf (including $\hfoneeighttwo$) will remain in the rocky mantle, but the moderately siderophile element ${\rm W}$ will mostly partition into the metallic core. 
Subsequent decay of ${}^{182}{\rm Hf}$ will augment the abundance of ${}^{182}{\rm W}$ in the mantle, but not in the core. 
The degree to which the mantle is enhanced in $\rm ^{182}W$ relative to the core  depends on whether core formation occurred early or late; core formation much later than the lifetime of $\hfoneeighttwo$ will result in no difference in $\epsilon^{182}{\rm W}$ between rocky mantle and core. 
Measurements of $\epsilon^{182}{\rm W}$ in samples of rock representing the mantle (for planets or achondrites) or the core (represented by iron meteorites) can be used to date core formation assuming chondrites  represent the initial materials from which the planets and planetesimals were built \citep{KleineWalker2017}.
From such analyses, bolstered by  improvements in the precision of ${\rm W}$ isotopic measurements and corrections for cosmic ray exposure, Hf-W chronometry has become an increasingly important tool.
For example, the time of Mars's core\index{Mars} formation has been placed at $\approx 1-3$ Myr after $t\!\!=\!\!0$ \citep{DauphasPourmand2011}, a result that has profoundly altered perceptions of how quickly planets, or planetary embryos, can grow, and has given great support for the growth of planetary embryos by pebble accretion, \citep[e.g.,][]{LambrechtsJohansen2012,SahijpalBhatia2017}. \index{Pebble accretion}

An even more profound use of ${}^{182}{\rm W}$ has been to date the timing of Jupiter's formation. 
It has been recognized that the bulk isotopic compositions of meteorites are dichotomous across a wide array of elements.  
This is especially evident in the $\epsilon^{50}{\rm Ti}$ (deviations of ${}^{50}{\rm Ti}/{}^{48}{\rm Ti}$ from a standard) and $\epsilon^{54}{\rm Cr}$ (deviations of ${}^{54}{\rm Cr}/{}^{52}{\rm Cr}$ from a standard) stable isotope anomalies in Solar System materials \citep{TrinquierEtal2009, Warren2011}.
The isotopic compositions of meteorites fall either into the ``CC" camp with carbonaceous chondrites\index{Meteorites!chondrites}, or into the ``NC" (non-carbonaceous chondrites) camp along with ordinary and enstatite chondrites, Earth, Mars\index{Mars}, etc.
\citet{KruijerEtal2017} subsequently showed that this dichotomy extends to ${\rm Mo}$ and ${\rm W}$ isotopes in iron meteorites, and attributed this dichotomy to the opening of a gap\index{Protoplanetary disk substructure!gaps} in the protoplanetary disk by the formation of Jupiter's $20-30 \, M_{\oplus}$ core\index{Jupiter}. 
If the stable isotope anomalies were carried on particles too large (roughly mm- to cm-sized) to cross the gap, then the isotopic compositions of the two reservoirs could evolve separately (by unspecified processes).
Moreover, they were able to show on the basis of $\epsilon^{182}{\rm W}$ anomalies, after accounting for the times of core formation within the iron meteorite parent bodies, that the two isotopic reservoirs began to evolve separately sometime between 0.4 and 0.9 Myr after $t\!\!=\!\!0$. 
This requires that Jupiter's core grew to $20-30 \, M_{\oplus}$ in $< 1$ Myr, demanding rapid growth by a mechanism like pebble accretion. \index{Pebble accretion}

\subsection{\textbf{Long-Standing Questions}}

These considerations of how SLR abundances are inferred are essential before meteoritic data can be used to address several long-standing questions, as follows.

{\it Can the SLRs be used to date solar nebula events?}

SLRs have long been used to date events in the solar nebula, based on the assumption that they were homogeneously distributed. 
If the initial $({}^{26}{\rm Al} / {}^{27}{\rm Al})_0$ in, say, an achondrite\index{Meteorites!achondrites} (a meteorite sampling a part of an asteroid that had completely melted and recrystallized) was $4.1 \times 10^{-7}$, 1/128 the initial abundance of $5.23 \times 10^{-5}$ in CAIs, then it could be inferred that the achondrite formed 7 half-lives of ${}^{26}{\rm Al}$, or 5.0 Myr, after the CAIs.
This exercise relies on the assumption that both the achondrite and the CAIs formed from the same isotopic reservoir of material, with 52 atoms of ${}^{26}{\rm Al}$ per million atoms of Al.
This assumption of homogeneity is frequently challenged. 
For example, \citet{BollardEtal2019} have argued that CAIs sampled a reservoir with $\alratio = 5.23 \times 10^{-5}$, but chondrules (and presumably chondrites and achondrites) sampled a reservoir that at the same time was characterized by $\alratio \approx 1.5 \times 10^{-5}$, a factor of 4 lower.
Any one SLR can be used as a chronometer only if it can be shown to have been homogeneously distributed at an early time in the solar nebula.

{\it Were SLRs inherited from the molecular cloud, or injected late and/or created in the solar nebula?}

Assessing the homogeneity of an SLR also bears on the question of its origin. 
If an SLR was uniformly distributed, it is likely that it was simply inherited from the Sun's molecular cloud, which contained the SLR as a result of ongoing stellar nucleosynthesis and contamination of molecular clouds.
If an SLR was not uniformly distributed, this could indicate any number of additional astrophysical events: a late injection\index{Late injection} of supernova\index{supernova} material into the Sun's protoplanetary disk \citep[e.g.,][]{SahijpalGoswami1998,OuelletteEtal2010}, time-varying accretion of spatially heterogeneous molecular cloud material \citep{NanneEtal2019}, or production within the solar nebula by irradiation by SEPs from the early Sun \citep{Lee1978,GounelleEtal2006}. 
Irradiation within the solar nebula\index{Solar nebula} is likely to lead to variations with time and heliocentric distance. 
Of course, even if most SLRs are inherited from the cloud, some might have a different source, and any one SLR might arise from multiple sources.

{\it If SLRs were inherited, what were their sources?}

Even if an SLR can be established to have been inherited from the Sun's molecular cloud, it remains to be determined how the molecular cloud acquired the SLR. 
For $\beten$, which is not created by typical stellar nucleosynthesis, this would practically require irradiation of the molecular cloud material by GCRs \citep{TatischeffEtal2014}.
Other SLRs would require stellar sources, including core collapse supernovae\index{supernova}, WR winds\index{Wolf-Rayet stars}, Asymptotic Giant Branch (AGB) stars, neutron star mergers, etc. (see review by \citealt{LugaroEtal2018}).
Distinguishing the relative contributions of these sources requires simultaneous consideration of the abundances of all the inherited SLRs: WR stars may contribute to $\altwosix$ but not $\fesixty$, while a supernova might contribute more $\fesixty$, relatively, than $\altwosix$ \citep{DiehlEtal2021}. 
A successful model will find a combination of all the sources that produces all the SLRs in their observed proportions and does not overproduce any.
It is also important to consider the astrophysical likelihood of a molecular cloud being contaminated by any of these sources. 
Supernovae and WR stars are the end stages of rapidly evolved, massive stars, and therefore associated with star-forming regions \citep[e.g.,][]{HesterDesch2005}. 
In contrast, AGB stars and neutron star mergers are the progeny of long-lived stellar objects and therefore are not at all associated with young stars.
The probability that the Sun's molecular cloud encountered a single AGB star and was polluted by its ejecta enough to explain its SLR inventory is $< 10^{-6}$ \citep{KastnerMyers1994}, possibly $< 10^{-9}$ \citep{OuelletteEtal2010}. Nevertheless, the Sun's molecular cloud formed from the Galactic ISM, which received contributions of SLRs from many such long-lived objects. 
A single, late-stage stellar object polluting the Sun's protoplanetary disk or molecular cloud is not likely; but many of these stellar objects combined left their signature in the Solar System's inventory of SLRs \citep{CoteEtal2021,LugaroEtal2014}.
% ,TruemanEtal2021}.
% SLR inventory  (cite Cote et al. 2021, Science, Lugaro et al. 2014 Science, Trueman et al. 2021, submitted)

{\it What was the Sun's star-forming environment?}

A large percentage of stars appear to form in clusters and high-mass star-forming regions, and this is the dominant mode of star formation in spiral arms of the Galaxy \citep[][and references therein]{HesterDesch2005}.
Because it touches on so many aspects of the formation of the Sun and planets, it is vital to determine whether the Sun formed in environments like these, similar to the Scorpius-Centaurus association, or in more isolated, low-mass star-forming regions like Taurus-Auriga. 
For example, protoplanetary disks in Taurus (e.g., HL Tau; \citealt{ALMAEtal2015}) often extend for hundreds of AU in size, while proplyds (protoplanetary disks) in Orion are more typically $< 50$ AU in radius \citep{McCaughreanODell1996}.
Scattering of planets and Oort cloud comets would have been common if the Sun formed in a massive cluster \citep[e.g.,][]{AdamsLaughlin2001}, but not if it formed in a more isolated region. 
If the Sun formed in proximity to massive stars, it is more likely to have contained SLRs that are the products of massive stars, produced in the previous tens of Myr. 
If instead it were established that the Solar System's SLRs were produced by irradiation within the solar nebula, this would possibly favor formation in a Taurus-like environment.
Determining the origins of the SLRs helps us to put the Sun's formation in an astrophysical context.

{\it How universal or unique is the Sun's inventory of SLRs?}

As predicted by \citet{Urey1955}, asteroids\index{asteroids} in the Solar System melted because they grew to large sizes while they contained live $\altwosix$ \citep{GrimmMcSween1993}.
The melting and differentiation of asteroids was a major milestone in the creation of planets, possibly even helping to set their volatile inventories
\citep{DeschLeshin2004,CieslaEtal2015,LichtenbergEtal2019}.
To assess the frequency of Earth-like planets in the Galaxy, and how well the insights gained about formation of the Sun's planets can be applied to exoplanetary systems, it must be established how common or rare it is to have acquired the Solar System's inventory of SLRs, especially $\altwosix$.
If SLRs are attributable to irradiation within the solar nebula, then at some level all planetary systems might contain the same SLRs, with possible variations with host star spectral type. 
If instead SLRs are attributable to inheritance from the molecular cloud, then the Sun's SLR abundances would be common for all stars born in spiral arms, but not for the fraction of stars not born in such regions. 
%(ED question) is it worth saying the predicted fraction of stars not born in high SFR environments?
% Good point, but we eventually get to it in S 4.5.

There are reasons to suspect that $\alratio \approx 5 \times 10^{-5}$ could be universal. 
\citet{JuraEtal2013} noted that the mass of $\altwosix$ in the Galaxy (determined by its $\gamma$ ray emission) should be compared with the mass of H$_2$ in order to estimate the concentration of $\altwosix$ in star-forming regions.  For the total mass of $\altwosix$ of $1.7 \pm 0.2 \, M_{\odot}$ \citep{Martin2009},  and the mass of molecular ${\rm H}_{2}$ gas of $5 \times 10^8 \, M_{\odot}$ \citep{RomanDuvalEtal2016}, and assuming the Al/H number ratio in the Galaxy is like that in the solar nebula, $\approx 3.5 \times 10^{-6}$ \citep{Lodders2003},  the average value in star-forming regions of the Galaxy should be $\alratio \approx 4 \times 10^{-5}$, remarkably close to the value inferred for the solar nebula, although considerable uncertainty exists: \citet{TangDauphas2012} derived $3 \times 10^{-6}$ using a much greater mass of gas. 
Nevertheless, further identification of the stellar sources is necessary, for example, to determine whether the $\altwosix$ in the Galactic environment can be incorporated rapidly enough into newly forming systems. Notably, if the same argument was applied to $\fesixty$, this would lead to an estimate for this isotope orders of magnitude higher than observed in the Solar System (\S 2.3.1).

These outstanding questions essentially boil down to whether the SLRs were homogeneous in the solar nebula. 
If so, they can be used as chronometers, and likely were inherited from the molecular cloud. 
Identification of their stellar sources could then be used to put the Sun's formation in a Galactic context and assess their universality.
If the SLRs were heterogeneous in the solar nebula, that would imply additional sources like local irradiation by SEPs, or a single, late stellar source, with different implications about the universality of SLRs and the Sun's birth location. 

\subsection{\textbf{Outline of this Chapter}} 

In this chapter we synthesize the rapid developments in meteoritic data in the last few decades and address the long-standing questions outlined above.

In \S 2 we review the meteoritic evidence.
We discuss the evidence that $\beten$ appears to have been homogeneously distributed in the solar nebula at a level $\beratio \approx 7.1 \times 10^{-4}$ at $t\!\!=\!\!0$. 
We discuss the meteoritic evidence that the other SLRs, especially $\altwosix$, $\mnfivethree$, and $\hfoneeighttwo$, were also homogeneously distributed in the solar nebula, with abundances at $t\!\!=\!\!0$ of $\alratio \equiv 5.23 \times 10^{-5}$, $\mnratio \approx 7.8 \times 10^{-6}$, and $\hfratio \approx 1.04 \times 10^{-4}$.
The SLR ${}^{129}{\rm I}$ appears to have been homogeneous as well, at a level ${}^{129}{\rm I}/{}^{127}{\rm I} = 1.7 \times 10^{-4}$. 
All of the SLRs appear consistent with homogeneous distributions at $t\!\!=\!\!0$ except for the SLR $\clthirtysix$, which is evidenced only in CAIs that were aqueously altered on their parent bodies, and which appears to have been present at variable levels in these CAIs.
We also review how recent improvements in the inferred initial abundances of SLRs can be used for in conjunction with Pb-Pb dating to improve chronometry of events in the solar nebula.

In \S 3 we discuss models for the origins of the SLRs.
After reviewing the astrophysics of star formation in the Galaxy, we present the arguments that $\beten$ was created by GCR spallation in the Sun's molecular cloud, and not by SEP irradiation within the solar nebula.
We review models for how the other SLRs, which arise from stellar nucleosynthesis, could have been produced and introduced into the Sun's molecular cloud. 
Both lines of evidence strongly suggest the Sun formed in a spiral arm of the Galaxy, in a region of higher-than-average star formation rate.  

In \S 4 we use these insights to revisit the long-standing questions raised above (\S 1.3). 
We critically reassess evidence previously used to argue for heterogeneity of $\altwosix$ and for late injection of SLRs into the disk. 
In the context of $\clthirtysix$, we explore the possible timing and extent of SLR production by SEP irradiation in the solar nebula.
We use these findings to place the Sun's formation in an astrophysical context, and address the universality of the Sun's SLR abundances. 
Finally, we explore future directions for the field.

%--------------------------------------------------------------------
%
% SECTION II. METEORITIC EVIDENCE  
%
\section{\textbf{METEORITIC EVIDENCE}}

\bigskip

\subsection{\textbf{Uniformity of \boldmath${}^{10}{\rm Be}$}}
 
\subsubsection{Initial $(\beratio)_0$ ratios of CAIs}

The SLR $\beten$\index{Radionuclides!Beryllium-10}, which decays to ${}^{10}{\rm B}$ ($t_{1/2} \!\!= \!\!1.39$ Myr) is unique among SLRs in that it is produced almost exclusively by non-thermal nuclear reactions induced by GCRs and/or possibly SEPs. 
If any SLRs are produced in the solar nebula by SEP irradiation, $\beten$ likely is one of them, and irradiation would contribute to the majority of $\beten$ in the early Solar System 
\citep[e.g.,][]{GounelleEtal2001,WielandtEtal2012}. 
In that case, the abundances and initial $\beratio$ ratios would show variations.
Models of $\beten$ production by SEP irradiation predict a decrease with heliocentric distance $r$ and an increase with time $t$ \citep{Jacquet2019}; CAIs forming across the solar nebula would show variations of factors of 2 or more in their initial $(\beratio)_0$ ratios (\S 3.2).
Thus the role of in situ irradiation in producing the SLRs can be assessed by quantifying the heterogeneity in the $(\beratio)_0$ ratios recorded by CAIs.

The first evidence for the one-time presence of $^{10}$Be in CAIs was discovered two decades ago by \citet{McKeeganEtal2000}.
Since then, researchers have used ion probes (CAMECA IMS-6f, IMS-1270, IMS-1290, and NanoSIMS) to measure  $^{10}$Be$-^{10}$B isotope systematics in CAIs to produce 67 isochron regressions (some isochron regressions include data from multiple CAIs).
The first CAIs\index{Calcium-rich, aluminum-rich inclusions (CAIs)} measured were primarily from CV3 chondrites\index{Meteorites!chondrites}, due to their relatively high abundance and large sizes. 
To mitigate against the possibility of disturbance by aqueous alteration of CAIs in more altered chondrites (e.g., Allende\index{Meteorites, individual!Allende}, an oxidized CV3.6 chondrite), these studies were expanded to CAIs from more pristine chondrite types such as CO3 and CR2 \citep{FukudaEtal2019,DunhamEtal2022}. 
The reported initial $({}^{10}{\rm Be}/{}^{9}{\rm Be})_0$ ratios of all measured CAIs range from $\sim 3-100$ $\times 10^{-4}$ 
\citep{McKeeganEtal2000, SugiuraEtal2001, MarhasEtal2002, MacPhersonEtal2003, ChaussidonEtal2006, LiuEtal2009, LiuEtal2010, WielandtEtal2012, GounelleEtal2013, SrinivasanChaussidon2013, SossiEtal2017, FukudaEtal2019, MishraMarhas2019, FukudaEtal2021, DunhamEtal2020, BekaertEtal2021, DunhamEtal2022}.
%(McKeegan et al., 2000; Sugiura et al., 2001; Marhas et al., 2002; MacPherson et al., 2003; Chaussidon et al., 2006; Liu et al., 2009; Liu et al., 2010; Wielandt et al., 2012; Gounelle et al., 2013; Srinivasan and Chaussidon, 2013; Sossi et al., 2017; Fukuda et al., 2019; Mishra and Marhas, 2019; Fukuda et al., 2021b; Dunham et al. 2020, Dunham et al. submitted).
At first glance, $^{10}$Be in CAIs would seem to be distributed heterogeneously; however, upon a closer look, most CAIs  record a $({}^{10}{\rm Be}/{}^{9}{\rm Be})_0$ value around 7 to 8 $\times 10^{-4}$. 
To assess the heterogeneity of $\beten$, the isochron for each CAI needs to be considered carefully, to make sure that it is definitely uncompromised and reflects the $({}^{10}{\rm Be}/{}^{9}{\rm Be})_0$ at $t\!\!=\!\!0$.
Then the initial $(\beratio)_0$ values must be statistically evaluated to determine if they adhere to a single, homogeneous population or multiple populations. 

\begin{figure*}
    \centering
    \includegraphics[width=0.4\textwidth]{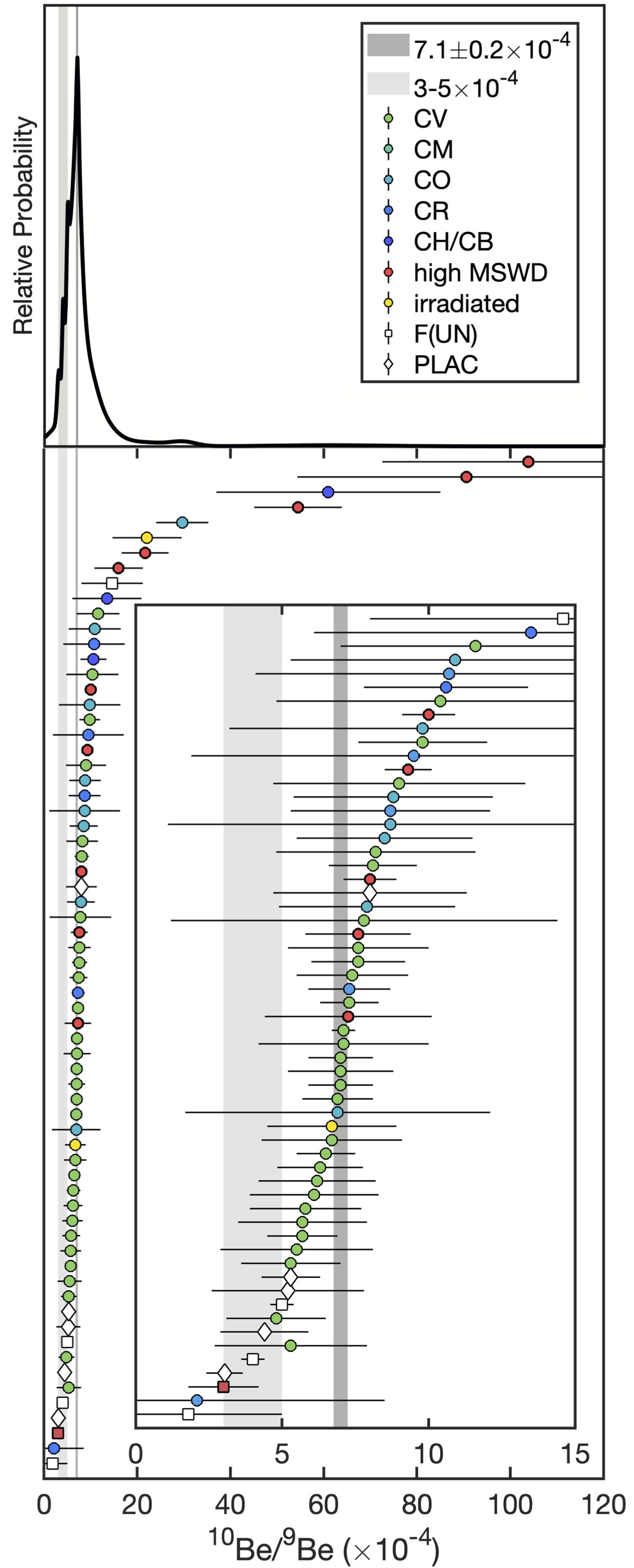}
    \caption{Initial $(\beratio)_0$ ratios from 67 CAI isochron regressions (n$>$2) from different carbonaceous chondrites 
    \citep{McKeeganEtal2000, SugiuraEtal2001, MarhasEtal2002, MacPhersonEtal2003, ChaussidonEtal2006, LiuEtal2009, LiuEtal2010, WielandtEtal2012, GounelleEtal2013, SrinivasanChaussidon2013, SossiEtal2017, FukudaEtal2019, MishraMarhas2019, FukudaEtal2021, BekaertEtal2021, DunhamEtal2022}.
%    (McKeegan et al., 2000; Sugiura et al., 2001; Marhas et al., 2002; MacPherson et al., 2003; Chaussidon et al., 2006; Liu et al., 2009; Liu et al., 2010; Wielandt et al., 2012; Gounelle et al., 2013; Srinivasan and Chaussidon, 2013; Sossi et al., 2017; Fukuda et al., 2019; Mishra and Marhas, 2019; Fukuda et al., 2021; Dunham et al., submitted).
The upper probability density distribution shows that the data center around a single peak at $^{10}{\rm Be}/{}^{9}{\rm Be} = 7.1 \times 10^{-4}$, with a few small peaks $< 7 \times 10^{-4}$, the FUN CAIs and PLACs. \index{Calcium-rich, aluminum-rich inclusions (CAIs)!FUN CAIs}
The lower plot shows the CAI isochron regression $\beratio$ data points. The inset highlights isochron regressions with  $(\beratio)_0 <  15 \times 10^{-4}$. The different symbols and colors designate which carbonaceous chondrite the CAI was found in, if the CAI is a F(UN) (white square) or PLAC (white diamond), or if the ${}^{10}{\rm Be}-{}^{10}{\rm B}$ regression is compromised and does not  likely reflect the initial $^{10}$Be/$^{9}$Be at the beginning of the Solar System (red = high MSWD, yellow = CAI boron affected by irradiation on the parent body). The shaded gray horizontal bars indicate the range of typical $^{10}$Be/$^{9}$Be ratios for F(UN) CAIs and PLACs, $^{10}$Be/$^{9}$Be = 3 $-$ 5 $\times 10^{-4}$, and for all uncompromised $^{10}$Be/$^{9}$Be isochron regressions (circles which are not red or yellow) in normal CAIs (i.e., not FUN or PLAC), with weighted mean $^{10}$Be/$^{9}$Be = (7.1 $\pm$ 0.2) $\times 10^{-4}$ (SD = 1.9). Uncertainties are $2\sigma$.}
    \label{fig:10Be}
\end{figure*}

\citet{DunhamEtal2022}
 reevaluated their own and literature $^{10}$Be data consistently, using appropriate standardized techniques to allow consideration of all the CAI isochron regressions together, with the goal of determining whether the distribution of $\beten$ among CAIs was statistically homogeneous or heterogeneous.
The results are displayed in {\bf Figure~\ref{fig:10Be}}.
Reported data were re-analyzed using consistent instrumental relative sensitivity factors, and York regressions applied consistently to calculate MSWD.
Isochrons were considered valid only if they adhered to the criterion  ${\rm MSWD} < 1 + 2 \sigma_{\rm MSWD}$ using the number of data points regressed in interpreting the robustness of regressions.
Eleven CAI regressions were found to have high MSWD $> 1 + 2 \sigma_{\rm MSWD}$ (red data points in Figure~\ref{fig:10Be}):  
one from a CH/CB chondrite, six from CV3$_{\rm ox}$ chondrites, three from CV3$_{\rm red}$ chondrites, and one from a CO3 chondrite.
Six of these eleven regressions show inferred $\beratio$ ratios inconsistent with $\beratio \approx 7 \times 10^{-4}$, with four of these showing $\beratio \gg 7 \times 10^{-4}$.
This suggests that inferred ($\beratio)_0 \gg 7 \times 10^{-4}$ is associated with disturbed isochrons with high MSWD, affected possibly by aqueous alteration of the CAIs or analytical artifacts. 
%Four CAI $^{10}$Be$-^{10}$B regressions were found to be consistent with a mixing between two distinct B reservoirs, and not necessarily decay of $\beten$ (orange circles in Figure~ \ref{fig:10Be}).
Two CAIs\index{Calcium-rich, aluminum-rich inclusions (CAIs)} show evidence that they incorporated spallogenic B resulting from parent-body irradiation, which can artificially increase their inferred $(\beratio)_0$ ratios (yellow circles in Figure~\ref{fig:10Be}). These two CAIs from CV3 NWA 6991, \textit{Lisa} and \textit{B4}, are known to have existed within a few meters of their parent-body surface because they have elevated $^{150}$Sm, due to interaction of ${}^{149}{\rm Sm}$ with high-energy neutrons generated by GCR irradiation 
\citep{ShollenbergerEtal2018}.
These same GCRs would also create spallogenic B within the CAI, affecting the  $^{10}$Be$-^{10}$B regression, especially as the B concentration is very low (down to a few ppb); reasonable GCR fluences could have increased the slope of the regression of CAI \textit{Lisa} without increasing the MSWD 
%In similar fashion, CAI 3529-41 appears not to have been directly irradiated, but to have received additional spallogenic Li and B during aqueous alteration 
\citep[][\S~\ref{sec:beryllium7}]{DunhamEtal2020,DeschOuellette2006}.
%steve can you add a reference or more explanation?)
Setting aside the regressions with high MSWD and those that likely experienced parent body irradiation, there remain 54 CAIs that appear to have faithfully recorded an initial $(\beratio)_0$ ratio. 

%---------------------------
% FIGURE 2 
%------------------------------------

Among the 54 uncompromised CAIs, two groups stand out: 46 normal CAIs with primarily $({}^{10}{\rm Be}/{}^{9}{\rm Be})_0$ $\sim 7-8 \times 10^{-4}$, and 8 FUN and PLAC inclusions (described below) with  $({}^{10}{\rm Be}/{}^{9}{\rm Be})_0$ $\sim 3-5 \times 10^{-4}$. 
The first group consists primarily of multi-minerallic CAIs with melilite $\pm$ hibonite $\pm$ spinel $\pm$ Al-Ti diopside $\pm$ anorthite, that do not display isotope fractionations or stable isotope anomalies. 
CAI isochron regressions in this group have a weighted mean $(\beratio)_0 = (7.1 \pm 0.2) \times 10^{-4}$, and likely recorded the $\beratio$ ratio in the Solar System at $t\!\!=\!\!0$.
This weighted mean is corroborated by a probability density distribution with a single primary peak at $(\beratio)_0 = 7.1 \times 10^{-4}$.
There are only two CAIs within the normal CAI group that record high $\beratio$ ratios \citep{FukudaEtal2019, FukudaEtal2021}, %they do not influence the weighted mean significantly due to 
with large uncertainties; future work is needed to understand these unique CAIs and what they imply about the origin of $^{10}Be$. 
%The distribution of $(\beratio)_0$ values about their mean has MSWD=1.36, lower than the maximum allowable value of 1.45 for $n\!=\!40$, consistent with variations about a single value due solely to measurement uncertainties.
The distribution shows visually that the group of normal CAIs defines a population with a common $(\beratio)_0$ value.

There is almost no overlap in terms of $(\beratio)_0$ ratios between the normal CAI population and the second group with lower $({}^{10}{\rm Be}/{}^{9}{\rm Be})_0$. 
This group includes objects that are distinctive in terms of petrology or isotopic compositions:   Fractionations and Unidentified Nuclear effects (FUN) CAIs\index{Calcium-rich, aluminum-rich inclusions (CAIs)!FUN CAIs}; Unidentified Nuclear effects (UN) CAIs; as well as PLAty hibonite Crystals (PLACs)\index{Calcium-rich, aluminum-rich inclusions (CAIs)!PLACs}, which exhibit UN effects. 
This group defines a poorly understood but clearly separate population.
One way to reconcile this group with a homogeneous $\beratio$ in the solar nebula is if these objects formed or were reset after $t\!\!=\!\!0$.
A second explanation might be that the unique thermal histories of these objects affected their ${}^{10}{\rm Be}$-${}^{10}{\rm B}$ isochron regressions in a way that mimicked a late resetting. 
Here we explore these potential explanations.

\subsubsection{FUN CAIs, UN CAIs, and PLACs}

Evaporation experiments suggest FUN CAIs that are coarse-grained igneous types (type A, type B, or forsterite-bearing type B) melted in a low-pressure environment in which evaporation is more likely than for normal CAIs
\citep{KrotEtal2014,MendybaevEtal2017}.
Isotopic fractionations of major elements like O, Mg, and Si,  due to intense evaporation, are a defining characteristic of FUN CAIs\index{Calcium-rich, aluminum-rich inclusions (CAIs)!FUN CAIs}.
PLACs are hibonite inclusions tens to hundreds of microns in size, that likewise exhibit large stable isotope anomalies and may or may not exhibit fractionations.\index{Calcium-rich, aluminum-rich inclusions (CAIs)!PLACs} 
These include some objects like \textit{HAL} and \textit{SHAL} that  are primarily composed of hibonite and which may have formed via melt distillation
 \citep{IrelandEtal1992}.
About 20\% of PLACs and related objects show significant mass fractionation effects
\citep{KoopEtal2016, KoopEtal2018a}.
The unusual object CAI {\it 31-2} (DOM 08006 CO3.0) exhibits ``UN" effects without the fractionation, like PLACs, and is a hibonite core surrounded by other condensed minerals \citep{SimonEtal2019}.
Many of the objects associated with low $(\beratio)_0$ are associated with intense evaporation.

Evaporation may have affected the ${}^{10}{\rm Be}$-${}^{10}{\rm B}$ isochrons of most of these inclusions. 
B is relatively volatile compared to Be [50\% condensation temperatures of 906 K and 1445 K, respectively; \citet{Lodders2003}], so extensive fractionation would have allowed escape of B.
If evaporative loss of B occurred at $\sim \!\!1$ Myr, after some decay of $\beten$, then the data points in the isochron would move to higher ${}^{9}{\rm Be}/{}^{11}{\rm B}$ ratios (to the right) and to lower ${}^{10}{\rm B}/{}^{11}{\rm B}$ (down) (see Fig.~\ref{fig:regressions}). 
Both effects would lower the slope of the isochron, making it appear that the inclusion formed with lower initial $(\beratio)_0$ than it did.
A comparison between FUN CAIs\index{Calcium-rich, aluminum-rich inclusions (CAIs)!FUN CAIs} and the UN CAI {\it 31-2} strengthens the case for such evaporation as the cause of low $(\beratio)_0$ ratios.
For example, both FUN CAI {\it CMS-1} and UN CAI {\it 31-2} are $\altwosix$-poor inclusions with nucleosynthetic isotopic anomalies. 
{\it CMS-1} clearly shows evidence of melting and evaporation events \citep{WilliamsEtal2017}.
In contrast, UN CAI {\it 31-2} has no mass-dependent fractionations, and based on its petrology could not have experienced significant thermal processing after it condensed \citep{SimonEtal2019}.
Like other FUN CAIs, {\it CMS-1} has $(\beratio)_0 \approx 5 \times 10^{-4}$, whereas UN CAI {\it 31-2} has $(\beratio)_0 = (14.6 \pm 6.6) \times 10^{-4}$, more consistent with the value recorded by the normal population of CAIs. 
Low $(\beratio)_0$ ratios therefore seem associated with evaporation.

Alternatively, lower $(\beratio)_0$ values may simply be attributable to late formation or thermal resetting in the solar nebula, at $t\!\!\sim\!\!1$ Myr, after which $\beten$ would have decayed to levels $\beratio \approx 4 \times 10^{-4}$.
The coarse-grained (type A and B) FUN CAIs\index{Calcium-rich, aluminum-rich inclusions (CAIs)!FUN CAIs} are associated with lower $(\alratio)_0$ that is broadly consistent with having been reset in the solar nebula these times
\citep{DunhamEtal2020lpsc}.
This would also be consistent with them experiencing lower nebular pressures.
However, hibonite-bearing FUN CAIs, and especially the hibonite-dominated PLACs\index{Calcium-rich, aluminum-rich inclusions (CAIs)!PLACs}, often have initial $(\alratio)_0$ ratios so low they could not have been reset in the nebula before incorporation into their chondrite parent bodies.
These are interpreted to have formed very early in the solar nebula. 
We discuss possible explanations for the decoupling between $\beten$ and $\altwosix$ in \S~\ref{sec:lateinjection}. 
Here, we merely point out that thermal resetting of the ${}^{10}{\rm Be}-{}^{10}{\rm B}$ system in a FUN CAI or a hibonite grain during its long residence in the protoplanetary disk would not be unexpected.

\subsubsection{Summary}

There are many reasons to suspect that the physically distinctive FUN CAIs, UN CAIs, and PLACs that define a second population with low $(\beratio)_0$ do not record the $\beratio$ ratio in the solar nebula at $t\!\!=\!\!0$.
During the roughly 3 Myr these objects spent in the protoplanetary disk before incorporation into their chondritic parent bodies, there were many opportunities for the $^{10}{\rm Be}$--${}^{10}{\rm B}$ isochron to be thermally reset and/or altered by evaporative loss of B. 
The lower $(\alratio)_0$ ratios of some, and the strong association with evaporation in many, of these inclusions make clear that they experienced conditions not experienced by the bulk of CAIs. 
These can be regarded as a separate population, with the 46  
other normal CAIs forming a population that very probably samples the $(\beratio)$ ratio in the solar nebula at $t\!\!=\!\!0$. 
The weighted mean value of this population, $(\beratio)_{\rm SS} = (7.1 \pm 0.2) \times 10^{-4}$, defines the canonical ratio for the Solar System.
The $(\beratio)_0$ values of the main population of CAIs cluster around this mean exactly as expected according to measurement uncertainties, indicating that $\beten$ was distributed uniformly at $t\!\!=\!\!0$ in the CAI forming region.

\subsection{\textbf{Concordancy of Pb-Pb and SLR-derived ages}} 

The SLRs $\altwosix$, $\mnfivethree$, and $\hfoneeighttwo$ can be used as chronometers if they were homogeneous in the early Solar System. 
One of the best ways to assess whether the abundances of these SLRs 
%like ${}^{26}{\rm Al}$, ${}^{53}{\rm Mn}$, and ${}^{182}{\rm Hf}$
were uniformly distributed in the solar nebula is to assess if they predict  consistent times of formation across the board for appropriate meteoritic samples, often bulk measurements of achondrites.
The SLR $\altwosix$ existed in the solar nebula at the time most CAIs were forming, at a level close to the ``canonical" ratio $\alratio = 5.23 \times 10^{-5}$ \citep{JacobsenEtal2008}.
This is not to say that all CAIs record this exact value; indeed, many CAI minerals appear to have condensed from the nebula with $(\alratio)_0$ values 5.4 to $4.9 \times 10^{-5}$, suggesting formation over a timespan $\sim 0.1$ Myr \citep{LiuEtal2019}.
For practical purposes, it is convenient to define a ``time zero" ($t\!\!=\!\!0$) to be the time when Al in the Solar System was characterized by
$\alratio = (\alratio)_{\rm SS} \equiv 5.23 \times 10^{-5}$, and acknowledge that CAIs were forming {\it around} this time.

Assuming $\altwosix$ was uniformly distributed with this value in the solar nebula at $t\!\!=\!\!0$, then a lower
ratio $(\alratio)_0$ in an inclusion or achondrite indicates it formed at a time after $t\!\!=\!\!0$ of
\begin{equation}
\Delta t_{26} = \tau_{26} \, \ln \left[ (\alratio)_{\rm SS} / (\alratio)_0 \right],
\end{equation}
where $\tau_{26} \! \approx \! 1.03$ Myr is the mean life of $\altwosix$.
This time of formation can be compared to times of formation as inferred from other isotopic
systems, including the Pb-Pb chronometer, or from the $(\mnratio)_0$ or $(\hfratio)_0$ ratios.
If $\mnfivethree$ and $\hfoneeighttwo$ were distributed homogeneously in the solar nebula, and if one
knew the initial ratios in the Solar System, $(\mnratio)_{\rm SS}$ or $(\hfratio)_{\rm SS}$,
one could calculate the time of formation of an inclusion in the same way:
\begin{equation}
\Delta t_{53} = \tau_{53} \, \ln \left[ (\mnratio)_{\rm SS} / (\mnratio)_0 \right],
\end{equation}
and
\begin{equation}
\Delta t_{182} = \tau_{182} \, \ln \left[ (\hfratio)_{\rm SS} / (\hfratio)_0 \right].
\end{equation}
Likewise, if one could hypothetically measure a Pb-Pb age of objects that recorded the time $t\!\!=\!\!0$, $t_{\rm SS}$,
then one could infer a time of formation after $t\!\!=\!\!0$ by measuring the Pb-Pb age of a sample, $t_{\rm Pb}$:
\begin{equation}
\Delta t_{\rm Pb} = t_{\rm SS} - t_{\rm Pb}.
\end{equation}
If $\altwosix$, $\mnfivethree$, and $\hfoneeighttwo$ isotopes were all homogeneously distributed
in the solar nebula at $t\!\!=\!\!0$, and if a sample cooled fast enough so that all the isotopic systems achieved closure\index{isotopic closure} simultaneously (at times differing by $< 10^5$ years), then the times of formation recorded by
each isotopic system should match: $\Delta t_{26} \!\! = \!\! \Delta t_{53} \!\! = \!\! \Delta t_{182} \!\! = \!\! \Delta t_{\rm Pb}$.
More precisely, a weighted average of these values could provide the best estimate of the actual time of formation of the inclusion or achondrite, $\Delta t$. 
If each of the ages $\Delta t_{26}$, etc., are consistent within $2\sigma$ measurement uncertainties of $\Delta t$, they could be considered to match.

If all the ages do match---are `concordant'---then that sample would provide strong support for uniformity and against heterogeneity of the SLRs in the solar nebula.
If the ages do not match for a given inclusion or achondrite, that could indicate that the SLRs were not homogeneously distributed, {\it or} it could indicate that the isotopic systems in that inclusion or achondrite did not achieve closure simultaneously.
Calculating these derived ages, assuming homogeneity, is a powerful technique for assigning times after $t\!\!=\!\!0$ to events in the solar nebula; but flipping the problem around, to see if ages are concordant, is a strong test of whether the SLRs were uniformly distributed.

\citet{DeschEtal2022b} recently tested concordancy by comparing 38 times of formation for 13 achondrites\index{Meteorites!achondrites}.
%Achondrites are meteorites sampling parts of asteroids that completely melted and recrystallized.
Any achondrite could be used to test homogeneity, with the caveat that any one achondrite may have been disturbed, or taken too long to cool for the different systems to close at the same time.
Quenched angrites are a particular type of volcanic achondrite that cooled rapidly enough that the different isotopic systems almost certainly achieved closure simultaneously.
An example is D'Orbigny\index{Meteorites, individual!D'Orbigny}, which has been measured repeatedly for initial $(\alratio)_0$, $(\mnratio)_0$, $(\hfratio)_0$, and Pb-Pb ages.
It is therefore possible to derive $\Delta t_{26}$, $\Delta t_{53}$, $\Delta t_{182}$ and $\Delta t_{\rm Pb}$ (and weighted mean $\Delta t$) for D'Orbigny, and the uncertainties in those formation times, provided $(\mnratio)_{\rm SS}$, $(\hfratio)_{\rm SS}$, $t_{\rm SS}$ are known (the definition of $(\alratio)_{\rm SS}$ fixes the time $t\!\!=\!\!0$).
Because D'Orbigny and the other quenched angrites---Sahara (SAH) 99555 and Northwest Africa (NWA) 1670---all likely cooled rapidly, lack of concordancy in these achondrites in particular would likely rule out homogeneity of the SLRs.
The volcanic achondrites NWA 7325\index{Meteorites, individual!NWA 7325} and Asuka 881394 also are believed to satisfy the condition for concordancy.
The case is unclear for other samples.

\begin{figure*}[t!]
    \centering
    \includegraphics[width=0.7\textwidth]{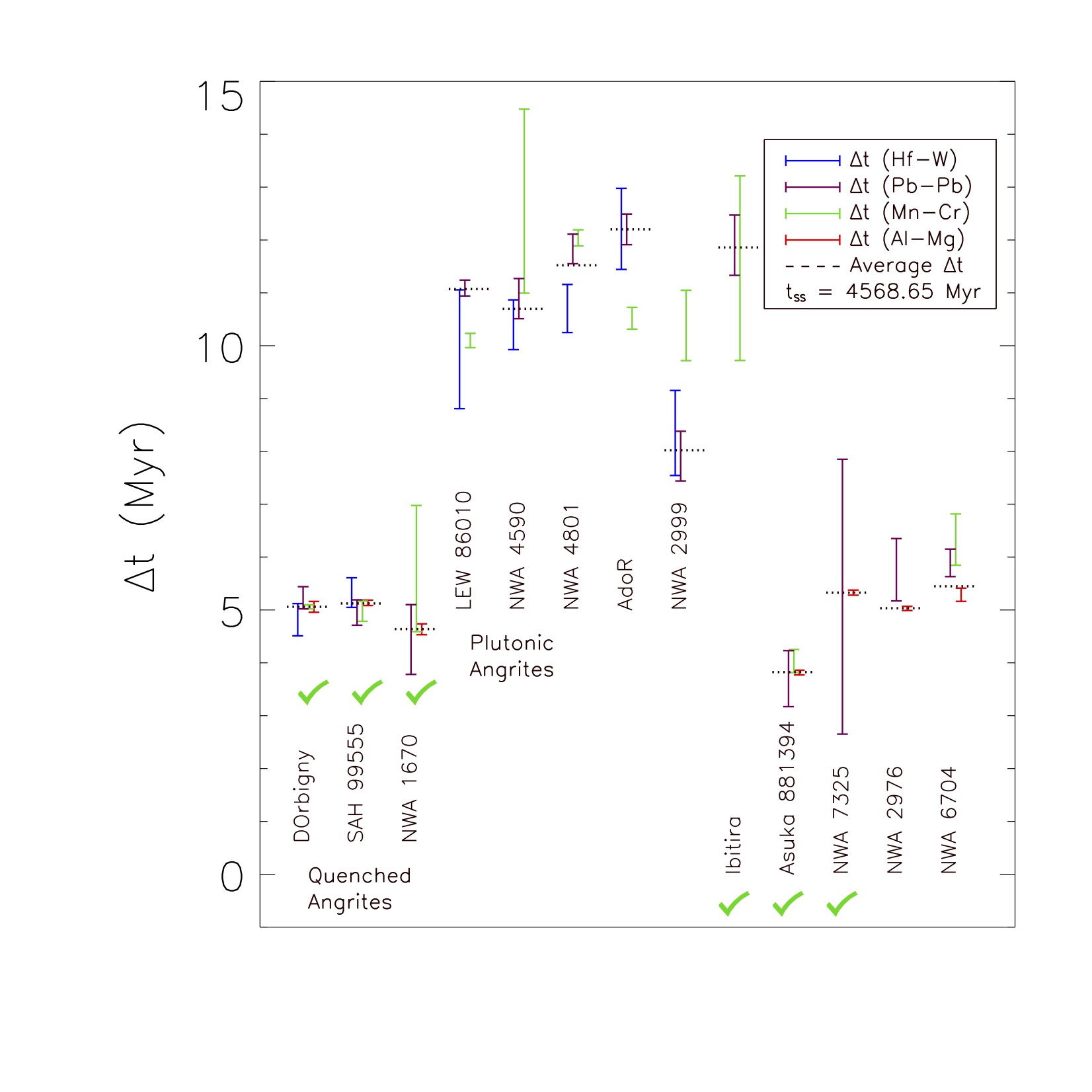}
    \caption{
    Times of formation $\Delta t_{26}$, $\Delta t_{53}$, $\Delta t_{182}$, and $t_{\rm Pb}$, plus the weighted mean, $\Delta t$, for 13 achondrites, as determined by \citet{DeschEtal2022b}, using compiled data and their best-fit values for $(\mnratio)_{\rm SS}$, $(\hfratio)_{\rm SS}$, and $t_{\rm Pb,CAI}$. Checkmarks denote concordancy among all systems, reconciling 18 ages among six achondrites with a statistically significant $\chi_{\nu}^{2} = 1.45$. The Hf-W and Pb-Pb systems, but not the Mn-Cr system, are concordant in the plutonic angrites. The carbonaceous achondrites NWA 2976 and NWA 6704 are not concordant.  The ability to reconcile all the formation times in the volcanic achondrites that should be concordant  is strong evidence that $\altwosix$, ${}^{53}{\rm Mn}$, and ${}^{182}{\rm Hf}$ were homogeneously distributed at early times in the disk. From \citet{DeschEtal2022b}.
    }
    \label{fig:concordancy}
\end{figure*}

The essence of the approach of \citet{DeschEtal2022b} was to determine if there were single values of $(\mnratio)_{\rm SS}$, $(\hfratio)_{\rm SS}$ and $t_{\rm SS}$ that can lead to concordancy, especially in the volcanic achondrites and to constrain those values.
Obtaining the $(\mnratio)_{\rm SS}$ value entailed the additional difficulty that the half-life of ${}^{53}{\rm Mn}$ is poorly known \citep[$3.7 \pm 0.37 (1\sigma)$ Myr][]
{HondaImamura1971}.
\citet{DeschEtal2022b} allowed these four quantities to be free parameters, then determined what values would minimize the discrepancies between $\Delta t_{26}$, $\Delta t_{53}$, $\Delta t_{182}$ and $\Delta t_{\rm Pb}$ in each achondrite.
They defined a goodness-of-fit metric $\chi_{\nu}^{2}$ that included information about the uncertainties in each formation time, which arise from  measurement uncertainties from the isochrons.
Minimizing $\chi_{\nu}^{2}$ for the volcanic achondrite data, they constrained key quantities.
The minimization favored a long half-life for $\mnfivethree$ ($\approx 4.7$ Myr), but experimental measurements favor one closer to 3.7 Myr; the range of values consistent with both achondrite concordancy and measurements was found to be $3.98 \pm 0.22$ Myr. 
Based on this, \citet{DeschEtal2022b} favored the following best-fit parameters and their ($2\sigma$) uncertainties:\index{Radionuclides!Manganese-53}\index{Radionuclides!Hafnium-182}
\[
(\mnratio)_{\rm SS} = (7.80 \pm 0.36) \times 10^{-6},
\]
\[
\tau_{53} = (3.98 \pm 0.22) / (\ln 2) \, {\rm Myr},
\]
\[
(\hfratio)_{\rm SS} = (10.41 \pm 0.12) \times 10^{-5},
\]
and
\[
t_{\rm SS} = 4568.65 \pm 0.10 \, {\rm Myr}.
\]
Using these optimal values, the formation times $\Delta t_{26}$, $\Delta t_{53}$, $\Delta t_{182}$ and $t_{\rm Pb}$ could be calculated for each achondrite and concordancy assessed.
It was found that all derived formation times were concordant  (differed by $< 2\sigma$) for all three quenched angrites considered (D'Orbigny\index{Meteorites, individual!D'Orbigny}, Sahara 99555, and NWA 1670) as well as for NWA 7325\index{Meteorites, individual!NWA 7325}, Asuka 881394, and Ibitira, representing 18 formation times for six achondrites that have been reconciled, with a statistically significant $\chi_{\nu}^{2} = 1.46$.
The plutonic angrites were found to not be concordant, but almost entirely because the Mn-Cr system appeared to close separately from the Hf-W and U-Pb systems; except for NWA 4801, the Hf-W and Pb-Pb ages were concordant in the plutonic angrites using the same parameters above.
The two `carbonaceous achondrites' NWA 2976 and NWA 6704 were not concordant, possibly because their oxidized chemical composition drove isotopic closure\index{isotopic closure} of the Pb-Pb system to lower temperatures.
%\citep{SanbornEtal2019} formed by melting and recrystallization of more volatile-rich materials, which could lead to different diffusion rates of isotopes and/or prolong the cooling time until isotopic closure is achieved.
These results, displayed graphically in {\bf Figure~\ref{fig:concordancy}} \citep[from][] {DeschEtal2022b}, give substantial support for homogeneity of these SLRs in the early Solar System.

%---------------------------
% FIGURE 3 
%------------------------------------

One of the most surprising results of this analysis is that the Pb-Pb age of CAIs is predicted to be $4568.65 \pm 0.10$ Myr, if the Pb-Pb system in them closed at $t\!\!=\!\!0$.\index{Solar nebula!age}
There have been only four U-corrected Pb-Pb ages reported for CAIs in the refereed literature, and they are all younger than this:
$4567.18 \pm 0.50$ Myr for Allende\index{Meteorites, individual!Allende} CAI\index{Calcium-rich, aluminum-rich inclusions (CAIs)} \textit{SJ101} \citep{AmelinEtal2010}; and
$4567.35 \pm 0.28$ Myr, $4567.23 \pm 0.29$ Myr, and $4567.38 \pm 0.31$ Myr for the three Efremovka
CAIs \textit{22E}, \textit{31E}, and \textit{32E} \citep{ConnellyEtal2012}.
In addition, \citet{BouvierEtal2011} found $4567.94 \pm 0.31$ Myr for CAI \textit{B4} from NWA 6999, and  \citet{BouvierWadhwa2010} found $4568.22 \pm 0.18$ Myr for CAI \textit{B1} from NWA 2364.
The result for \textit{B4} was not published in the refereed literature, and a proxy method was used to derive the U isotope ratios for \textit{B1}.
The only way to reconcile these ages with the value of $t_{\rm SS}$ found by \citet{DeschEtal2022a,DeschEtal2022b} is if the Pb-Pb system achieved isotopic closure\index{isotopic closure} some $\sim \!\!1$ Myr after closure in the Al-Mg system was achieved at $t\!\!=\!\!0$.
\citet{BouvierWadhwa2010} suggested that some perhaps Pb-Pb ages may have been reset.
\citet{DeschEtal2022a} calculated that transient heating events like those experienced by chondrules and type B CAIs (peak temperatures $\approx 1700$ K, cooling rates $\sim 500 \, {\rm K} \, {\rm hr}^{-1}$) would indeed lead to resetting of the Pb-Pb system in pyroxenes, but would not reset Al-Mg system in unmelted pyroxene and spinel grains, which would continue to record the original $(\alratio)$ ratio from $t\!\!=\!\!0$. 
If this interpretation is correct, all the CAIs studied so far were heated while free-floating in the protoplanetary disk, and all achieved closure of the Pb-Pb system at about 1 Myr after $t\!\!=\!\!0$.
If the Pb-Pb system was reset in CAIs $\sim 1$ Myr after the Al-Mg system closed, as the data from achondrites suggests, then this would potentially resolve many of the apparent paradoxes that have led to interpretations of $\altwosix$ being heterogeneous in the solar nebula (\S~\ref{sec:lateinjection}). 
As discussed below, chronometry also allows an assessment of the homogeneity of other SLRs, especially ${}^{129}{\rm I}$ and $\fesixty$.

\subsection{\textbf{Abundances of other Radionuclides}}

Establishing the initial Solar System abundances of ${}^{53}{\rm Mn}$ and  ${}^{182}{\rm Hf}$ %$(\mnratio)_{\rm SS}$ and $(\hfratio)_{\rm SS}$ 
and the Pb-Pb age of $t\!\!=\!\!0$ also allowed \citet{DeschEtal2022b} to assess the heterogeneity of other SLRs.
%, besides $\altwosix$, $\mnfivethree$ and $\hfoneeighttwo$.
Below we review the data for the well-established SLRs ${}^{129}{\rm I}$, $\fesixty$, and ${}^{36}{\rm Cl}$.
We do not discuss the homogeneity of other SLRs, because they are not measured precisely in as many objects, so it is more difficult to tell whether they are discordant or not.
We close with a discussion of the controversial SLRs ${}^{135}{\rm Cs}$, and ${}^{7}{\rm Be}$. 

\subsubsection{Iron-60}

The SLR\index{Radionuclides!Iron-60} $\fesixty$, if abundant at a high level $(\feratio)_{\rm SS} \sim 10^{-6}$, would demand an origin in a nearby supernova, possibly injected\index{Late injection} late into protoplanetary disk.
While the one-time existence of $\fesixty$ is established, the controversy over its abundance has only recently been settled.
\citet{TachibanaHuss2003} reported $\feratio = 1-2 \times 10^{-7}$ based on internal isochrons of ${}^{60}{\rm Ni} / {}^{61}{\rm Ni}$ vs.\ ${}^{56}{\rm Fe} / {}^{61}{\rm Ni}$ for troilite (FeS) grains in the Chainpur and Krymka ordinary chondrites.
These data were obtained using Secondary Ion Mass Spectrometry (SIMS), involving direct measurements of analysis spots in the sample.
After extrapolating backward in time using other radiometric dating, \citet{TachibanaHuss2003} inferred a lower limit $\feratio > 5 \times 10^{-7}$ at $t\!\!=\!\!0$.
SIMS analyses were subsequently conducted on chondrules, metal grains, etc., by a large number of authors.
An analysis of a CAI yielded $\feratio = (4.2 \pm 1.3) \times 10^{-7}$ \citep{QuitteEtal2007}, and extrapolated to $t\!\!=\!\!0$ yielded a Solar System initial  $\feratio \approx 1.8 \times 10^{-6}$.
But this initially clear picture of $(\feratio)_{\rm SS} \sim 10^{-6}$ has been overturned by several developments.
First, the half-life was revised from 1.5 Myr to 2.62 Myr \citep{RugelEtal2009}.
Second, it was found that previously common data reduction techniques, when combined with the low count rates associated with SIMS analyses, had led to overestimates of the $\feratio$ ratios \citep{OglioreEtal2011,TelusEtal2012}.
Third, it has also been established that open-system mobilization of Fe and Ni isotopes is pervasive \citep{TelusEtal2018}.
\citet{TelusEtal2018} argued that persistent excesses in ${}^{60}{\rm Ni}$ indicate $(\feratio)_{\rm SS} \approx (0.85 - 5.1) \times 10^{-7}$, but without an undisturbed isochron or a model for the disturbance, this number cannot be used with confidence.
Fourth, and most importantly, very sensitive Resonance Ionization Mass Spectrometry (RIMS) measurements on chondrules have revealed instrumental isotopic fractionation effects that arise during SIMS analyses, negating the excesses seen in earlier SIMS analyses \citep{TrappitschEtal2018}.
More recently, \citet{CookEtal2021} argued for $(\feratio)_{\rm SS} \approx (6.4 \pm 2.0) \times 10^{-7}$, based on ${}^{60}{\rm Ni}$ excesses in iron meteorites rather than internal isochrons. 
This result is based entirely on the assumption that the iron meteorite parent bodies had initial compositions like CI carbonaceous chondrites; if they were more like CV carbonaceous chondrites (much more likely, based on their similar ${}^{62}{\rm Ni}$ excesses), then $(\feratio)_{\rm SS} < 10^{-8}$ would be inferred instead.   

Meanwhile, the other technique for analyzing inclusions has been wet chemistry separation of Ni followed by multi-collector inductively coupled plasma mass spectrometry (MC-ICP-MS).
This is applied to mineral or size separates from bulk achondrites or to individual chondrules within a chondrite to produce data that may array along an isochron.
\citet{TangDauphas2012} and \citet{TangDauphas2015} analyzed Semarkona chondrules\index{chondrules}, magnetic/size separates from the ordinary chondrite NWA 5717, the quenched angrites D'Orbigny\index{Meteorites, individual!D'Orbigny} and SAH 99555, among other samples.
From those samples dated with other systems, they inferred $(\feratio)_{\rm SS} = (1.01 \pm 0.27) \times 10^{-8}$.
Likewise, \cite{QuitteEtal2010} used MC-ICP-MS to obtain $(\feratio)_0$ for D'orbigny and SAH 99555, and \citet{SpivakBirndorfEtal2011} for D'Orbigny.

The MC-ICP-MS analysis technique also is vulnerable to disturbance, as open-system behaviors can also affect Fe and Ni isotope ratios; but \citet{TangDauphas2015} argued that their samples were not disturbed, and the argument that a very wide array of planetary materials appear consistent with $(\feratio)_{\rm SS} \sim 10^{-8}$ is compelling.
This strongly supports the homogeneity of $\fesixty$, as concluded also by \citet{TangDauphas2012}.
\citet{DeschEtal2022b} synthesized information pertaining to the isotopic closure of the angrite and eucrite parent bodies, to derive $({}^{60}{\rm Fe} / {}^{56}{\rm Fe})_{\rm SS} = (0.92 \pm 0.14) \times 10^{-8}$.

\subsubsection{Iodine-129}

The I-Xe system\index{Radionuclides!Iodine-129} is an often-used chronometer, but only in limited cases, as retention of radiogenic ${}^{129}{\rm Xe}$ is sensitive to the temperature and the specific mineral.
Different phases even within the same meteorite might record different events: either the initial abundance of ${}^{129}{\rm I}$ at formation, or the amount when the meteorite was shocked by an impact on the parent body.
There are very few minerals, in which I-Xe systematics have been measured, for which U-corrected Pb-Pb ages also have been obtained; and practically no examples where Al-Mg, Mn-Cr, or Hf-W systematics have been measured.
A quirk of the I-Xe system is that excesses of ${}^{129}{\rm Xe}$ are not measured against I/Xe ratios, but instead different minerals are irradiated by neutrons, transmuting ${}^{127}{\rm I}$ to radioactive ${}^{128}{\rm I}$, which decays in hours to ${}^{128}{\rm Xe}$. 
Then Xe is driven out of the mineral and analyzed, and an isochron of ${}^{129}{\rm Xe}/{}^{132}{\rm Xe}$ vs.\ ${}^{128}{\rm Xe}/{}^{132}{\rm Xe}$ is constructed. 
This approach allows times of isotopic closure of the I-Xe system to be determined relative to an anchor, almost always the enstatite achondrite Shallowater\index{Meteorites, individual!Shallowater}; but it is not as easy to derive the ${}^{129}{\rm I}/{}^{127}{\rm I}$ ratio itself.

I-Xe chronometry allows a test of the homogeneity of ${}^{129}{\rm I}$. 
\citet{PravdivtsevaEtal2017} and \citet{GilmourCrowther2017} both took averages of various objects for which both I-Xe and Pb-Pb ages were measured, to derive Pb-Pb ages of Shallowater of $4562.4 \pm 0.2$ Myr (n=8) and $4562.7 \pm 0.3$ Myr (n=11), respectively.
%The latter age was obtained using a greater number of samples.
%(ED comment) idk if this is useful, but instead of saying "latter ages...greater number" you could write the number of samples studied after each age like "4562.4±0.2 (n=10)"
% thanks!
Adopting the latter, and assuming a Pb-Pb age of $t\!\!=\!\!0$ of $4568.65 \pm 0.10$ Myr, it would appear that Shallowater\index{Meteorites, individual!Shallowater} formed $5.9 \pm 0.3$ Myr after $t\!\!=\!\!0$.
The Pb-Pb and I-Xe ages of the samples employed in this fit (e.g., Ibitira, NWA 7325\index{Meteorites, individual!NWA 7325}, Kernouve phosphate, etc.) appear concordant, but an independent test of the concordancy can be made by applying the result to other systems.
For example, chondrules\index{chondrules} in the CB chondrite Gujba were found to have concordant formation times according to the Pb-Pb, Mn-Cr, and Hf-W systems, $\Delta t_{\rm Pb} = 6.14 \pm 0.22$ Myr 
after $t=0$, or an absolute age of $4562.51 \pm 0.24$ Myr \citep[][and references therein]{DeschEtal2022b}.
The I-Xe systematics of chondrules from the CB chondrite Hammadah al Hamra 237 (which is believed to date the same event) were found to imply a formation time $0.29 \pm 0.16$ after Shallowater \citep{PravdivtsevaEtal2017}, implying a formation time $6.24 \pm 0.27$ Myr after $t\!\!=\!\!0$.
This is compatible at the $< 0.6 \sigma$ level with the formation time inferred from the other systems, and $\ionetwonine$ appears to have been homogeneous.

As for the actual abundance of ${}^{129}{\rm I}$, \citet{PravdivtsevaEtal2021} recently were able to calibrate the neutron fluence and infer ${}^{129}{\rm I}/{}^{127}{\rm I} = 1.35 \times 10^{-4}$ at the time of Shallowater's formation.
Extrapolating backward $5.9 \pm 0.3$ Myr, \citet{DeschEtal2022b} found $({}^{129}{\rm I}/{}^{127}{\rm I})_{\rm SS} \approx 1.74 \times 10^{-4}$.

\subsubsection{Chlorine-36}

The SLR\index{Radionuclides!Chlorine-36} with the strongest evidence for not being uniformly distributed in the nebula at $t\!\!=\!\!0$ is $\clthirtysix$.
This isotope decays with a half-life of $0.30$ Myr (mean life $\tau_{36} = 0.43$ Myr) to ${}^{36}{\rm S}$, with branching ratio 1.9\% (and to ${}^{36}{\rm Ar}$ with branching ratio 98.1\%).
Because Cl-rich minerals are required for this SLR to be measured, evidence for ${}^{36}{\rm Cl}$ comes from sodalite [${\rm Na}_{8}({\rm Al}_{6}{\rm Si}_{6}{\rm O}_{24}){\rm Cl}_{2}$] or wadalite [${\rm Ca}_{6}{\rm Al}_{5}{\rm Si}_{2}{\rm O}_{16}{\rm Cl}_{3}$], minerals that are formed by aqueous alteration of CAIs by Cl-rich fluids while on the parent body.
Isochrons of ${}^{36}{\rm S}/{}^{34}{\rm S}$ vs.\ ${}^{35}{\rm Cl}/{}^{34}{\rm S}$ in these minerals
leave little doubt that live $\clthirtysix$ existed at the time of alteration, with inferred values
$({}^{36}{\rm Cl}/{}^{35}{\rm Cl})_0$ in the range $\sim 2 \times 10^{-6}$ to $\sim 2 \times 10^{-5}$ \citep{LinEtal2005, HsuEtal2006, UshikuboEtal2007, NakashimaEtal2008, JacobsenEtal2011, TangEtal2017}.
% (Lin et al.\ 2005; Hsu et al.\ 2006; Ushikubo et al.\ 2007; Nakashima et al.\ 2008; Jacobsen et al.\ 2011; Tang et al.\ 2017).
It is not clear, however, what the value of ${}^{36}{\rm Cl}/{}^{35}{\rm Cl}$ was at $t\!\!=\!\!0$.
The sodalite and wadalite in which ${}^{36}{\rm Cl}/{}^{35}{\rm Cl}$ is inferred also admit measurements
of ${}^{26}{\rm Al}/{}^{27}{\rm Al}$, from which the time of aqueous alteration, $\Delta t_{26}$, can be fixed.
Most of these record a time of Al-Mg isotopic closure at several Myr, consistent with the consensus that parent bodies like Allende\index{Meteorites, individual!Allende} took 3$-$4 Myr to accrete \citep{DeschEtal2018}, a prerequisite for parent-body processes).
%(ED comment) perhaps reference Desch et al. 2018 for the 3-4 Myr of Allende accretion?
% twist my arm

One exception is the Allende\index{Meteorites, individual!Allende} CAI ``\textit{Curious Marie}," which apparently records $({}^{26}{\rm Al}/{}^{27}{\rm Al})_0 = (6.2 \pm 0.9) \times 10^{-5}$, implying formation no later than 0.05 Myr after $t\!\!=\!\!0$.
If true, a complicated history for this CAI is required to allow it to be aqueously altered on an icy body, then incorporated later into a different parent body \citep{TangEtal2017}.
If we assume that the Cl-S system  closed at the same time as the alteration and closure of the Al-Mg system, then the value $({}^{36}{\rm Cl}/{}^{35}{\rm Cl})_{\rm SS}$
$= ({}^{36}{\rm Cl}/{}^{35}{\rm Cl})_0 \, \exp( +\Delta t_{26} / \tau_{36})$ at $t\!\!=\!\!0$ that would yield the measured value can be inferred.
For the CAI \textit{Curious Marie}, for which $({}^{36}{\rm Cl}/{}^{35}{\rm Cl})_0 = (2.3 \pm 0.6) \times 10^{-5}$, a Solar System level $({}^{36}{\rm Cl}/{}^{35}{\rm Cl})_{\rm SS} = (1.7 - 3) \times 10^{-5}$ \citep{LugaroEtal2018} is implied, demanding that ${}^{36}{\rm Cl}$ was present at this level in the solar nebula at $t\!\!=\!\!0$.
This amount of $\clthirtysix$ can be inherited from the molecular cloud (\S~\ref{sec:irradiation}). 
For the other five CAIs, which appear to have been altered several Myr after $t\!\!=\!\!0$, this exercise implies $({}^{36}{\rm Cl}/{}^{35}{\rm Cl})_{\rm SS}$ $\sim 10^{-2} - 10^{-1}$.
These values are too high to have been plausibly inherited \citep{JacobsenEtal2011}, suggesting a separate, late addition of $\clthirtysix$ in these inclusions. 

It is currently difficult to distinguish between all possible explanations,  because evidence for $\clthirtysix$ necessarily comes only from inclusions that have been altered and disturbed.
One possibility is that the Al-Mg system in \textit{Curious Marie} was disturbed but it actually was altered at $t\!\!\approx\!\!3$ Myr like the other CAIs with $\clthirtysix$. 
Another possibility is that the Cl-S system is not recording $({}^{36}{\rm Cl}/{}^{35}{\rm Cl})_0$ at the time of aqueous alteration, instead retaining memory of the value from an earlier time; 
it is difficult to envision how this would occur, but it is notable that no other CAIs record $({}^{36}{\rm Cl}/{}^{35}{\rm Cl})_0 > 2.3 \times 10^{-6}$.
A likely possibility appears to be that some $\clthirtysix$, the amount recorded by \textit{Curious Marie}, was inherited from the molecular cloud, but that the rest was created by irradiation in the solar nebula. 
The model of \citet{JacobsenEtal2011} posits that as the surface density of the protoplanetary disk decreased during the latter stages of its evolution, SEPs induced nuclear reactions [e.g., ${}^{34}{\rm S}({}^{3}{\rm He},p){}^{36}{\rm Cl}$] that created $\clthirtysix$ in the gas phase. 
The $\clthirtysix$ would have condensed as HCl, been incorporated into ices accreted by asteroids\index{asteroids}, and then would have participated in aqueous alteration on the parent body.
These ideas are discussed further in \S~\ref{sec:irradiation}.

\subsubsection{Cesium-135}

For completeness, we discuss two other SLRs whose existence is controversial.
The SLR\index{Radionuclides!Cesium-135} ${}^{135}{\rm Cs}$ ($t_{1/2} = 2.3$ Myr)
may have existed in the solar nebula, but its presence is difficult to confirm
because while the daughter product, ${}^{135}{\rm Ba}$, is refractory, Cs itself is relatively volatile
and would not be incorporated into CAIs. Therefore the one-time presence of ${}^{135}{\rm Cs}$ must
be inferred from a deficit of ${}^{135}{\rm Ba}$ in CAIs, relative to the bulk composition of meteorites.
Previous analyses have suggested its presence, at levels
$({}^{135}{\rm Cs}/{}^{133}{\rm Cs})_0 \approx 1.4 \times 10^{-4}$ \citep{McCullochWasserburg1978}; 
$({}^{135}{\rm Cs}/{}^{133}{\rm Cs})_0 \approx 6.8 \times 10^{-4}$ \citep{HidakaYoneda2013};
and an estimate
$({}^{135}{\rm Cs}/{}^{133}{\rm Cs})_0 \approx 2.8 \times 10^{-4}$ \citep{BerminghamEtal2014}.
On the other hand, other analyses have only placed upper limits, sometimes strict limits:
$({}^{135}{\rm Cs}/{}^{133}{\rm Cs})_0 < 10^{-5}$ (Ranen 2006); and most recently and compellingly,
$({}^{135}{\rm Cs}/{}^{133}{\rm Cs})_0 < 2.8 \times 10^{-6}$ \citep{BrenneckaKleine2017}.
The abundance and even presence of ${}^{135}{\rm Cs}$ are not yet firmly established.

\subsubsection{Beryllium-7}
\label{sec:beryllium7}

Likewise, the SLR\index{Radionuclides!Beryllium-7} ${}^{7}{\rm Be}$ has twice been suggested to have existed in the early Solar System.
With a half-life of only $t_{1/2} = 53$ days, it would demand production within the solar nebula,
presumably by SEP irradiation of gas.
\citet{ChaussidonEtal2006} claimed the one-time existence of ${}^{7}{\rm Be}$ in Allende\index{Meteorites, individual!Allende} CAI \textit{3529-41}, from an isochron of ${}^{7}{\rm Li}/{}^{6}{\rm Li}$ vs.\ ${}^{9}{\rm Be}/{}^{6}{\rm Li}$, after correcting for spallogenesis of light isotopes by GCRs.
\citet{DeschOuellette2006} disputed this, demonstrating that the correlation was not linear, that
\citet{ChaussidonEtal2006} had over-corrected for spallogenesis, and that there was evidence for introduction of spallogenic B into the sample \citep[but see the reply by][]{ChaussidonEtal2006b}.
\cite{DeschOuellette2006}  demonstrated that the linear correlation had ${\rm MSWD} = 3.94$, far greater than the limit of 1.48 for $n\!\!=\!\!37$ data points.
They concluded there was no evidence that the CAI incorporated live ${}^{7}{\rm Be}$.

More recently, \cite{MishraMarhas2019} presented similar isochrons for Efremovka CAI \textit{E40}.
Although they found a slope ${}^{7}{\rm Be}/{}^{9}{\rm Be} = (1.2 \pm 1.0) \times 10^{-3}$, marginally resolved from zero, for their isochron of \textit{E40}, they also found ${\rm MSWD} = 1.9$, which exceeds the limit of 1.6 for n = 22  points.
This indicates that the correlation is not linear and, therefore, that this is not a valid isochron. 
As with \textit{3529-41}, it is likely that the isotopes in \textit{E40} have been disturbed, probably by the introduction of spallogenic Li and B into the CAI.
This is corroborated by the high ${\rm MSWD} = 3.1$ in the isochron of ${}^{10}{\rm B}/{}^{11}{\rm B}$ vs.\ ${}^{9}{\rm Be}/{}^{11}{\rm B}$ in \textit{E40}, well above the limit of 2.15 for $n\!\!=\!\!8$, indicating that Be and B isotopes have been disturbed.

To date there is no compelling evidence for the one-time existence of ${}^{7}{\rm Be}$ in CAIs.

\subsection{\textbf{Summary}} 

The picture that has emerged over the last decade is that almost all the SLRs were homogeneously distributed.
The SLR $\beten$ shows no evidence for heterogeneity: all 40 of the CAIs that can be said with confidence to record $(\beratio)_0$ in the solar nebula at $t\!\!=\!\!0$ record a common value $(\beratio)_{\rm SS} \approx (7.1 \pm 0.2) \times 10^{-4}$, with their $(\beratio)_0$ ratios statistically distributed about this mean as expected given their measurement uncertainties. 
Of the nine other samples, the FUN CAIs were very plausibly affected by late thermal resetting and/or evaporation, and the PLACs plausibly record a chemical heterogeneity in the disk.
Likewise, the concordancy of formation times derived from Pb-Pb dating with the formation times derived from the Al-Mg, Mn-Cr, Hf-W and even I-Xe systems demonstrates that $\altwosix$, $\mnfivethree$, $\hfoneeighttwo$, and likely  $\ionetwonine$ and $\fesixty$, were present at $t\!\!=\!\!0$ and were quite homogeneous across the reservoirs where achondrites and CAIs formed.
These findings are consistent with inheritance of all these SLRs from the molecular cloud. 

Irradiation appears to have played a limited role in producing SLRs.
As discussed below (\S 3.2), SEP irradiation in the solar nebula is more likely to produce $\beten$ than any other SLR, leading to larger heterogeneities in $\beratio$ ratios than any other SLR, yet $\beten$ appears homogeneous.
Evidence presented for the one-time existence of ${}^{7}{\rm Be}$, which would demand production by irradiation, instead appears more consistent with disturbance of the Li isotopes in CAIs.
The one exception appears to be $\clthirtysix$. 
Sodalite and wadalite in altered CAIs seem to record $({}^{36}{\rm Cl}/{}^{35}{\rm Cl})_0 \sim 2 \times 10^{-6} - 2 \times 10^{-5}$ in the fluids in chondrite parent bodies at late times $t > 3$ Myr.
An additional component $({}^{36}{\rm Cl}/{}^{35}{\rm Cl}) \sim (1.7 -3) \times 10^{-5}$ at $t\!\!=\!\!0$ might have been inherited from the molecular cloud. 

Astrophysical theories for the origins of the SLRs are considered in more detail below. 

%--------------------------------------------------------------------
%
% SECTION III. ASTROPHYSICAL ORIGINS OF THE SLRs   
%

\section{\textbf{ASTROPHYSICAL ORIGINS OF THE SLRs}}

\bigskip

\subsection{\textbf{Star Formation in the Galaxy}}

%---------------------------------------
% FIGURE 4
% Steve edited this caption on 8/31.
\begin{figure*} [t!]
    \centering
	\includegraphics[width=1.0\textwidth]{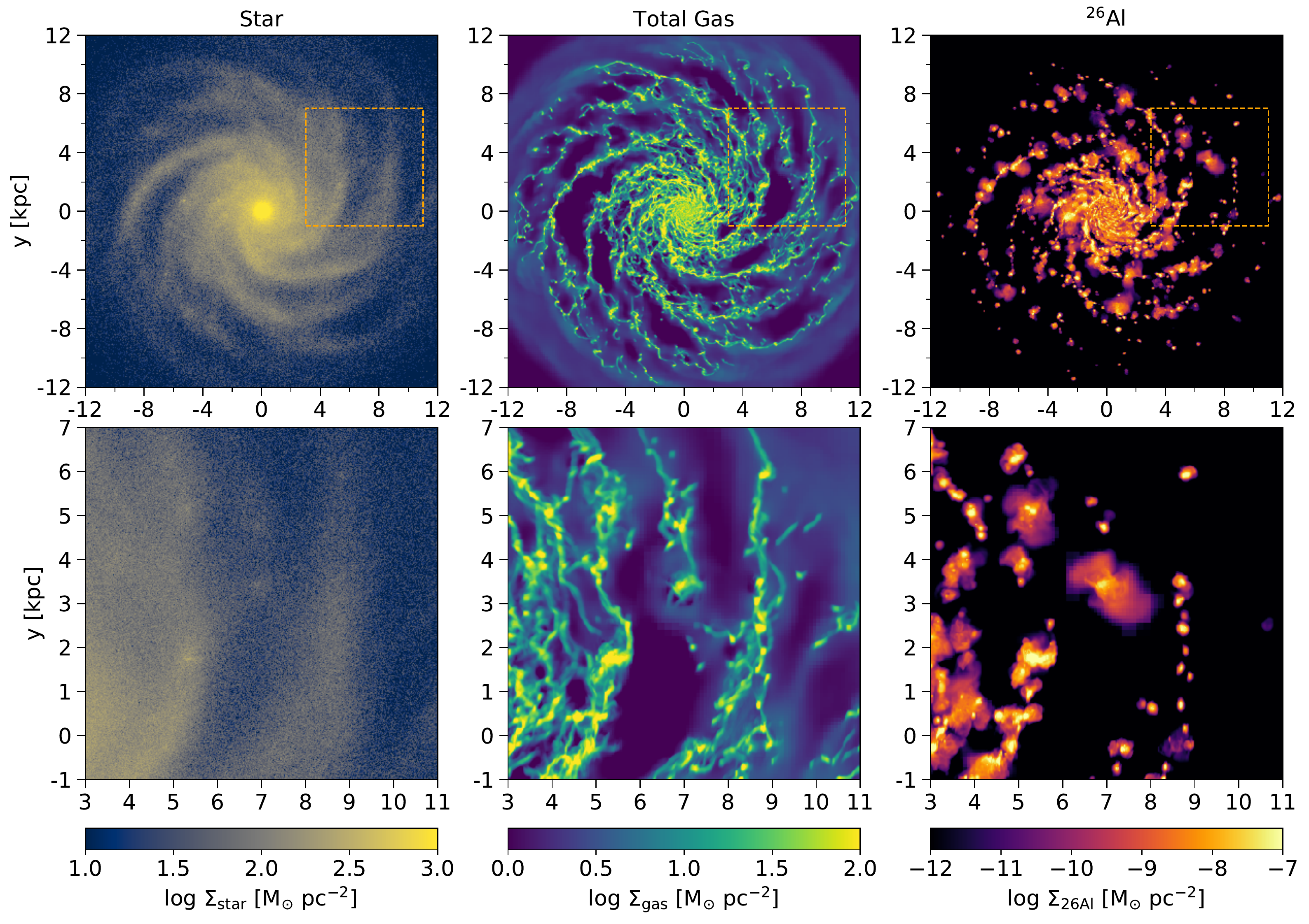}
    \caption{Numerical simulations of the structure of a Milky Way-like galaxy. Panels from left to right show the surface density of stars, total gas, and ${}^{26}{\rm Al}$. The top row shows the entire face-on galactic disk; the bottom row displays zoomed-in images of the region denoted by boxes in the top row centered on a point with galactocentric distance of 8 kpc, current Solar circle in the Milky Way where the Sun would orbit. The galaxy rotates clockwise. From \citet{Fujimoto2020a}.}
    \label{galaxy_simulation}
\end{figure*}
%---------------------------------------

%---------------------------------------
% FIGURE 5
% Steve edited this caption 8/31 
\begin{figure}[h]
    \centering
	\includegraphics[width=0.5\textwidth]{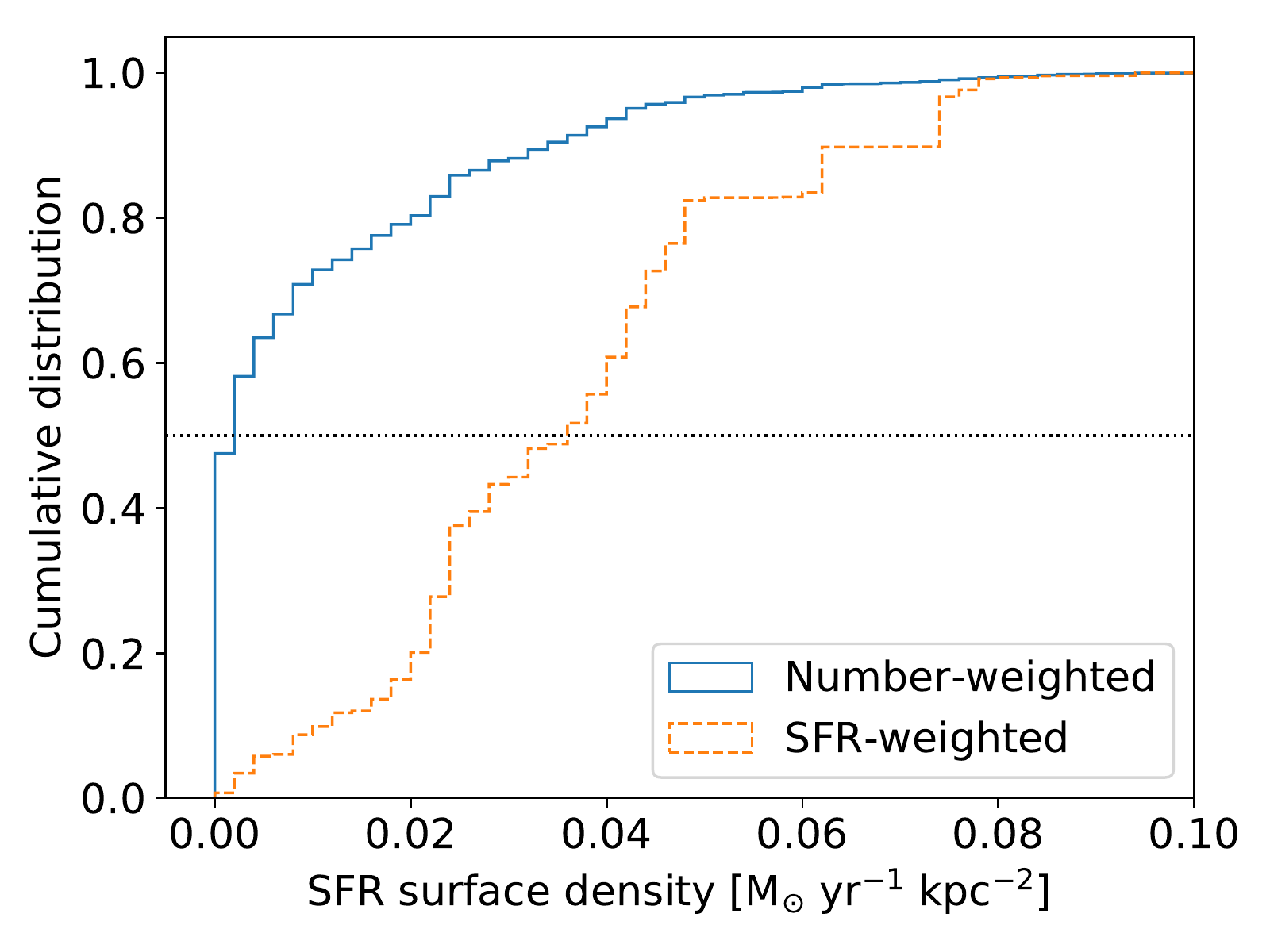}
    \caption{Cumulative distribution functions (CDFs) of the SFR in the simulated galaxy of \citet{Fujimoto2018}. The solid blue curve is weighted by number of {$(1.3 \, {\rm kpc})^2$ area patches in the galactic disk}, and represents the spatial average of the SFR in the {patches}; the median SFR in any area of the galactic disk is $< 0.01 \, M_{\odot} \, {\rm yr}^{-1} \, {\rm kpc}^{-2}$. The dashed orange curve is weighted by number of area patches times SFR, i.e., by {newly} formed stars; the median SFR seen by new stars in the disk is $\approx 0.04 \, M_{\odot} \, {\rm yr}^{-1} \, {\rm kpc}^{-2}$. Newly formed stars see higher SFR than the galaxy-wide average, because they preferentially form in regions of high SFR.}  
    \label{SFR_Histogram}
\end{figure}
%---------------------------------------

There appears to be increasing evidence that SLRs in the solar nebula were the products of ongoing stellar nucleosynthesis in the Sun's local environment, enriching the molecular cloud.
To understand their origins, we review star formation\index{Star formation} in a galactic context. 

Spiral galaxies like the Milky Way evolve in a complicated fashion, with global stellar and gas dynamics (e.g., spiral arms) and local material circulation caused by star formation and its feedbacks---mainly from massive stars in the form of stellar winds, H{\sc ii} regions, and core-collapse supernovae. 
Observations of nearby galaxies have shown that at the galactic scale star formation obeys a power-law relation between the gas surface density and the surface density of the star formation rate (SFR), the so-called Kennicutt-Schmidt (KS) relation\index{Star formation!scaling relations} \citep{Kennicutt1989}; and that this relation holds not only for averages of galaxies but also for kpc-sized patches within spiral galaxies \citep[e.g.,][]{Bigiel2008, Leroy2013}. 
The KS relation shows that star formation in galaxies does not occur in the interstellar medium (ISM\index{Interstellar medium}) at random. 
Instead, star formation is hierarchical and triggered in global-scale potential wells shaped with galactic structures such as spiral arms that consist of old stellar populations. 
A similar conclusion is reached from the correlation between the timescales % (ED: timescales of what?) {\Yusuke of star formation} 
of star formation and the sizes of star-forming regions, extending from kpc-scale spiral arms down to 0.1 pc-scale dense molecular cores \citep[e.g.,][]{EfremovElmegreen1998}. 

Numerical galaxy simulations have succeeded in reproducing these observational features \citep[e.g.,][]{Hopkins2011, Agertz2013, Goldbaum2016}.
\citet{Fujimoto2018} performed hydrodynamical simulations of a Milky Way-like galaxy with a recent standard recipe for star formation and stellar feedback  including stellar winds, H{\sc ii} regions and core-collapse supernovae, as used in the similar galaxy simulations cited above. 
Similar simulations by \citet{Fujimoto2020a} are displayed in {\bf Figure~\ref{galaxy_simulation}}. 
These simulations show excellent agreement with all observables averaged on $>$ kpc scales, including total galactic SFRs, radial profiles of the ISM through the galactic disc, mass ratios of the ISM components, both whole-galaxy and kpc-scale KS relations, and molecular cloud properties such as mass, size, and gravitational boundness.
{\bf Figure~\ref{SFR_Histogram}} shows the cumulative distribution functions (CDFs) of the surface density of the SFR in $(1.3 \, {\rm kpc})^2$ patches of the simulated galaxy of \citet{Fujimoto2018}. 
The spatial resolution of the simulation is 8 pc, and the averaging area to calculate the surface density is $(1.3\ \rm{kpc})^2$. 
The blue line shows the number-weighted CDF, and the orange line shows the SFR-weighted CDF. 
The mean of the SFR-weighted CDF is 3.7 times larger than that of the number-weighted CDF, indicating a contrast between star-forming regions and random regions in the galactic disk.
In other words, the average newly formed star sees a local SFR that is 3.7 times higher than the average rate in the galactic disk overall, because most stars form in regions of high star formation.
This result demonstrates the hierarchical structure of star formation in the Galaxy.

Hierarchical star formation and subsequent stellar feedback play an important role in distributing SLRs in the Galaxy.
\citet{Fujimoto2018} performed chemo-hydro-dynamical simulations of the entire Milky Way Galaxy, including SLR injection\index{Late injection} of $\altwosix$ and $\fesixty$ from massive stars and their transport in the ISM. 
They showed that the distributions of SLRs are more extended, with regions of SLR enrichment typically exceeding 1 kpc in size.
SLRs are associated with massive stars and areas of enhanced star formation.
Star formation in a galaxy is highly correlated in time and space \citep[e.g.,][]{EfremovElmegreen1998,Gouliermis2010}, triggered in kpc-scale large structures such as spiral arms and associated gaseous filaments, 
in multi-generation star-forming cycles lasting for a hundred Myr or more, until the local galactic spiral arm disappears.
Within a spiral arm, the gas density will peak within a large giant molecular cloud\index{Giant molecular clouds} (GMC) complex consisting of multiple molecular clouds and embedded star-forming regions, typically over scales $\sim\!\!100$ pc.
The ejecta from stellar winds and supernovae\index{supernova} are not confined to individual GMCs, however; large SLR-bearing gas bubbles can extend as far as $\sim\!\!1$ kpc from the GMCs.
Because GMCs are not closed boxes, but instead continually accrete surrounding atomic gas throughout their star-forming lives \citep[e.g.,][]{FukuiKawamura2010}, this pre-enriched gas has a high chance of being incorporated into nearby existing GMCs or forming new GMCs along with the galactic spiral arms. 

\citet{Fujimoto2020b} found that the lifetimes of SLR-bearing gas bubbles are comparable to the lifetimes of galactic spiral arms, a result of continuous fueling of gas and star formation by galactic-scale spiral flows.
Their isolated galaxy simulation showed that galactic spiral arms are transient structures that continually form, merge, and shear away, on timescales of $\sim\!\!10^2$ Myr, due to local instabilities in the combined gas-star fluid.
The observed lifetime of the Local Arm in which the Sun currently resides is slightly longer, $\sim 300$ Myr \citep{Kounkel2019}, possibly because the Milky Way's arms are strengthened by companion dwarf galaxies and/or the Galactic bar.
Regardless, the residence time of SLR-enriched gas in a molecular cloud complex ($\tau_{\rm cloud}$ in Equation 13 below) is $\sim 10^2$ Myr.

\subsection{\textbf{Origin of \mbox{\boldmath${}^{10}{\rm Be}$}}}

As discussed in \S 2.1, a subset of 49 (non-FUN) CAIs is thought to faithfully record the initial $(\beratio)$ ratio of the solar nebula at the time of their formation.
These cluster tightly around two values.
There are 9 FUN CAIs and PLACs that are mineralogically distinct from other CAIs; these record $(\beratio)_0 \approx 3-5 \times 10^{-4}$.
These may have been thermally reset or experienced evaporation of B, after about 1 Myr of disk evolution.
The remaining population of 40 ``normal" CAIs are distributed as expected due to their measurement errors (MSWD $= 1.36$) around a mean value $(\beratio)_{\rm SS} \approx 7.1 \times 10^{-4}$.
Among this group there is no evidence for heterogeneity of $\beten$.\index{Radionuclides!Beryllium-10}

This homogeneity strongly suggests inheritance of $\beten$ from the molecular cloud, and argues against irradiation in the solar nebula as a source.
Nevertheless, $(\beratio)_{\rm SS} \approx 7.1 \times 10^{-4}$ exceeds the predictions of many early models for how much $\beten$ could be created in the molecular cloud due to GCR irradiation of CNO nuclei within it \citep{GounelleEtal2001, TatischeffEtal2014}.
This has driven the idea among one camp that most $\beten$ was created in the solar nebula by SEP irradiation \citep{GounelleEtal2001,GounelleEtal2006,WielandtEtal2012, Jacquet2019}.
It has also prompted a second camp to search for alternative interstellar sources of $\beten$, including trapped $\beten$ GCRs \citep{DeschEtal2004}, or neutrino-driven nucleosynthesis during supernovae \citep{BanerjeeEtal2016}. 
More recently it has been proposed that the GCR flux in the Sun's molecular cloud simply may have been greater than expected \citep{DunhamEtal2022}. 
Here we discuss the models driving these different ideas.

%the initial $({}^{10}{\rm Be}/{}^{9}{\rm Be})_0$ ratios of measured CAI regressions range from $\sim 3$ to $\sim 100$ $\times 10^{-4}$ (McKeegan et al., 2000; Sugiura et al., 2001; Marhas et al., 2002; MacPherson et al., 2003; Chaussidon et al., 2006; Liu et al., 2009; Liu et al., 2010; Wielandt et al., 2012; Gounelle et al., 2013; Srinivasan and Chaussidon, 2013; Sossi et al., 2017; Fukuda et al., 2019; Mishra and Marhas, 2019; Fukuda et al., 2021b; Dunham et al. 2020, Dunham et al. submitted), however, upon a closer look, essentially all CAIs that 

\subsubsection{Production of ${}^{10}{\rm Be}$ in the Solar Nebula}

The discovery of $\beten$ was taken as the ``smoking gun" for production of SLRs by SEP irradiation of solar nebula materials \citep{GounelleEtal2001}, because $\beten$ was understood to form only from spallation of (mostly) CNO nuclei by energetic particles, and is not a product of normal stellar nucleosynthesis. 
\citet{GounelleEtal2001} interpreted $\beten$ as having been created by SEP irradiation of rock vapor very close ($< 0.1$ AU) to the early Sun, but this model was challenged on multiple physical grounds by \citet{DeschEtal2010} (see \S~\ref{sec:irradiation}).
Today, production of $\beten$ by irradiation is considered mainly in the context of the gaseous protoplanetary disk.

The only quantitative model for $\beten$ production in the disk is that of \citet{Jacquet2019}. 
The conditions that produce $\beten$ in its measured abundance are not expected to produce the observed amounts of other SLRs (e.g., $\altwosix$), but the irradiation model still needs to be tested for $\beten$.
One robust prediction of the irradiation model is that $(\beratio)$ ratios in the disk would vary greatly, decreasing with heliocentric distance $r$, and increasing with time $t$ (for both irradiation of gas and solids), so that the $\beratio$ ratio recorded by CAIs would scale as $r^{-3/2} \, t^{+1}$. 
The actual value recorded by CAIs depends on when and where they formed; it can be expected that such a location in the disk ($r_{\rm CAI}$), where midplane temperatures are $> 1300$ K and CAI minerals can condense would vary with time.
As $r_{\rm CAI}$ decreases with time, recorded $(\beratio)_0$ ratios easily could vary by an order of magnitude.
Importantly, CAIs formed at $t\!\!=\!\!0$ must diffuse through the protoplanetary disk for $> 3$ Myr before being incorporated into carbonaceous chondrite (CC) parent bodies.
CAIs in other types of chondrites (CV, CR, and CO) should each record a similarly diverse range of initial $(\beratio)_0$ values.

\citet{YangCiesla2012} found that temperatures in an evolving disk exceed $\approx\!\!1300$ K only for $r < r_{1300}(t) \approx 1.4 \, \left( t / 0.3 \, {\rm Myr} \right)^{0.63} \, {\rm AU}$ for $t < 0.3$ Myr, and $r < r_{1300}(t) \approx 1.4 \, \left( t / 0.3 \, {\rm Myr} \right)^{-1} \, {\rm AU}$ for $t > 0.3$ Myr. 
Considering CAIs to form between, say, 0.5 AU and $r_{1300}(t)$, they would record $\beratio$ values that varied by at least a factor of 2.
Specifically, if the mean of all CAIs was $\beratio \approx 7 \times 10^{-4}$, those formed at $t < 0.3$ Myr would record values between 3 and 13 $\times 10^{-4}$, and those formed at $t > 0.3$ Myr would record values between 6 and $21 \times 10^{-4}$. 
Based on this, \citet{DunhamEtal2022} estimated that a fraction $> 50\%$ of CAIs would have $\beratio < 5 \times 10^{-4}$ or $> 10 \times 10^{-4}$, resolvably different from the mean value of $\approx 7 \times 10^{-4}$ for typical $2\sigma$ uncertainties. 
This is in marked contrast to the observed distribution of 40 CAIs that confidently record $(\beratio)_0$ ratios at $t\!\!=\!\!0$; these statistically conform to a single value, with only $< 4\%$, if any, differing from the mean.
Even among the full set of 49 CAIs faithfully recording a $\beratio$ at the time of their formation, $< 20\%$ differ significantly from the mean. 
These findings, in addition to other challenges (\S~\ref{sec:irradiation}) would seem to rule out SEP irradiation as a significant source of $\beten$ in the solar nebula.
Note that the model of \citet{Jacquet2019}, with the interpretation by \citet{DunhamEtal2022}, is the only one for which the spread of values has been quantitatively predicted, but similar variations in $(\beratio)_0$ are a hallmark of irradiation models.

\subsubsection{A Supernova Source for ${}^{10}{\rm Be}$?}

With the difficulties reconciling $\beratio$ ratios with production of $\beten$ within the solar nebula, alternative external contributions to ${}^{10}{\rm Be}$ were explored.  
One such model is that of \citet{BanerjeeEtal2016}: using improvements in the ${}^{12}{\rm C}(\nu,\nu'pp){}^{10}{\rm Be}$ cross section, they calculated the amount of $\beten$ that would be produced by neutrino spallation during a core-collapse supernova.\index{supernova} 
These authors demonstrated that for the plausible dilution factors and time delays,
%associated with getting the meteoritic abundance of ${}^{182}{\rm Hf}$, 
a single supernova from a $< 12 \, M_{\odot}$ progenitor could produce $\beten$ with abundance $\beratio \approx 6 \times 10^{-4}$. 
However, $\fesixty$ would be overproduced in such a model, and it is also not clear how a single supernova taking $> 20$ Myr to explode could contaminate the early Solar System.
Nevertheless, neutrino-driven nucleosynthesis is thought to be a major source of ${}^{7}{\rm Li}$ and ${}^{11}{\rm B}$ in the Galaxy \citep{Prantzos2012}, and the contributions from supernovae to $\beten$ merit further exploration.

A high fraction of the GCRs that are accelerated by supernovae are themselves $\beten$ nuclei (produced by spallation of GCRs that were ions of oxygen and other species). 
\citet{DeschEtal2004} proposed that low-energy ($< 10$ MeV/nucleon) GCRs, including $\beten$ nuclei, would be stopped and trapped as they passed through the Sun's molecular cloud.
Extrapolating the GCR energy spectrum to low energies, they predicted $\beratio \approx 7 \times 10^{-4}$ in the early Solar System. 
Subsequent measurements of the low-energy GCR spectrum \citet{CummingsEtal2013} showed a much steeper dropoff than assumed by \citet{DeschEtal2004}. 
Trapping of $\beten$ GCRs as modeled by \citet{DeschEtal2004} fails to explain the solar nebula $\beten$ abundance by one or two orders of magnitude \citep{TatischeffEtal2014}.

\subsubsection{Production of ${}^{10}{\rm Be}$ in the Molecular Cloud}

Production of $\beten$ by GCR irradiation of CNO nuclei in the Sun's molecular cloud is the remaining viable model, but was discounted as the only source because it was thought to produce $\beten$ at levels no higher than $\beratio \approx 1.3-2 \times 10^{-4}$ \citep{TatischeffEtal2014}. 
This assessment was based on the GCR flux in the Sun's molecular cloud matching the GCR flux at the Sun's location today, multiplied by a factor $\Psi \approx 1.5 - 2.0$ to account for the higher average SFR in the Galaxy 4.57 Gyr ago \citep[e.g.,][]{Ruiz-Lara2020}.

The time variation of the GCR flux is well-understood.
Once accelerated, GCRs tend to diffuse from their point of origin for on average $\sim\!\!15$ Myr before they are thermalized or escape the Galaxy, during which time they travel $\sim\!\!1$ kpc \citep{YanasakEtal2001}. 
Massive stars that will explode as supernovae\index{supernova} generally take a comparable 4 to 50 Myr to evolve.
Therefore, the GCR flux in any location reflects the star formation rate (SFR) averaged over the nearest $\sim\!\!1$ kpc, over the last $\sim\!\!15$ Myr.
As reviewed by \citet{DeschEtal2004} and \citet{TatischeffEtal2014}, the spatially averaged SFR in the Galaxy 4.57 Gyr ago was probably a factor $\Psi \approx 1.5 - 2.0$ higher than today, based on such diverse lines of evidence as ages of G dwarfs and white dwarfs, and galactic chemical evolution models.

However, as discussed above (\S 3.1), the SFR is not spatially homogeneous in the Galaxy, and so the GCR flux 4.57 Gyr ago would not necessarily be exactly $\Psi$ times greater than today's flux near the Sun.
%.
As it happens, the GCR flux near the Sun today appears close to the spatially averaged GCR flux at the Sun's galactocentric radius today.
The Sun is in a minor spiral arm (the so-called Orion Arm, or Local Arm), with a higher SFR than an interarm region, but less than in a major spiral arm like the Perseus or Sagittarius arms in which many prominent star-forming regions exist.
As reviewed by \citet{DeschEtal2004}, a comparison of the measured GCR flux with the inferred GCR ionization rates in nearby molecular clouds suggests the Sun sees a GCR flux very close to the spatial average.
The Sun 4.57 Gyr ago, in contrast, likely did not see a GCR flux matching the spatial average.
As seen from Figure 4, the SFR is highly variable across the Galaxy.  
According to the CDFs displayed in Figure 5, the GCR flux at a random point in a spiral galaxy---proportional to the average SFR within the nearest $(1.3 \, {\rm kpc})^2$---is smaller than the average GCR flux sampled by a newly formed star, by a factor of 3.7.
A range of lower and higher SFRs is sampled by new stars with varying probabilities.
New stars see greater GCR fluxes because they form in spiral arms\index{Spiral arms} nearer to supernovae.\index{supernova}
The Sun most likely saw a SFR and GCR flux a factor of 3.7$\times$ higher than the Galaxy-wide average 4.57 Gyr ago.\index{Star formation!history}  

Multiplying all the factors determined above, \citet{DunhamEtal2022} found that the most likely value of $\beratio$ in the Sun's molecular cloud due to GCR irradiation was 
$(0.9 \times 10^{-4}) \times$ $\Psi$ $\times 3.7$ $\approx (5.0 - 6.6) \times 10^{-4}$, with lower (or higher) values possible depending on whether the Sun formed outside of a spiral arm or in a region of especially high SFR. 
From the cumulative distributions (Figure 5), they calculated a 20\% probability that a new star will inherit $^{10}$Be/$^{9}$Be $\geq$ 7 $\times$ 10$^{-4}$ if $\Psi$ = 1.5, and a 40\% probability of $^{10}$Be/$^{9}$Be $\geq$ 7 $\times$ 10$^{-4}$ if $\Psi$ = 2.0.
Thus, the apparent initial value $(\beratio)_{\rm SS} = 7.0 \times 10^{-4}$ is a highly probable value for a star born in a spiral arm, i.e., a region of high SFR, 4.57 Gyr ago. 

\subsection{\textbf{Origins of the other Radionuclides}}

The evidence from concordancy of dates found using the Al-Mg, Mn-Cr, Hf-W and I-Xe systems along with Pb-Pb ages (\S 2.2) strongly suggests that the SLRs $\altwosix$, $\mnfivethree$, $\hfoneeighttwo$, and $\ionetwonine$ were homogeneously distributed at $t\!\!=\!\!0$ and inherited from the Sun's molecular cloud. 
Only $\clthirtysix$ shows evidence for an irradiation origin, although some $\clthirtysix$ also may have been inherited from the molecular cloud. 
These findings raise the question of how these SLRs were incorporated into the Sun's molecular cloud in the first place. 
The SLR $\beten$ probably formed there by spallation of cloud material, but the rest must have arisen from stellar nucleosynthesis. 
What were their stellar sources? 

Models for how radionuclides produced in stars could enter the Sun's molecular cloud fall into one of two broad categories \citep{Young2014, Young2018}. 
One category emphasizes the ``granular" nature of delivery of nuclides \citep{Wasserburg96}, especially for the SLRs. 
In these models, referred to as ``punctuated" delivery by \cite{Young2018}, specific nuclides can be traced to individual events near the Sun's parent molecular cloud.  
These events could include encounters between the parental cloud and core-collapse supernovae\index{supernova}, AGB stars, WR stars\index{Wolf-Rayet stars}, or kilonovae (merger of two neutron stars or a neutron star and a black hole)  \citep[e.g.,][]{Wasserburg96, Lugaro2014, Dwarkadas2017,Bartos2019}.
The likelihood of these scenarios is tied to the probability of the close encounter considered. 

The second category is based on the concept of regional enrichment of the SLRs, combined with a persistent background of the longer-lived radionuclides.
The enrichment of molecular clouds by SLRs is due to the spatial and temporal correlation of star formation on kpc scales over timescales of $10^8$ yr \citep{Elmegreen2007, Elmegreen2010}.
While individual molecular clouds have lifetimes of $\sim 10$ Myr \citep{lada03}, giant molecular cloud complexes persist for hundreds of Myr.
Below we discuss the ability of both classes of models to explain the SLR abundances.

To quantify this ability, models predict the parameter $\alpha(R) = (N_{\rm R}/N_{\rm S}) / (P_{\rm R}/P_{\rm S})$, where $N_{\rm R}$ and $N_{\rm S}$ are the number abundances of a radionuclide R and its stable reference nuclide S, and $P_{\rm R}$ and $P_{\rm S}$ are their stellar production rates.
Dating back at least to the study by \cite{Schramm70}, it was recognized that the ratio $N_{\rm R}/P_{\rm R}$ is the most meaningful parameter for interpreting the significance of the relative abundances of the radionuclides, and by dividing abundances by their production rates, the effect of variations in production rates among the nuclides is accounted for.  
Similarly, the influence of differences in the chemistry of nuclides is accounted for by dividing the radionuclide abundance by that for a stable reference nuclide S, preferably a stable isotope of the same element.  
For these reasons, the parameter $\alpha({\rm R})$ has been adopted as the best quantity to compare relationships between the radionuclides \citep[e.g.,][]{Wasserburg2006, Huss2009, Jacobsen2005, Young2014}.

\subsubsection{Punctuated Delivery of SLRs} 

In the case of ``granular" scenarios involving accumulation of products from individual nucleosynthesis events, the process can be described by an equation for the sum of individual nucleosynthesis events with an average temporal spacing $\delta t$, followed by a final event with a well-defined actual free decay time $\Delta t$  \citep{Wasserburg2006,Lugaro2014, Young2016}.
After a time $T$, the age of the Galaxy at the time of the Solar System's birth, the abundance of a stable nuclide is $N_{\rm S} = P_{\rm S} \, T$, while the abundance of the radionuclide immediately after the last event is found from the sum of a very large number of events:
\begin{equation}
N_{\rm R} = \sum_{j=0}^{\infty} \, P_{\rm R} \, \delta t \, e^{-j \delta t / \tau_{\rm R}} = P_{\rm R} \, \delta t \, \frac{ 1 }{ 1 - e^{-\delta t / \tau_{\rm R}} },
\label{eqn:sum}
\end{equation}
where $\tau_{\rm R}$ is the mean-life of radionuclide R.  The ratio of R and S is therefore

\begin{equation}
 \frac{N_{\rm R}}{N_{\rm S}} = K \frac{P_{\rm R}}{P_{\rm S}}\frac{\delta t }{T}\frac{1}{ 1 - e^{-\delta t / \tau_{\rm R}} }.
\label{eqn:granular_ratio}
\end{equation}

\noindent Here we have included the parameter $K = (k+1)$ introduced by \cite{Clayton1985b} to account for the effects of addition of low-metallicity gas on the galactic chemical evolution (GCE) of the Galaxy.  The GCE effects on $N_{\rm R}$ and $N_{S}$ are different due to their disparate lifetimes in the ISM.  These effects include preferential dilution of the stable nuclides due to infall of low-metallicity gas, and an enhancement in astration of the stable nuclides relative to the radionuclides caused by episodes of rapid star formation triggered by infall.  Both factors should lead to an apparent increase in the effective, time-integrated production ratios $P_{\rm R}/P_{\rm S}$.  This enhancement is accommodated using $K P_{\rm R}/P_{\rm S}$ with $K > 1$.  Estimates for  $K$ values range from 1 to $\sim 3$. A recent study found a value for $K$ of $2.3\substack{+3.6 \\ -0.7}$ \citep{Cote2019}. In most cases  $N_{\rm R}$ and $N_{S}$ both refer to secondary nuclides, but in those cases where $N_{\rm R}$ and $N_{S}$ refers to a primary and a secondary nuclide, different values of $K$ may apply \citep{Huss2009}.  Therefore, uncertainties in the values for $K$ translate into uncertainties in the effective production ratios of order $\sim 2$. Including the GCE effects embodied by $K$ leads to a refinement in the definition of $\alpha$ such that  $\alpha(R) = (N_{\rm R}/N_{\rm S}) / (K(P_{\rm R}/P_{\rm S}))$.

Including a time of isolation and decay of $\Delta t$ prior to Solar System formation results in the
final result for punctuated delivery: 
\begin{equation}
\alpha(R) = \left( \frac{ \delta t / T }{ 1 - e^{-\delta t / \tau_{\rm R}} } \right) \, \exp \left( -\Delta t / \tau_{\rm R} \right).
\label{eqn:granularalpha}
\end{equation}
The interval $\Delta t$ can be thought of as the final sequestration time, or equivalently a free decay time, for the nuclides following their build up in the interstellar medium.
While it has been suggested that these equations apply only when the individual nucleosynthetic events producing R also produce S \citep{LugaroEtal2014}, the derivation above shows that the equations are entirely general for a single-phase ISM, as long as $P_{\rm R}$ and $P_{\rm S}$ are defined appropriately. 
Where the same events produce both R and S (e.g., ${}^{238}{\rm U}$ and ${}^{232}{\rm Th}$), production ratios are determined solely by the nucleosynthetic yields; where R and S have different origins (e.g., ${}^{26}{\rm Al}$ and ${}^{27}{\rm Al}$), other factors such as GCE and mixing volumes must be included.
In the limit of small $\delta t$ and $\Delta t$, corresponding to continuous (not punctuated) enrichment, Equation~\ref{eqn:granularalpha} reduces to $\alpha({\rm R}) =  \tau_{\rm R}/ T$, coinciding with the steady-state one-phase ISM\index{Interstellar medium} models of \citet{Jacobsen2005}, \citet{HussEtal2009}, and \citet{Young2014}, which do not explain the solar system abundances of radionuclides.
The case of punctuated delivery is more complicated.

%---------------------------------------------
% FIGURE 6
\begin{figure*} [t!]
    \centering
    \begin{minipage}{0.49\textwidth}
	\includegraphics[width=0.95\textwidth]{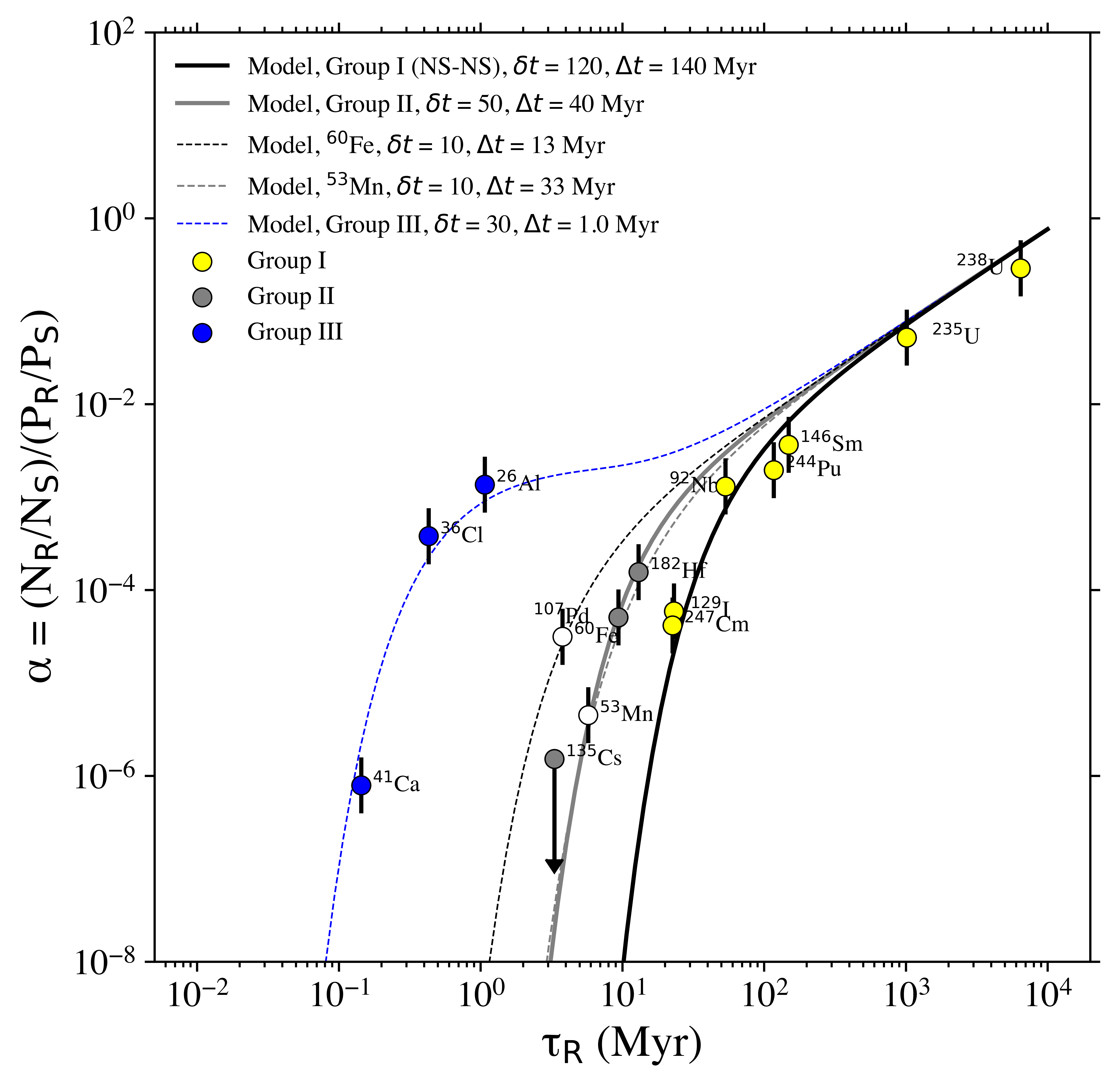}
    \label{fig:granular}
    \end{minipage}
    \hfill
    \begin{minipage}{0.49\textwidth}
    \includegraphics[width=0.95\textwidth]{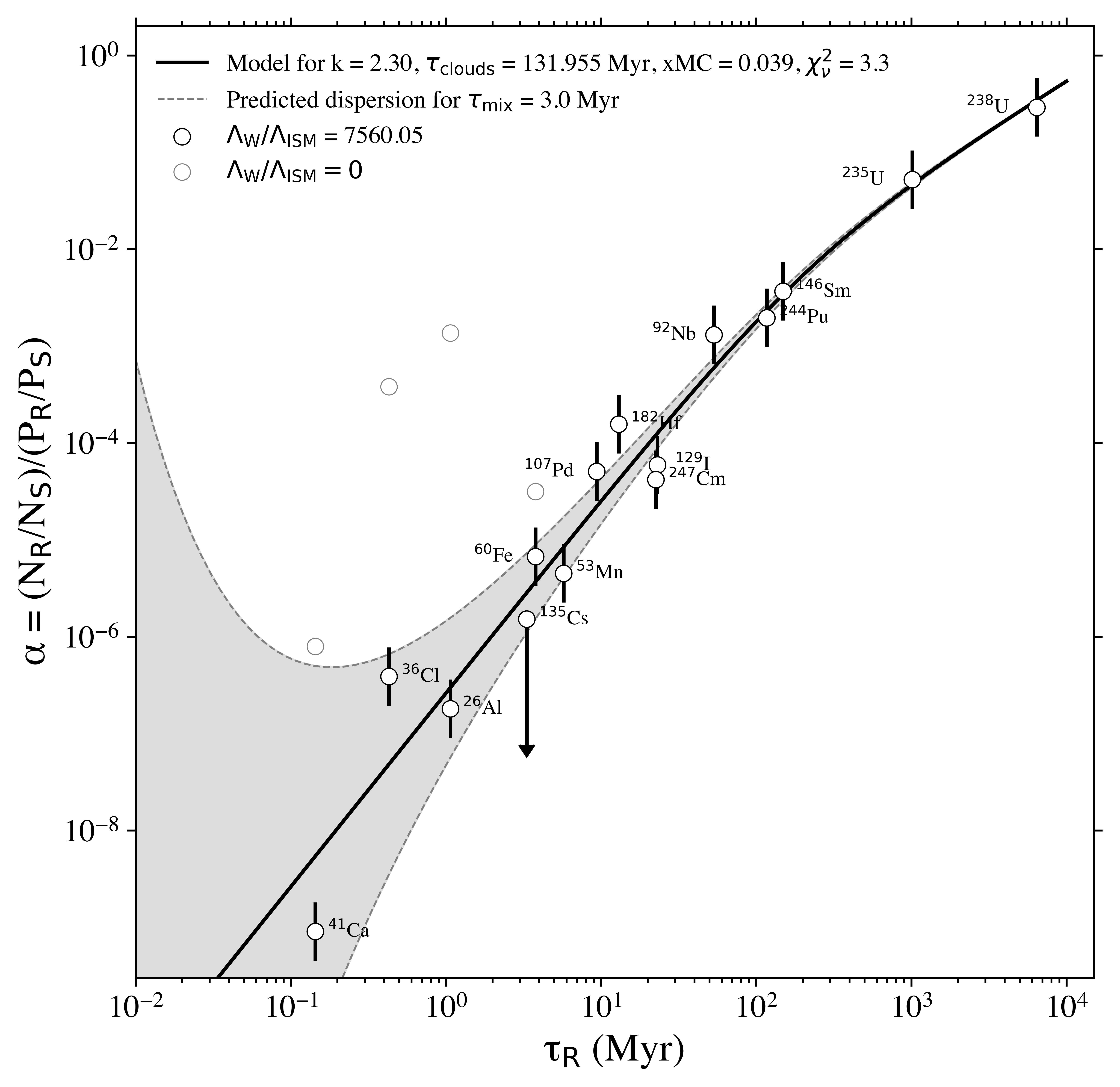}
    \label{fig:enrichment}
    \end{minipage}
        \caption{({\it Left}). Plot of $\alpha(R)$ vs.\ mean life $\tau$ for fifteen radionuclides and their stable reference isotopes.  The isotope and production ratios for the SLRs are listed in Table~\ref{table:one}.  A value for $K$ of $2.3$ is used. Ratios of longer-lived isotopes shown here include the r-process nuclides $\rm ^{238}U/^{232}Th$, $\rm ^{235}U/^{232}Th$, and $\rm ^{244}Pu/^{232}Th$, where ${}^{232}{\rm Th}$ serves as a surrogate for the stable reference isotope for the actinides, and the p-process nuclides and their stable partners $\rm ^{146}Sm/^{144}Sm$ and $\rm ^{92}Nb/^{93}Nb$. Error bars refer to $\pm 100\%$ relative error (i.e., factor of 2) for $\alpha$ values.  Various curves based on Equation \ref{eqn:granularalpha} are discussed in the text. ({\it Right}). MCMC best-fit for 
        $\log{\alpha(R)}$ vs. $\log{\tau_{\rm R}}$ for fifteen radionuclides and their stable reference isotopes using the two-phase ISM model that includes enrichment by WR winds. The fit is obtained with optimal values for three independent parameters, including including $x_{\rm MC} = 0.04$, $\tau_{\rm cloud} = 132$ Myr, and $\Lambda_{\rm W}/\Lambda_{\rm ISM} = 7560$.  Values for $\alpha({\altwosix})$, $\alpha({\fesixty})$, $\alpha({\clthirtysix})$, and $\alpha({\cafortyone})$ without enhancements by WR winds are shown as the open circles with grey outlines.  The values for $\alpha({\altwosix})$, $\alpha({\fesixty})$, $\alpha({\clthirtysix})$, and $\alpha({\cafortyone})$ depend on the relative efficiency factor $\Lambda_{\rm W}/\Lambda_{\rm ISM}$, characterizing winds relative to the surrounding ISM. }
\end{figure*}

Inspection of Equation~\ref{eqn:granularalpha} shows that given an average interval between nucleosynthesis events, $\delta t$, and free decay time $\Delta t$, the normalized relative abundances $\alpha({\rm R})$ should depend only on their radioactive mean-lives $\tau_{\rm R}$.  
The values for $\alpha({\rm R})$ plotted against $\tau_{\rm R}$ should form a single curve if the ISM source was well mixed and characterized by a single beat rate for nucleosynthesis $\delta t$.  
The steepness of the curve will reflect the final sequestration time $\Delta t$. 
Such a plot of $\alpha({\rm R})$ vs. $\tau_{\rm R}$ for the 15 radionuclides for which we have measurements of initial isotope ratios at the birth of the Solar System, estimates of production ratios for the average ISM (Table 1), and a value for $K$ of 2.3, is shown in {\bf Figure 6a}. 
It is immediately evident that a single (monotonic) curve based on Equation \ref{eqn:granularalpha} cannot fit all of the radionuclides.  \cite{Young2016}, as reviewed here, noted that in this scenario of ``punctuated" addition to the ISM the radionuclides can be considered to define at least three groups, and require five separate curves to explain their abundances.

The first group is composed of the r-process products (actinides) and can be referred to as ``Group I".  
The position of the curve is anchored by independent constraints on the production ratio for $\rm ^{238}U/^{232}Th$ based on the probable estimates for the age of the Galaxy (and/or its long-lived nuclides) of approximately 10 Gyr to 13.7 Gyr.  
This constrains the production ratio to lie between $0.55$ and $0.65$.  
Previous studies have shown that the Group I nuclides can be fit with a beat rate for the introduction of the r-process nuclides to the ISM of $10$ Myr, appropriate for production in core-collapse supernovae \citep{Meyer2000}, and a free decay time $\Delta t$ of $\sim 100$ Myr.
However, in the Bayesian sense, the most likely prior probability for a model based on Equation~\ref{eqn:granularalpha} would be where $\delta t \sim \Delta t$, and in Figure 6a a curve based on $\delta t = 120$ Myr and $\Delta t = 140$ Myr is shown to fit the Group I nuclides. 
Since the site of the r process is now thought to be mainly kilonovae from neutron star-and-neutron star/black hole mergers \citep[e.g.,][]{Kasen2017, Smartt2017}, the longer interval between injections of $\sim 100$ Myr is probably appropriate.  
Whether a $\delta t$ of $\sim 100$ Myr is consistent with observationally informed estimates for injection of kilonovae material at any given location in the Galaxy remains to be determined.  

In Figure 6a, the p-process nuclides ${}^{146}{\rm Sm}$ and ${}^{92}{\rm Nb}$ are also included in Group I.\index{Radionuclides!Samarium-146}\index{Radionuclides!Niobium-92}
Little is known definitively about the astrophysical environment of the p process and so whether it is coincidence or not that these isotopes are fit by the same  curve as the r-process species is unknown.

The second group of radionuclides, Group II, consists of the s-process nuclides $\rm ^{182}Hf$ and $\rm ^{107}Pd$. \index{Radionuclides!Hafnium-182}\index{Radionuclides!Palladium-107}
These are fit with a curve defined by $\delta t = 50$ Myr and $\Delta t = 40$ Myr \citep[see also][]{Lugaro2014}.  
The injection interval of $50$ Myr is considered to be of the right magnitude for AGB encounters with random positions in the Galaxy. 
The SLR ${}^{135}{\rm Cs}$ is\index{Radionuclides!Cesium-135} included in this group because it is also an s-process nuclide and is consistent with the model curve, especially since the value given by \cite{BrenneckaKleine2017} is an upper limit.

The third group, Group III nuclides, consists of the shortest-lived nuclides: $\altwosix$, $\clthirtysix$, and $\cafortyone$. \index{Radionuclides!Aluminum-26}\index{Radionuclides!Chlorine-36}\index{Radionuclides!Calcium-41}
For these, according to Equation~\ref{eqn:granularalpha}, the relative abundances are dictated almost entirely by the timing of the last event, $\Delta t$, because of their rapid decay. 
A free decay time of $1.0$ Myr combined with an arbitrarily long $\delta t$ fits the data in Figure 6a.
These SLRs are all ejected by core-collapse supernovae and/or WR stellar winds. 
Finally, the two supernova products $\fesixty$ and $\mnfivethree$ require their own evolution curves with similar $\delta t$ appropriate for supernova injections and somewhat different free decay times $\Delta t$ of 13 and 33 Myr, respectively (Figure 6a).  
While the curve for the Group II, s-process, nuclides appears to fit the $\mnfivethree$ $\alpha$ datum, $\mnfivethree$ has\index{Radionuclides!Manganese-53} a different nucleosynthetic origin than the s-process nuclides.  

The punctuated delivery model described by Equation~\ref{eqn:granularalpha} therefore requires five evolution curves to explain the relative abundances of 15 radionuclides, their reference stable partners, and typical ISM production ratios.
In this interpretation, the solar complements of radionuclides represent a confluence of different, discrete last events with different lead times prior to the formation of the Solar System, including one or more kilonovae (heavy r-process nuclides), AGB stars (s-process nuclides), core-collapse supernovae ($\fesixty$ and $\mnfivethree$), and WR stars ($\altwosix$, $\clthirtysix$, and $\cafortyone$). The precise values for $\delta t$ and $\Delta t$ depend on the value adopted for $K$ but do not vary much over the range in $K$ values from 1 to 3 (e.g., $\Delta t$ for the Group II nuclides shifts from 40 Myr to 35 Myr when $K$ varies from 2.3 to 1).  
% While the sources of these isotopes are generally agreed upon, the assertion that their abundances in the early Solar System represent discrete events that can be dated by their decay is debatable. 

\subsubsection{Regional Self-Enrichment of SLRs} 

%---------------------------------------------
% FIGURE 7
%\begin{figure*} [t!]
%\centering
%\includegraphics[width=0.50\textwidth]{alpha_vs_meanlife_Desch_vals.png}
%\caption{Plot of $\alpha(R)$ vs mean life $\tau$ for fifteen radionuclides and their stable reference isotopes compared with the two-phase ISM model that includes enrichment by WR winds. Values for $\alpha({\altwosix})$, $\alpha({\fesixty})$, $\alpha({\clthirtysix})$, and $\alpha({\cafortyone})$ without enhancements by WR winds are shown as the open circles with grey outlines. The model curve relies on a cloud residence time of 200 Myr. The  Values for $\alpha({\altwosix})$, $\alpha({\fesixty})$, $\alpha({\clthirtysix})$, and $\alpha({\cafortyone})$ depend on the relative efficiency factor $\Lambda_{\rm W}/\Lambda_{\rm ISM}$ of about 4000, characterizing winds relative to the surrounding ISM.  The fit between the data and the model is thus achieved using two fit parameters. }
%\label{fig:enrichment}
%\end{figure*}

The other end-member scenario, the alternative to ``punctuated" delivery (Equation~\ref{eqn:granularalpha}), is quasi-continuous self-enrichment of the Sun's natal star-forming region.  
A critical distinction between models of this type and those based on discrete events is that they account explicitly for the distinction between molecular clouds and environs, and the diffuse ISM.
The consequences of a two-phase ISM\index{Interstellar medium} have been investigated by \cite{Jacobsen2005}, \cite{Huss2009}, and \cite{Young2014}, among others.
The concept employed is that the local ISM in a star-forming region is in a steady state between semi-continuous production of nuclides, their sequestration into cloud material, and radioactive decay. 
At first it may seem counter-intuitive that such a model could have anything to do with the abundances of the SLRs in the early Solar System; however, simple box models \citep{Young2014, Young2016} and hydrodynamic simulations of SLR buildup in star-forming regions \citep{Fujimoto2018} demonstrate that the frequency and rate of SLR production in star-forming regions leads to steady-state enrichments even if the typical residence time in molecular clouds is on the order of $10^8$ yr, orders of magnitude longer than the mean lives of these isotopes.
These studies show that even if the timescale for delivery is many times longer than the half-life for decay, a finite steady-state concentration is expected over tens to hundreds of Myr, depending on the scale of the star formation.

To quantify this scenario, \cite{Jacobsen2005} derived an analytical expression for $\alpha(R)$ in molecular clouds for the steady state arising from exchange of material between the surrounding ISM and the clouds, as well as from  radioactive decay.  Rederivation based on simultaneous solution of the four ordinary differential equations describing exchange of radionuclides and stable nuclides between molecular clouds and the surrounding ISM  at steady state leads to 
\begin{equation}
\alpha({\rm R})  = \frac{{{\tau _{\rm{R}}}^2(1 - {x_{{\rm{MC}}}})}}{{\tau _{{\rm{cloud}}}^2({\tau _{\rm ISM}} \, {x_{{\rm{MC}}}} + {\tau _{\rm{R}}})}},
\label{eqn:continuous}
\end{equation}
 
\noindent where the abundances of radionuclides and their stable partners in $\alpha({\rm R})$ now refer to those in the molecular cloud phase and the proximal intercloud space comprising the star-forming region, %$N_{\rm{R,MC}}$ and $N_{\rm{S,MC}}$, respectively, 
as opposed to the diffuse ISM outside of the star-forming region.  
Here, $x_{\rm MC}$ is the mass fraction of the Galaxy contained in molecular clouds, $\tau_{\rm cloud}$ is the residence time of material in clouds against either sequestration into newborn stars or dispersal to the diffuse ISM, and $\tau_{\rm ISM}$ is the residence time in the diffuse interstellar medium that can be taken as the age of the Galaxy (at the time of Solar System formation in this application). The physics of exchange of material between the two phases of the ISM are contained within the values for $\tau_{\rm cloud}$ and $\tau_{\rm ISM}$, and to first order apply to all nuclides.
Hierarchical star formation means that individual clouds exist for shorter time spans than the star-forming regions as a whole. 
Therefore, time spent passing from one cloud to another through inter-cloud space in the region is included in $\tau_{\rm cloud}$. 

 \cite{Young2014} showed that this expression for the steady-state values for $\alpha(R)$ in molecular clouds could be used to investigate the effects of enrichment of cloud material by WR winds\index{Wolf-Rayet stars}. 
 Because the initial solar $\alratio$ is similar to the ratio observed in star-forming regions today (\S~\ref{sec:usual_abundances}), one may posit that the production rate for $\altwosix$ is enhanced in star-forming regions compared with the average ISM.  
 In this case, the denominator in $\alpha(\altwosix)$ must increase, decreasing $\alpha(\altwosix)$.
 Appealing to the concept that all of the radionuclide $\alpha({\rm R})$ values should be controlled by the same combination of residence times in clouds %and mean-life
 (Equation~\ref{eqn:continuous}), one can deduce the enhancement in production due to winds necessary to bring $\altwosix$ in line with its longer-lived cloud cohabitants. 
 For this purpose, the radionuclide production terms relevant to the molecular cloud setting are modified to include not only production via inheritance from the ISM beyond the star-forming region, ($P^{\rm{ISM}}_{\rm{R}}$), but also production from proximal WR winds, ($P^{\rm{W}}_{\rm{R}}$): 

\begin{equation}
\frac{{{P_{\rm{R}}}}}{{{P_{\rm{S}}}}} = \frac{{{\Lambda _{{\rm{ISM}}}}P_{\rm{R}}^{{\rm{ISM}}}}}{{{\Lambda _{{\rm{ISM}}}}{P_{\rm{S}}}}} + \frac{{{\Lambda _{\rm{W}}}P_{\rm{R}}^{\rm{W}}}}{{{\Lambda _{{\rm{ISM}}}}{P_{\rm{S}}}}} ,
\label{eqn:windproduction}
\end{equation}

\bigskip
\noindent where $\Lambda_{\rm W}$ and $\Lambda_{\rm ISM}$ are the relative efficiencies for trapping the two sources of nuclides in the star-forming region.
The adjustable parameter $\Lambda_{\rm W}/\Lambda_{\rm ISM}$ will affect $\alpha({\rm R})$ for each SLR that is a product of WR winds differently.  The degree to which all of them, including $\alpha({}^{26}{\rm Al})$, $\alpha(\clthirtysix)$, $\alpha(\cafortyone)$, $\alpha(\fesixty)$, and $\alpha(^{107}{\rm Pd})$, are fit by a single value for $\Lambda_{\rm W}/\Lambda_{\rm ISM}$ serves  as a test of the validity of the hypothesis.  
One can use the yield ratios for a $60 M_\odot$ progenitor WR star\index{Wolf-Rayet stars} to obtain these ratios from the literature \citep{Arnould2006, Gounelle2012}:  
$P_{\rm Ca41}^{\rm W}/P_{\rm Al26}^{\rm W}=0.0114$, $P_{\rm Cl36}^{\rm W}/P_{\rm Al26}^{\rm W}=0.010$, $P_{\rm Fe60}^{\rm W}/P_{\rm Al26}^{\rm W}=4.29 \times 10^{-5}$, and $P_{\rm Pd107}^{\rm W}/P_{\rm Al26}^{\rm W}=1.29 \times 10^{-5}$ \citep{Young2014}, as listed in Table 1.  
Clearly, the relative abundances of $\altwosix$, $\clthirtysix$, and $\cafortyone$ will be most affected by WR winds, while $\fesixty$ and $^{107}{\rm Pd}$  will be only minimally affected.  

Markov-chain Monte Carlo (MCMC) sampling can be used to obtain a best fit to the meteoritical data in $\alpha({\rm R})$ vs. $\tau_{\rm R}$ space by optimizing $x_{\rm MC}$, $\tau_{\rm cloud}$, and  $\Lambda_{\rm W}/\Lambda_{\rm ISM}$.  The result for $K = 2.3$ is shown in {\bf Figure 6b}. The first-order characteristic of the trend of the data is the slope of $\sim 2$ in $\log{\alpha(R)}$ vs. $\log{\tau}$ space that is characteristic of the two-phase ISM case (Equation~\ref{eqn:continuous}). The reduced chi squared for the fit, including the factor-of-$2$ uncertainty in the values for $\alpha$ and the expected divergence in $\alpha({\rm R})$ values as a function of mean decay life, is $3$.  The best-fit values for $x_{\rm MC}$, $\tau_{\rm cloud}$, and  $\Lambda_{\rm W}/\Lambda_{\rm ISM}$ are $0.04\pm 0.01$, $132$ Myr, and 7561, respectively. The value for $x_{\rm MC}$ and its uncertainty are independent of the value for $K$.  The value for $\Lambda_{\rm W}/\Lambda_{\rm ISM}$ is also robust, although we ascribe no significance to the small associated uncertainty of $< \pm 0.1$ in best fit.  The value for $\tau_{\rm cloud}$ depends on $K$, and ranges from 87 Myr for $K=1$ to 151 Myr for $K=3$.    

All of the nuclides are within a factor of 2 of the shaded region in Figure 6b that represents the expected spread in values based on the increases in dispersion as mean-lives get progressively smaller. To quantify this expected dispersion,  
\cite{Young2014} used the mixing expression $\sigma_{\log{\alpha}} = 0.44 (\tau_{\rm mix}/\tau)^{1/2}$, where $\tau_{\rm mix}$ is the characteristic timescale for mixing the parental cloud complex and $\sigma_{\log{\alpha}}$ is the dispersion in the logarithm of the $\alpha$ values.
Using the turbulent viscosity $\nu_{\rm T}=3.3\times 10^{-7} {\rm pc}^2/{\rm yr}$ from \cite{Xie1995} and a linear dimension of 1 pc, $\tau_{\rm mix} = 3$ Myr.  
The shaded region in Figure 6b corresponds to this dispersion, and shows that the scatter about the curve evidenced by  $\alpha(\cafortyone)$ and $\alpha(\clthirtysix)$ is consistent with expectations.
The quasi-continuous enrichment model provides a satisfactory fit to all 15 radionuclides.

The best-fit residence time is broadly consistent with estimates for the highly inefficient rate of conversion of cloud material into stars.  
Dividing the total mass of molecular cloud H$_2$ in the Galaxy today, $5.0 \times 10^{8} M_{\odot}$, by the present star formation rate of $\sim 3 \, M_{\odot} \, {\rm yr}^{-1}$ \citep{Kennicutt2012} yields an average residence time in clouds against loss to star formation of $165$ Myr. The value for $x_{\rm MC}$ is also  consistent with recent estimates \citep[e.g.][]{Nakanishi2020}.

The relative efficiency of accumulating nuclides from WR winds compared with accumulation from the surrounding diffuse ISM, represented by $\Lambda_{\rm W}/\Lambda_{\rm ISM}$ of  $\approx 7600$, reflects the requirement that contributions of ${}^{26}{\rm Al}$ from WR\index{Wolf-Rayet stars} winds outweigh contributions of ${}^{60}{\rm Fe}$ from supernova ejecta: the ${}^{60}{\rm Fe}/{}^{26}{\rm Al}$ ratio in the early Solar System was 0.002, much lower than the Galactic average value of 0.55 indicated by $\gamma$ ray observations \citep{Young2014}, which at face value implies $\Lambda_{\rm W}/\Lambda_{\rm ISM}$ of at least $\approx 300$.
The question of how WR winds can dominate contributions of SLRs in molecular clouds is therefore related to the question of how ${}^{26}{\rm Al}$ and ${}^{60}{\rm Fe}$ are distributed in the Galaxy.

\subsection{\boldmath${}^{26}{\rm Al}$ and \boldmath${}^{60}{\rm Fe}$ in a Galactic Context}

%-------------------------
% FIGURE 8
\begin{figure}[h]
    \centering
	\includegraphics[width=0.5\textwidth]{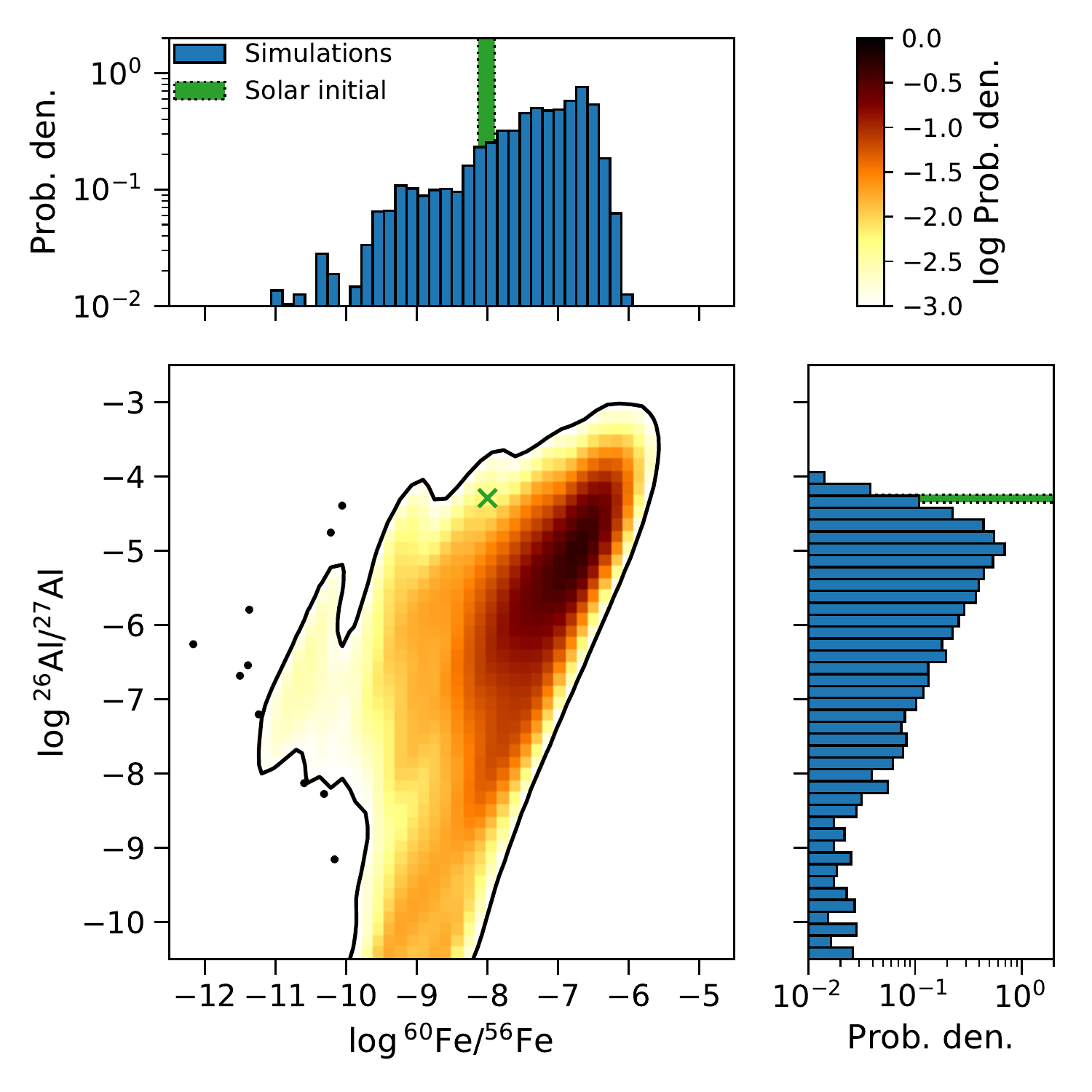}
    \caption{Abundance ratios of SLRs in newly formed stars in a chemo-hydrodynamical galaxy simulation. The central panel shows the joint PDF of $\feratio$ and $\alratio$, with colors showing probability density and black points showing individual stars in sparse regions. The top and right-hand panels show the PDFs of $\feratio$ and $\alratio$ individually\index{Radionuclides!Iron-60}\index{Radionuclides!Aluminum-26}, with simulations shown in blue. Green bands and cross show the solar initial abundances estimated from meteorites. This is a revised version of Figure 5 of \citet{Fujimoto2018}, {using a new galaxy simulation with $N$-body live stellar disk and self-consistent spiral arms}\index{Spiral arms}.}
    \label{abundance_ratios_Fujimoto2022}
\end{figure}

Numerical simulations of the Galaxy (described in \S 3.1) support the idea of inheritance of $\fesixty$ and $\altwosix$ in the early Solar system from its parent GMC, after it had been enriched with SLRs by ongoing star formation in a spiral arm.
The self-enrichment scenario described in \S 3.3.2 appears consistent with the specific abundances of nearly all the SLRs, including $\altwosix$ and $\fesixty$, if WR\index{Wolf-Rayet stars} winds contribute significantly more than supernovae to the inventories of SLRs in GMCs (i.e., $\Lambda_{\rm W} \sim 4000$). 
This result is somewhat surprising, in the sense that a population of massive stars including WR stars should also include supernovae that would contribute to $\fesixty$ as well. 
Numerical simulations of the production of $\altwosix$ and $\fesixty$ in massive stars, and their inheritance by newly forming stars, provide an independent assessment of the plausibility of the quasi-continuous self-enrichment scenario.
% --

\citet{Fujimoto2018} modeled the abundances of $\fesixty$ and $\altwosix$ in newly formed stars using their global-scale chemo-hydrodynamical simulation of the ISM of a Milky Way-like galaxy, with 8 pc resolution. 
This simulation included hydrodynamics, self-gravity, radiative cooling, photoelectric heating, stellar feedback in the form of photoionization, stellar winds, plus core-collapse supernovae, to represent the dynamical evolution of the turbulent multi-phase ISM, and SLR injection from supernovae\index{supernova} and WR\index{Wolf-Rayet stars} winds, {separated in time}. 
In the simulation, when self-gravity and radiative cooling cause the gas to collapse below the resolution limit, the mass is replaced by  300 $M_{\odot}$ star particles that represent stochastically generated stellar populations drawn star-by-star from the initial mass function (IMF).
Each massive star in these populations evolves individually until it produces and returns to the ISM a mass-dependent yield of $\fesixty$ and $\altwosix$ during its lifetime. 
The yields are motivated by stellar nucleosynthetic models, calibrated by the $\gamma$-ray fluxes of $\altwosix$ and $\fesixty$ in the Galaxy.
They subsequently tracked the transport in the ISM and decay of these isotopes, and their incorporation into new stars. 

{The galaxy model used by \citet{Fujimoto2018} assumed a smooth distribution of old stars and dark matter, using a fixed axisymmetric galactic potential. Although flocculent spiral\index{Spiral arms} structures spontaneously formed anyway in the simulation due to the galactic shear and gas self-gravity, the gravitational potential did not include explicit spiral perturbations.
\citet{Fujimoto2020a} improved on this galaxy model with an $N$-body live dark matter halo and stellar disk, allowing the galactic disk to form spiral arms self-consistently via gravitational interactions among dark matter, stars, and gas. 
The spatial resolution of \citet{Fujimoto2020a} was 20 pc.
Here we show a restart of that simulation with 10 pc spatial resolution, comparable to the 8 pc resolution obtained by \citet{Fujimoto2018}.}
%and here we restarted the simulation with the resolution of 10 pc, which is comparable to the 8 pc resolution in \citet{Fujimoto2018}.}
This newer, more realistic simulation was used to update the probability distribution functions (PDFs) of $\feratio$ and $\alratio$ ratios for newborn stars in the Galaxy, in particular for stars formed between Galactocentric radii of 5.5 and 6.5 kpc thought to represent the Sun, which has migrated outward through the Galactic disk over 4.6 Gyr \citep[e.g.,][]{Wielen1996}.
In {\bf Figure~\ref{abundance_ratios_Fujimoto2022}}, we display an updated version of Figure 5 of \citet{Fujimoto2018}, but revised with a factor-of-3 reduction in the $\fesixty$ production rate to better match the observed $\gamma$-ray emission, as described by \citet{Fujimoto2020a}.

The numerical results displayed in Figure~\ref{abundance_ratios_Fujimoto2022} support the underlying assumptions of the quasi-continuous self-enrichment model, and also suggest that the Sun's $\altwosix$ and $\fesixty$ abundances are not atypical. 
The distributions of $\altwosix$ and $\fesixty$ are tightly correlated, and stars forming in spiral arms will receive high abundances of both. 
The meteoritic $\alratio$ ratio is slightly higher than the most probable value $\sim 1 \times 10^{-5}$, but remains probable at the tens of percent level.
Likewise, the meteoritic $\feratio$ ratio is lower than the most probable value, $\sim 2 \times 10^{-7}$, but also remains probable at the tens of percent level.

Moreover, inclusion of physics neglected in these simulations may make the meteoritic values more probable.
One improvement to these simulations would be to include fallback of supernova ejecta onto a central black hole during the explosion of the most massive stars.
This would prevent the escape of a large fraction of the $\fesixty$ from the most massive stars, with progenitor masses $> 20 \, M_{\odot}$ \citep{Fryer1999}.
$\fesixty$ would then only arise from the lower-mass supernovae, which would disperse farther from star-forming regions before they explode. 
This would be likely to shift the $\fesixty$ PDF to the left in Figure~\ref{abundance_ratios_Fujimoto2022}.

The match also might improve with better pre-supernova feedback recipes.
\citet{Fujimoto2019} found that their galaxy simulations succeeded in reproducing all observables on kpc- or larger scales, but failed to reproduce the decorrelation between tracers of gas and star formation on $\leq 100$ pc scales, observed by \citet{Kruijssen2019}.
They concluded the discrepancy between observations and the simulation was due to an insufficiently strong pre-supernova feedback in the simulation and a failure in the model for massive stars to disperse surrounding gas before they go supernova.
Neglect of this physics may lead to overpredictions of the abundance of $\fesixty$, which arises from the supernovae, and underpredictions of the abundance of $\altwosix$, which arises from both the supernova and the earlier stellar winds.

Improved stellar models may also alter the results. 
The simulations on which Figure 7 is based used chemical yields as a function of stellar mass from the mass-dependent yield tables of \citet{Sukhbold2016}; but other models include additional physics, such as the model of \citet{LimongiChieffi2018}, which took into account rotational mixing within the star, among other things.
Higher spatial resolution, to zoom into individual cloud cores, is also desired to build up better statistics.

\subsection{\textbf{Summary}}

The homogeneity of SLRs and their inheritance from the molecular cloud, at the levels observed from meteorites, appear attributable to ongoing stellar nucleosynthesis. 
Their abundances are naturally explained if the Sun formed in a spiral arm of the Milky Way Galaxy and received a much higher proportion (than the Galactic average) of material from WR winds than supernova materials. 
Simulations of the Galaxy \citep{Fujimoto2018} reveal how populations of massive stars enrich the surrounding ISM and GMCs with $\altwosix$, $\fesixty$, and other SLRs.
The Solar System's particular $\alratio$ and $\feratio$ ratios are reasonably probable outcomes. 
The simulations also predict that the GCR flux in the Sun's molecular cloud was greater than the Galaxy-wide spatial average by a factor $\approx\!4$, and therefore the Solar System's $\beratio$ is also a highly probable outcome.
The excellent fit to 15 radionuclide abundances (including $\altwosix$ and $\fesixty$) with essentially two parameters, plus the match to the $\beten$ abundance in the same context, makes the semi-analytical quasi-continuous self-enrichment model \citep{Young2014}
a very likely scenario for the origins of the SLRs in the Solar System. It obviates the need to invoke in situ irradiation within the solar nebula to explain the abundances of SLRs, which the homogeneity of $\beten$ indicates must be very limited. 
It also obviates the need to invoke stochastic events such as a late, nearby supernova.

%--------------------------------------------------------------------
%
% SECTION IV.    
%

\section{IMPLICATIONS}

The picture that has emerged from meteoritics over the last decade is that the most-studied SLRs ($\altwosix$, $\mnfivethree$, $\hfoneeighttwo$, ${}^{129}{\rm I}$) were homogeneously distributed in almost all materials, from the birth of the Solar System, as attested to by the concordancy of formation times within volcanic achondrites and other components that should be concordant.
The homogeneity of $\beten$ places severe constraints on the production of SLRs by SEP irradiation, which may have contributed significantly only to ${}^{36}{\rm Cl}$.
While some evidence exists for a chemical heterogeneity of $\altwosix$ associated with hibonite, there is little evidence for late injection\index{lateinjection} of $\altwosix$ into the disk. 
Meanwhile, astrophysical models over the last decade have made a stronger case for inheritance of the Solar System's SLRs from the molecular cloud, after enrichment by massive stars in the neighboring star-forming environment in a spiral arm.
The abundance of $\beten$ in the solar nebula likewise points to the Sun's formation in spiral arm of the Galaxy.

In this section we answer some of the questions raised in \S 1.3, in the context of these new findings as reviewed in this Chapter.
Can the SLRs be used to date solar nebula events?
In \S 4.1 we show they can be, and discuss new chronological interpretations.
Were the SLRs inherited from the molecular cloud, or injected late and/or created in the solar nebula? 
If inherited, what were their sources? 
In \S 4.2 we assess the origin of $\altwosix$, which appears mostly inherited from the molecular cloud, after enrichment primarily from WR winds.
In \S 4.3 we assess the role of irradiation in producing $\clthirtysix$, and discuss other SLRs.
What was the Sun's birth environment? 
How universal or unique was the Sun's inventory of SLRs? 
In \S 4.4 we discuss the implications from SLRs, that the Sun had a fairly typical history, born in a spiral arm of the Galaxy.

%Meteorites were witnesses to the birth of the Sun and its planets. 
%Their most powerful testimony lies in the isotopic signatures of a dozen or more SLRs, now extinct but once present in the Sun's protoplanetary disk.
%Measurements yielding their former abundances enable dating of early events such as formation of CAIs and chondrules, and the formation and differentiation of planets. 
%Assessments of their heterogeneities allow us to decipher their origins, informing models of where in the Galaxy the Sun and planets formed.
%
%The picture that has emerged in the last decade is that the SLRs in the solar nebula were spatially distributed very homogeneously, and with few exceptions were present in all materials from the birth of the Solar System.
%For example, there is no strong evidence for distinct reservoirs with different $\alratio$ ratios in the inner or outer Solar System.
%The solar nebula does not appear to have been temporally heterogeneous, either. 
%There is little evidence for late injection of SLRs into the protoplanetary disk. 
%Production of SLRs by SEP irradiation appears to have been very limited, perhaps restricted essentially to the SLR $\clthirtysix$. 
%These findings strongly imply that almost all the SLRs appear to have been inherited from the Sun's molecular cloud, which itself must have been contaminated with ejecta from massive stars, mostly Wolf-Rayet stars, due to the Sun's formation within a spiral arm.  

\subsection{Chronometry}
\label{sec:chronometryimplications}

One of the primary reasons to study SLRs in meteorites is to determine the timing of events in the solar nebula such as the formation of CAIs and chondrules, the accretion of planetary embryos, and core formation.
Inferences of the initial $(\alratio)_0$ in a sample, from a valid isochron, can yield a time of formation $\Delta t_{26}$ after $t\!\!=\!\!0$ because the Solar System initial value $(\alratio)_{\rm SS}$ at that time is known to be $\approx 5 \times 10^{-5}$ (and can be defined to be $\equiv 5.23 \times 10^{-5}$). 
Likewise, the times of formation of a sample using Mn-Cr or Hf-W systematics can be determined if the initial Solar System values $(\mnratio)_{\rm SS}$ or $(\hfratio)_{\rm SS}$ at $t\!\!=\!\!0$ are known. 
These are difficult to obtain using direct measurements of CAIs, but recent comparisons of these systems with times of formation found by Pb-Pb dating of achondrites has allowed them to be determined more precisely (Table 1). 

This technique, and indeed all chronometry using SLRs, is based on the assumption that the SLRs were homogeneously distributed at an early time in the solar nebula. 
Because the times of formation of most systems in most achondrites, including all systems (Al-Mg, Mn-Cr, Hf-W, U-Pb, and even apparently even I-Xe)  in volcanic achondrites are found to be mostly concordant (\S 2.2), this strongly suggests the abundances of the SLRs $\altwosix$, $\mnfivethree$, $\hfoneeighttwo$ and $\ionetwonine$ were homogeneous, and their initial abundances can be used to find times of formation. 
A key result from this work is that the age of the Solar System (more precisely: the Pb-Pb age that would be obtained, using currently accepted U half-lives, on a sample for which Pb reached isotopic closure at the time $t\!\!=\!\!0$ when $\alratio = 5.23 \times 10^{-5}$) is 4568.7 Myr. \index{Solar nebula!age}

Chronometry using the Hf-W system is possible even if an internal isochron cannot be constructed, by determining $\epsilon ^{182}{\rm W}$ in a bulk sample. 
This technique tends to date the time of metal-silicate separation. 
Key results from such studies include: the dating of Mars\index{Mars} to about 1-3 Myr after $t\!\!=\!\!0$ 
\citep{DauphasPourmand2011}; and the dating of when Jupiter's $\approx 20-30 \, M_{\oplus}$ core\index{Jupiter} opened a gap\index{Protoplanetary disk substructure!gaps} in the protoplanetary disk, to about 0.4-0.9 Myr after CAIs
\citep{KruijerEtal2017}.
These results have had profound implications for planet formation, practically demanding growth of planetary embryos by pebble accretion.\index{Pebble accretion}
Among a number of studies applied to smaller bodies, measurements of $\epsilon^{182}{\rm W}$ have enabled timing of the accretion and differentiation of the asteroid\index{asteroids} 4 Vesta, to within $< 1$ Myr after CAIs \citep{Touboul2015}, when $\altwosix$ was sufficiently abundant to trigger global-scale melting. 
These results confirm Harold Urey's original insight that $\altwosix$ from outside the Solar System must have been the primary heat source for bodies during the first few Myr of the protoplanetary disk. 

\subsection{\textbf{\boldmath${}^{26}{\rm Al}$} and Late Injection}
\label{sec:lateinjection}

Immediately after live $\altwosix$ was discovered to have existed in the solar nebula \citep{LeeEtal1976}, \citet{CameronTruran1977}
argued that it was too short-lived to have been inherited during the course of Galactic evolution.
They postulated the supernova\index{supernova} trigger hypothesis, in which ejecta from a supernova both delivered $\altwosix$-rich supernova material to our Sun's molecular cloud, and triggered its collapse by shocking it. 
Simulations of this process indicated $\altwosix$ would be heterogeneous, at least in the molecular cloud \citep{BossEtal2008}. \index{Late injection}

The SLR $\altwosix$ has long been suspected of being spatially and/or temporally heterogeneous in the solar nebula.
\citet{FaheyEtal1987} measured a very low $\altwosix$ abundance in the FUN CAI\index{Calcium-rich, aluminum-rich inclusions (CAIs)!FUN CAIs} ``HAL" (named because it is a large Hibonite inclusion in the chondrite ALlende\index{Meteorites, individual!Allende}): $(\alratio)_0 = (5.2 \pm 1.7) \times 10^{-8}$.
% FUN CAIs are like other CAIs except they have suffered mass-dependent isotopic fractionation of major elements through extensive evaporation, and they display certain stable isotope anomalies (e.g., deviations in their ${}^{50}{\rm Ti}/{}^{48}{\rm Ti}$ ratios).
If HAL formed from a reservoir with the canonical $(\alratio)_{\rm SS} = 5.23 \times 10^{-5}$, it would have had to have formed or been reset roughly 7 Myr after $t\!\!=\!\!0$ in order to have such a low $(\alratio)_0$; but this inclusion could not have formed that late, as it is found in the Allende chondrite, which had accreted by 3 or 4 Myr after $t\!\!=\!\!0$ \citep[e.g.,][]{DeschEtal2018}.
Temperatures in the Allende parent body also were not high enough to thermally reset the Al-Mg system.
Therefore it was hypothesized that the inclusion formed at a time {\it before} $\altwosix$ was widespread in the solar nebula.
The large stable isotope anomalies (e.g., $\epsilon^{50}{\rm Ti})$ exhibited by FUN CAIs were interpreted as having not yet been ``homogenized" in the solar nebula. \index{Calcium-rich, aluminum-rich inclusions (CAIs)!FUN CAIs}
This idea was developed as the refined ``late injection" model\index{Late injection} by \citet{SahijpalGoswami1998}, in which it was posited that a nearby supernova injected $\altwosix$-rich (and perhaps ${}^{50}{\rm Ti}$-rich) material into the Sun's protoplanetary disk, in a manner that was initially spatially heterogeneous, but later mixed.

The late injection scenario has faced astrophysical challenges.
As demonstrated by \citet{OuelletteEtal2007},  supernova ejecta in the form of gas cannot penetrate into the disk, instead flowing around it after passing through a bow shock; only ejecta in the form of large ($> 0.1-1 \, \mu{\rm m}$) dust grains could enter the disk.
Condensation of such large grains is possible, but only if ejecta are clumpy \citep{FedkinEtal2010}.
It is also unlikely that a very young ($< 10^5$ year-old) protoplanetary disk would exist close enough to a $> 4$ Myr-old massive star in order to receive enough $\altwosix$ to match the Solar System value, which also demands that the ejecta explode in a heterogeneous, clumpy way.
This is observed in nearby supernovae\index{supernova} like Cassiopeia A\index[obj]{Cassiopeia A} \citep{HwangEtal2004}, but it is debated how common this is.
Even so, the probability of being several parsecs from a supernova and yet being hit by a clump of ejecta would be $\sim 0.1 - 1\%$ \citep{OuelletteEtal2010}.
Various scenarios for incorporation of material from evolved massive stars (either supernova ejecta or direct WR winds) into molecular cloud material have been considered, and the probability of contaminating a forming Solar System in this way is likely no more than $\sim 10\%$ \citep{GounelleEtal2009,PanEtal2012}.
Thus, these considerations do not rule out the possibility of late injection, but it is not an especially probable scenario, especially when compared to the alternative scenario in which $\altwosix$ and other SLRs in the solar nebula were already present in the molecular cloud before $t\!\!=\!\!0$ and were spatially well-mixed (\S 4.4), which should be more common in these same environments.

Despite the astrophysical challenges, much of the data collected over the last two decades regarding non-uniform $(\alratio)$ ratios have been interpreted as spatial or temporal heterogeneities in $\altwosix$, and as support for the late injection model.  
Much of this evidence was reviewed by \citet{KrotEtal2012}, some of whose arguments we discuss below.  
A review of these data, coupled with insights from the last few years, suggests that while some heterogeneities exist, they are not as prevalent as once thought. 
Rather than being consistent with the spatial and temporal heterogeneities posited by the late injection model, the data may be more consistent with a {\it chemical} heterogeneity in the solar nebula. 
% Rather than late injection or spatial separation of materials, $\altwosix$ was inherited from the molecular cloud and was always present and well mixed in the solar nebula. 

\subsubsection{Achondrites}

Many of the arguments for heterogeneity of $\altwosix$ have relied on discordancies between the Al-Mg and other systems. 
Improvements in precision and in interpretation have allowed updates to important parameters like $({}^{53}{\rm Mn}/{}^{55}{\rm Mn})_{\rm SS}$, allowing a reassessment in the discrepancies.
As described in \S 2.2, with these updates the formation times of at least the volcanic achondrites\index{Meteorites!achondrites} are found to be quite concordant, and the formation times determined by the Al-Mg system match the formation times found by the Mn-Cr, Hf-W, or Pb-Pb systems. 
This strongly indicates that the SLRs, including $\altwosix$, were homogeneously distributed among the reservoirs sampled by the achondrites, and by inference most solar nebula materials 
\citep{DeschEtal2022a,DeschEtal2022b}.

If there were spatial variations in the $\alratio$ ratios between two reservoirs, arguably the most noticeable difference would be between meteorites formed in the inner disk (inside Jupiter's orbit\index{Jupiter}) vs.\ the outer disk \citep[i.e., the ``NC" reservoir vs.\ the ``CC" reservoir][]{KruijerEtal2017}.
This dichotomy is associated with the stable isotope anomalies, and with supernova inputs in some models \citep{NanneEtal2019}. 
Unfortunately, only the two achondrites NWA 2976 and NWA 6704 from the CC reservoir have been dated using Al-Mg and at least one other system. 
As discussed by \citet{DeschEtal2022a,DeschEtal2022b}, it cannot be ruled out that these formed from a reservoir with $2\times$ the $\altwosix$ as the NC reservoir and CAIs had. 
Notably, though, models of large-scale $\altwosix$ heterogeneity often posit a very different scenario, that the CAI-forming region had $4\times$ the $\altwosix$ as the NC and CC regions \citep{BollardEtal2019}, as discussed below.
The achondrite NWA 6704 is discordant among its Al-Mg, Mn-Cr and Pb-Pb systems, suggesting that it cooled through the closure temperatures over millions of years, much more slowly than the volcanic achondrites.
\citet{DattaEtal2021metsoc} inferred that at lower temperatures, cooling rates in the angrite parent body slowed to $\approx 280$ K/Myr; if NWA 6704 cooled at similar rates, a difference in closure temperatures of just 280 K would mean the systems would close at times differing by 1 Myr, and therefore appear discrepant. 
Closure temperatures of the Pb-Pb system are probably lower in oxidized samples like NWA 6704, compared to other achondrites \citep{Cherniak2001,DeschEtal2022b}. 
More isotopic dating of CC achondrites is needed to better test these ideas, but  the slight discordancies in carbonaceous achondrites appear more easily explained by slow cooling rather than $\altwosix$ heterogeneity.

\subsubsection{${}^{26}{\rm Al}$ and ${}^{182}{\rm Hf}$ in the FUN CAI STP-1}

Another argument for $\altwosix$ heterogeneity involves the FUN CAI STP-1.\index{Calcium-rich, aluminum-rich inclusions (CAIs)!FUN CAIs} 
This inclusion was argued to have an initial $(\hfratio)_0$ value consistent with formation $\approx 0.33_{-1.47}^{+1.67}$ Myr after $t\!\!=\!\!0$ but an initial $(\alratio)_0$ consistent with formation $3.02 \pm 0.08$ Myr after $t\!\!=\!\!0$ \citep{HolstEtal2013}.
Because these formation times could not be reconciled with each other, it was argued that the distribution of $\altwosix$ was decoupled from that of $\hfoneeighttwo$.
As $\hfoneeighttwo$ was taken to be homogeneous, $\altwosix$ had to be heterogeneous, with a value $(\alratio)_0 = (2.94 \pm 0.21) \times 10^{-6}$ at $t\!\!=\!\!0$ in the reservoir from which STP-1 formed.
However, with updated initial values of $(\hfratio)_{\rm SS}$ and other considerations (\S 2.2), the Hf-W and Al-Mg times of formation are not actually that discordant, and the Al-Mg ($\Delta t_{26} = 2.98 \pm 0.07$ Myr) and Hf-W ($\Delta t_{182} = 1.56_{-1.45}^{+1.63}$ Myr) ages differ only at the $1.7\sigma$ level \citep{DeschEtal2022b}.

\subsubsection{${}^{26}{\rm Al}$ in Chondrules}

Another example of formation times being used to argue for $\altwosix$ heterogeneity comes from the comparison of Al-Mg and Pb-Pb ages in chondrules\index{chondrules} by \citet{BollardEtal2019}.
For a sample of eight chondrules from Allende\index{Meteorites, individual!Allende} and NWA 5697, they found initial $(\alratio)_0$ values and Pb-Pb ages, calculated the time since $t\!\!=\!\!0$ by subtracting the Pb-Pb ages from $4567.3 \pm 0.16$ Myr, then calculated the initial $\altwosix$ abundance of the Solar System, finding $(\alratio)_{\rm SS} \approx 1.5 \times 10^{-5}$. 
These authors argued that chondrules in Allende (a CC chondrite\index{Meteorites!chondrites}) formed from this reservoir with lower $\alratio$ than CAIs did. 
%[We note that since Allende formed from the CC reservoir, this reasoning yields the opposite trend as the achondrite data, with a lower implied $\alratio$ ratio in the CC reservoir.]
However, this apparent discrepancy is removed if one does not presume that the Pb-Pb age of $4567.3 \pm 0.16$ Myr derived by \citet{ConnellyEtal2012} dates $t\!\!=\!\!0$ as recorded by the Al-Mg system. 
This age is $\sim 1$ Myr younger than that of other CAIs \citep[e.g.,][]{BouvierWadhwa2010}, and is 1.4 Myr younger than the absolute age of $t\!\!=\!\!0$ inferred by \citet{DeschEtal2022a}, $4568.65 \pm 0.10$ Myr.
By extrapolating backwards only to a  $t\!\!=\!\!0$ at 4567.3 Myr, \citet{BollardEtal2019} may be neglecting as much as an extra two half-lives of decay of $\altwosix$, and their result  $(\alratio)_{\rm SS} \approx 1.5 \times 10^{-5}$ may be underestimating the true value by a factor of about 4. 
By allowing that the Al-Mg and Pb-Pb systems may not have achieved isotopic closure at the same time in CAIs (e.g., because of late-occurring transient heating events), the evidence for heterogeneity is largely removed; and Allende chondrules\index{chondrules} indicate that they, like CAIs, formed from a reservoir with  $(\alratio)_{\rm SS} \approx 5 \times 10^{-5}$.

\subsubsection{Bulk Mg Isotopic Compositions}

Other evidence for $\altwosix$ heterogeneity has come from measurements of ${}^{26}{\rm Mg}$ excesses in bulk chondrites.
The isotopic composition of CI chondrites, thought to best represent the bulk composition Solar System, includes ${}^{26}{\rm Mg}$ from the decay of all the initial $\altwosix$ that was present in the Solar System. 
Measurements of ${}^{26}{\rm Mg}/{}^{24}{\rm Mg}$ ratios in CI chondrites (corrected for isotopic fractionation, using ${}^{25}{\rm Mg}/{}^{24}{\rm Mg}$ measurements) can be compared to the initial ${}^{26}{\rm Mg}/{}^{24}{\rm Mg}$ ratios in CAIs or refractory olivine grains, obtained from the intercepts of the Al-Mg isochrons constructed for these objects. 
Because these inclusions formed before decay of ${}^{26}{\rm Al}$, this intercept should yield a lower ${}^{26}{\rm Mg}/{}^{24}{\rm Mg}$ ratio, manifesting itself as a deficit in ${}^{26}{\rm Mg}$ that correlates with the initial $(\alratio)_0$ in the inclusion (found from the slope of the isochron). 

\citet{LarsenEtal2011} found a deficit of 19.6 ppm in CAIs relative to CI chondrites; since the CAIs are defined by $(\alratio)_0 \approx 5.23 \times 10^{-5}$, they inferred a value $(\alratio) \approx 2.7 \times 10^{-5}$ in the Solar System overall to give CI chondrites their inferred ${}^{26}{\rm Mg}$ abundance.
This result would imply that CAIs sampled a reservoir with greater $\altwosix$ than the rest of the solar nebula. 
More recently, however,
\citet{GregoryEtal2020} have measured average deficits in refractory forsteritic inclusions of 38.5 ppm relative to CI chondrites, which is exactly consistent with CI chondrites sampling a reservoir with the same $\alratio = 5.23 \times 10^{-5}$ ratio (at $t\!\!=\!\!0$) as CAIs.
Although both the refractory forsteritic inclusions and CAIs both include minerals condensed from the gas, that probably sample similar reservoirs, the CAIs may have experienced exchange of Mg isotopes that the forsteritic inclusions did not.
Regardless of the details, the Mg isotopic compositions of bulk samples appear consistent with homogeneity of $\altwosix$.

\subsubsection{${}^{26}{\rm Al}$ in PLACs and FUN CAIs}

Most of the evidence from recent decades for $\altwosix$ heterogeneity does not withstand scrutiny, yet strong evidence remains from the HAL inclusion, and inclusions like it, such as PLACs (PLAty hibonite Crystals), other PLAC-like FUN CAIs, and some rare corundum grains. 
Many FUN CAIs also show anomalous abundances of the neutron-rich isotope ${}^{50}{\rm Ti}$, with $\epsilon^{50}{\rm Ti} = [ ({}^{50}{\rm Ti}/{}^{48}{\rm Ti}) / ({}^{50}{\rm Ti}/{}^{48}{\rm Ti})_{\rm std} -1]$ $\times 10^4$ values ranging from -200 to +200 (where ``std" refers to a terrestrial standard, and $\epsilon$ is measured in parts per 10,000).
These values far exceed the range of most CAIs, which cluster around $\epsilon^{50}{\rm Ti} \approx +8.4 \pm 1.3$ \citep[e.g.,][]{TorranoEtal2019}.
This scatter in isotope ratios has been taken as further evidence that they had formed early, before the carrier of the ${}^{50}{\rm Ti}$ anomaly was mixed in the solar nebula.
Following the measurement of very low $(\alratio)_0$ in the FUN CAI HAL, FUN CAIs as a class have been considered to have formed first in the solar nebula, before the late injection of $\altwosix$.

Over time, FUN CAIs\index{Calcium-rich, aluminum-rich inclusions (CAIs)!FUN CAIs} generally became associated with low $\alratio$ ratios and large scatter in $\epsilon^{50}{\rm Ti}$ values, but more recently it has been shown that only a subset of FUN CAIs are so characterized. 
Larger FUN CAIs with igneous petrology of type A or B (including STP-1) tend to have initial $(\alratio)_0$ ratios consistent with a thermal resetting in the solar nebula at 1-3 Myr after CAIs
\citep{DunhamEtal2020lpsc,DeschEtal2022b}.
%\citep{DunhamEtal2020lpsc,DunhamEtal2021}.
They also exhibit a tight clustering of their $\epsilon^{50}{\rm Ti}$ values around $-42.9 \pm 1.3$, just as tight as the clustering of normal CAIs around their mean \citep{TorranoEtal2021metsoc}. 
Since normal CAIs are interpreted to have formed in a well-mixed nebula (from the standpoint of ${}^{50}{\rm Ti}$), it also must be interpreted that these same large, type A/B FUN CAIs also formed in a well-mixed nebula, just at a later time than normal CAIs, after the $\epsilon^{50}{\rm Ti}$ value had evolved \citep{TorranoEtal2021metsoc}.
These FUN CAIs appear to have formed late in the evolution of the disk.
%Some FUN CAIs do still exhibit anomalously low $(\alratio)_0$ values and scattered $\epsilon^{50}{\rm Ti}$ values, suggesting an early formation; but many FUN CAIs clearly formed late in the evolution of the disk.

In fact, the only inclusions with $(\alratio)_0$ too low to form from a reservoir with canonical $\alratio$ are PLACs, PLAC-like FUN CAIs, and some corundum grains.
% Corundum, ${\rm Al}_{2}{\rm O}_{3}$ and hibonite ${\rm CaAl}_{12}{\rm O}_{19}$, are the two most refractory common minerals, are strongly associated with each other, and corundum readily reacts with hot CaO vapor to form hibonite.
\citet{MakideEtal2011} found that 52\% of corundum grains formed with $(\alratio)_0 < 2 \times 10^{-6}$ ($2\sigma$). 
PLACs\index{Calcium-rich, aluminum-rich inclusions (CAIs)!PLACs} (and the related inclusions known as Blue AGgregates of hibonite crystals [BAGs], and pyroxene-hibonite spherules) typically show no evidence for $\altwosix$, with $2\sigma$ upper limits $(\alratio)_0 < 5 \times 10^{-6}$ [\citet{KrotEtal2012} and references therein].
Not all hibonite or corundum grains were devoid of $\altwosix$: 
half of corundum grains show resolvable excesses of ${}^{26}{\rm Mg}$ \citep{MakideEtal2011}, and the inclusions known as SHIBs (Spinel-HIBonite CAIs) typically formed with near-canonical $(\alratio)_0 \approx 4-5 \times 10^{-5}$
[\citet{KrotEtal2012} and references therein].
But all the inclusions that have very low $(\alratio)_0$ values are dominated by the minerals hibonite (${\rm CaAl}_{12}{\rm O}_{19}$) or corundum (${\rm Al}_{2}{\rm O}_{3}$).
These Al-rich minerals are notable for being essentially the most refractory minerals that would condense from a gas of solar composition
\citep{EbelGrossman2000}.
Hibonite and corundum grains are capable of surviving temperatures $> 1400$ K that would vaporize all other minerals, even those containing Al.
These facts are suggestive of a chemical heterogeneity for $\altwosix$, rather than a spatial or temporal heterogeneity.
\citet{LarsenEtal2020} recently noted these same general trends and suggested that $\altwosix$-free inclusions formed from ancient interstellar dust in which $\altwosix$ had decayed, whereas most other inclusions sampled materials with recently created $\altwosix$.

\subsubsection{Origin of a Chemical ${}^{26}{\rm Al}$ Heterogeneity}

Consideration of the interstellar sources of ${}^{26}{\rm Al}$ suggests the following scenario for how a chemical---but not a spatial or temporal---heterogeneity could arise. 
Al in the ISM would be almost fully condensed into dust grains.
As suggested by \citet{LarsenEtal2020}, some of this dust may be relatively new, perhaps formed in supernovae or WR winds in the last $\approx 5$ Myr. 
This material would be characterized by $(\alratio)_0$ as low as $10^{-6}$, to as high as $\approx 0.08$ \citep{NittlerEtal2008}.
%\citet{ChoiEtal1999} analyzed two presolar hibonite grains that formed with $(\alratio)_0 \sim 10^{-2}$, and 
%\citet{ChoiEtal1998} found 30\% of presolar corundum grains formed with significant amounts of $\altwosix$. 
The presolar grains known as ``nanospinels," comprised mostly of the mineral spinel (${\rm Mg}_{2}{\rm Al}_{2}{\rm O}_{4}$), have been identified as a major carrier of the ${}^{54}{\rm Cr}$ isotopic anomaly in the Solar System \citep{DauphasEtal2010} and have been suggested as a carrier of $\altwosix$
\citep{LarsenEtal2011}.
In contrast to live $\altwosix$, most of the Al would reside in much more ancient dust grains, including perhaps hibonite and corundum grains, that had survived passage through the ISM, or were destroyed (e.g., by shocks) and reformed in the ISM; these would contain little to no live $\altwosix$.
These two populations of dust grains would be spatially mixed at all stages, as spatial mixing is efficient in the molecular cloud \citep{PanEtal2012}, during cloud core collapse \citep{KuffmeierEtal2017}, and within the disk \citep{Boss2013}.
If $\approx 5\%$ of the Al in the Sun's molecular cloud were contributed by high-$\alratio$ material formed in the last 5 Myr, the solar nebula could be characterized by $\alratio \approx 5 \times 10^{-5}$.

Among the Al-bearing materials in the solar nebula would be presolar grains of hibonite and corundum, of typical sizes $\sim 1 \, \mu{\rm m}$.
Perhaps because they are more refractory and resistant to destruction, most hibonite and corundum grains might be from the ancient dust population, and devoid of $\altwosix$. 
\citet{ZegaEtal2011} analyzed five presolar hibonite grains and found four to have come from AGB stars and one from a core-collapse supernova\index{supernova}. 
Unlike a supernova, which is associated with star-forming environments, the likelihood of an AGB star contaminating a star-forming region directly in the few Myr before stars form is exceedingly low \citep{KastnerMyers1994,OuelletteEtal2010}, so the AGB hibonite grains must have resided in the ISM for tens of Myr, and their $\altwosix$ decayed before entering the solar nebula.
It seems likely that the majority of the most refractory presolar grains were $\altwosix$-free.
\citet{ChoiEtal1998} found 30\% of presolar corundum grains had some $\altwosix$, but 70\% had no measurable amounts.

Most presolar grains would have been largely destroyed and reprocessed in the innermost regions of the protoplanetary disk, but there must have been some regions with temperatures $> 1397$ K (the vaporization temperature of spinel) but $< 1653$ K (the condensation temperature of hibonite; \citet{Lodders2003}), hot enough to vaporize all dust grains {\it except} those made of hibonite (or corundum, which in most cases would react with CaO in the gas to form hibonite).
In these environments, the only solid materials would have been either presolar hibonite grains with no live $\altwosix$; or hibonite grains condensed in the solar nebula, with canonical $\alratio \approx 5 \times 10^{-5}$. 
In these regions, %where the only solids were micron-sized hibonite grains, 
only hibonite grains would be able to coagulate into larger particles.
Coagulation of $\sim 10^3$ particles $\sim 1 \, \mu{\rm m}$ in size would yield a single agglomeration of hibonite $\sim 10 \, \mu{\rm m}$ in size. 
If such a particle were then transiently heated and melted, it would have a hibonite-dominated composition like PLACs. 
If it were to recrystallize into one or a few laths of hibonite, it would also resemble a PLAC in its morphology.

Presumably, the first hibonite inclusions in the solar nebula would form predominantly from presolar hibonite; but over time, the proportion of surviving presolar hibonite grains would decrease and the proportion of hibonite formed in the solar nebula would increase. 
If an early-formed PLAC derived from, e.g., $\approx\!\!95\%$ presolar hibonite grains and only 5\% solar nebula hibonite, it would form with $(\alratio)_0 \approx 2.5 \times 10^{-6}$ and ${}^{41}{\rm Ca}/{}^{40}{\rm Ca} \approx 2 \times 10^{-10}$. 
Because dust grains condensed in stellar outflows would not be expected to contain Be or B, as these are destroyed by stellar nucleosynthesis \citep{SackmannBoothroyd1999}, all the Be and B in the PLAC would derive from the component formed in the solar nebula, yielding an initial value $(\beratio)_0 \approx 7 \times 10^{-4}$.
If this PLAC were later thermally reset, e.g., at 1 Myr after CAIs but before it was incorporated into a parent body at $\approx 3-4$ Myr, these SLRs would decay to levels
$(\alratio)_0 \approx 1 \times 10^{-6}$ and ${}^{41}{\rm Ca}/{}^{40}{\rm Ca} \ll 10^{-12}$, and 
$(\beratio)_0 \approx 4 \times 10^{-4}$. 
In this environment, exchange of oxygen isotopes\index{Oxygen isotopes} with the surrounding gas presumably would be rapid, making them indistinguishable from other objects formed in the solar system, even if their Ca and Al largely derive from presolar grains.

All these properties are largely consistent with measurements of PLACs
\citep[e.g.,][and references therein]{LiuEtal2009,LiuEtal2010,LiuEtal2017,KrotEtal2012}.
This model also accommodates the very low $(\alratio)_0$ of HAL if it accreted mostly presolar hibonite and/or was reset later.
Regions with temperatures between 1400 and 1650 K, where hibonite grains were the only solids, also would exist at later times in the solar nebula; but at later times the hibonite grains in those regions are more likely to be those that had previously condensed from solar nebula gas. 
Coagulations of these objects would have $(\alratio)_0$ ratios closer to canonical. 
These late-forming PLAC-like inclusions might be similar to SHIBs.

\subsubsection{Summary}

The suggestion of \citet{LarsenEtal2020} that $\altwosix$-free inclusions preferentially sample ancient interstellar dust, and the scenario we describe above, indicate that chemical heterogeneities provide a plausible and satisfactory explanation for the existence of grains with $(\alratio)_0 \approx 0$, and is consistent with several observations.
First and foremost is the fact that all $\altwosix$-poor inclusions are associated with hibonite (and/or corundum), including PLACs, those FUN CAIs containing hibonite, some corundum grains, and even hibonite-bearing and corundum-bearing CAIs \citep{KrotEtal2012}.
Additionally, it has been noted that $\altwosix$ and $\cafortyone$ tend to be correlated: inclusions show evidence for both or neither
\citep{SahijpalGoswami1998}.
It also naturally explains the decoupling between $\beten$ and $\altwosix$, at least in PLACs and PLAC-like FUN CAIs. 
The arguments for widespread spatial or temporal heterogeneities in $\altwosix$, or late injection, are not necessarily favored over this chemical heterogeneity scenario.

\subsection{\boldmath${}^{36}{\rm Cl}$ and Irradiation in the Solar Nebula}
\label{sec:irradiation}

The idea that energetic ($> 10$ MeV) particles accelerated in solar flares might have induced nuclear reactions in meteoritic materials has a long pedigree.
\citet{FowlerEtal1961}  hypothesized the production of LiBeB isotopes by this process.
\citet{Lee1978} 
extended this to try to explain the recently discovered oxygen isotope\index{Oxygen isotopes} anomalies in the solar system \citep{ClaytonEtal1973} and the presence of $\altwosix$
\citep{LeeEtal1976}.
The idea was revived by \citet{ShuEtal1996}, and especially by \citet{GounelleEtal2001}, who took the discovery of $\beten$ in the early Solar System \citep{McKeeganEtal2000} as ``smoking gun" evidence for irradiation of solar nebula materials by SEPs.

Production of SLRs by SEP irradiation of solar nebula materials must have occurred at some level. 
T Tauri stars have X-ray luminosities indicating they are $\sim 10^5$ times more magnetically active than the Sun today. It's likely  $\sim 10^{48}$ protons and helium nuclei with energies $> 10$ MeV were generated in the first 1 Myr of the solar nebula.
If even a fraction of these SEPs impinged upon the disk, a significant number of nuclear reactions could have been induced. 

\subsubsection{Challenges to Producing SLRs by Irradiation}

Whether irradiation successfully explains SLRs depends on the answers to the following pertinent questions.
What fraction of SEPs actually induce nuclear reactions? 
Where do these reactions occur; do the SLRs produced make it into CAIs and other meteoritic materials? 
Are the SLRs produced by irradiation in the proportions observed in meteorites?
As research over the last few decades has answered these questions, irradiation of materials in the protoplanetary disk has diminished as a mechanism for producing most of the SLRs. 

The main obstacle to SLR production by irradiation is that only a small fraction ($\sim \!\!10^{-6}$) of the energy in SEPs actually goes into the nuclear reactions like ${}^{24}{\rm Mg}({}^{3}{\rm He},p){}^{26}{\rm Al}$ or ${}^{16}{\rm O}(p,x){}^{10}{\rm Be}$.
That is because the cross sections for loss of energy by ionization of ${\rm H}_{2}$ molecules (via the Bethe formula) far exceed (by factors $\sim \!\! 10^6$) the nuclear cross sections; SEPs give up almost all of their energy ionizing the gas rather than creating SLRs
\citep{ClaytonJin1995,DupratTatischeff2007}.
Despite the intense magnetic activity of the early Sun, the production of SLRs is actually energy-limited. 
A corollary of the large ionization cross section is that the SEPs only penetrate into the topmost $\sim 1 \, {\rm g} \, {\rm cm}^{-2}$ of the protoplanetary disk.
The actual penetration depth is sensitive to the geometry of the disk and the ability of SEPs to be scattered.
Any SLRs produced in this top layer must be mixed into the disk before being lost in a disk wind\index{Disk winds} or photoevaporative flow, if they are to be incorporated into CAIs or other inclusions.

\subsubsection{Production of Beryllium by Irradiation?}

The model of \citet{Jacquet2019} considers many of the obstacles to SLR production in its calculation of the $\beratio$ in the disk due to irradiation.
By assuming SEPs are scattered efficiently into the disk, to penetrate as deeply as possible, it finds abundances $\beratio \approx 7 \times 10^{-4}$ are possible at 1 AU, falling off steeply with heliocentric distance $r$, and starting off smaller but growing in time $t$, with $\beratio \propto r^{-3/2} \, t^{+1}$. 
Given the range of radii and times where CAIs would form, it is predicted that at least 25\% of CAIs would have $(\beratio)_0$ ratios measurably different from the average
\citep{DunhamEtal2022}.
This is contradicted by the observed distribution of $(\beratio)_0$ values in CAIs, of which $> 95\%$ are consistent with the canonical value $7 \times 10^{-4}$ (\S 2.1), suggesting a limited role for irradiation in producing $\beten$. 

Importantly, like other irradiation models \citep{GounelleEtal2009}, the other SLRs (e.g., $\altwosix$) are produced much less efficiently than $\beten$, and it is not possible to explain their abundances without overproducing $\beten$. 
To circumvent some of the issues with energy limitation, \citet{GounelleEtal2001} hypothesized that SEP irradiation occurred in an environment different from the disk, very close ($< 0.1$ AU) to the Sun.
This region is assumed to be devoid of hydrogen gas, which would be funneled onto the Sun by magnetically-regulated accretion.
SEPs would then irradiate only the rocky CAI particles or their vapor. 
\citet{DeschEtal2010} discussed many of the the challenges faced by such a model, including the inability of CAIs at $< 0.1$ AU to be transported out to several AU, and the fact that the oxygen fugacity in which CAI minerals formed demands the presence of ${\rm H}_{2}$ at  pressures like those found in a solar-composition gas in the protoplanetary disk.
Besides these issues, the model has difficulty reproducing the abundances of SLRs in their observed proportions, and has a tendency to overproduce $\beten$ relative to other SLRs like $\altwosix$. 

If the SLR ${}^{7}{\rm Be}$ existed\index{Radionuclides!Beryllium-7} in the solar nebula, this would be compelling evidence for irradiation of disk material, as its half-life is only 53 days. Like $\beten$, it must be produced by spallation of O and other nuclei; unlike $\beten$, it cannot be produced in the molecular cloud and survive to be incorporated into CAIs.
Despite claims of its one-time existence in Allende CAI 3529-41 \citep{ChaussidonEtal2006} and Efremovka CAI E40 \citep{MishraMarhas2019}, there is essentially no evidence for ${}^{7}{\rm Be}$ (\S~\ref{sec:beryllium7}).

\subsubsection{Production of ${}^{36}{\rm Cl}$ by Irradiation?}

Nevertheless, compelling evidence for SEP irradiation of solar nebula materials has come from more recent measurements indicating the presence of live $\clthirtysix$ in sodalite and wadalite in aqueously altered CAIs. 
As discussed above (\S 2.3.3), CAIs aqueously altered on their parent bodies record $({}^{36}{\rm Cl}/{}^{35}{\rm Cl})_0$ values in the range $\sim \!\! 2 \times 10^{-6}$ to $\sim \!\! 2 \times 10^{-5}$, from isochrons of ${}^{36}{\rm S}/{}^{34}{\rm S}$ vs.\ ${}^{35}{\rm Cl}/{}^{34}{\rm S}$.
Unless the Cl-S system somehow has recorded an earlier stage, this was the $({}^{36}{\rm Cl}/{}^{35}{\rm Cl})_0$ ratio at 3-4 Myr, at least in the fluids that altered the CAIs. 
If inherited from the molecular cloud, this would demand $({}^{36}{\rm Cl}/{}^{35}{\rm Cl})_{\rm SS} > 10^{-2}$ at $t\!\!=\!\!0$, which is much higher than can be supplied by plausible stellar sources. 
On this basis, \citet{JacobsenEtal2011} proposed that $\clthirtysix$ in these aqueously altered CAIs was created by irradiation.

The model of \citet{JacobsenEtal2011} posits that during the late stages of disk evolution, as the gas in the protoplanetary disk was clearing and the disk became optically thin, SEPs from the early Sun irradiated the disk, producing $\clthirtysix$. 
\citet{JacobsenEtal2011} pointed out that if the irradiated material had a solar composition, then production of sufficient $\clthirtysix$ would overproduce the SLRs $\beten$, $\altwosix$, and ${}^{53}{\rm Mn}$ relative to their typical abundances. 
They therefore hypothesized that the irradiation occurred in ices mantling grains, or in a volatile-rich gaseous reservoir, especially through reactions like ${}^{37}{\rm Cl}(p,pn){}^{36}{\rm Cl}$.
The $\clthirtysix$ so created formed HCl that condensed into ices that were subsequently accreted by the parent bodies in which the CAIs were altered, and Cl dissolved in the aqeuous fluid would have incorporated $\clthirtysix$ as it reacted with CAI minerals.

\citet{DeschEtal2011}  built on the model of \citet{JacobsenEtal2011}, hypothesizing that the SEPs did irradiate material of solar composition, and that significant $\clthirtysix$ was created by the reaction ${}^{38}{\rm Ar}(p,ppn){}^{36}{\rm Cl}$.
They also found that $\beten$ and ${}^{53}{\rm Mn}$ would be overproduced, but pointed out that Al and Be are not especially soluble, and therefore $\altwosix$ and $\beten$ may not be carried by the fluid to be incorporated into altered CAIs.
Conversely, Mn might and Li and B very likely would be.
Spallogenic Li would have ${}^{7}{\rm Li}/{}^{6}{\rm Li} \approx 9$ (lower than the chondritic ratio $\approx 12.0$) and spallogenic B would have ${}^{10}{\rm B}/{}^{11}{\rm B} \approx 0.44$ (higher than the chondritic ratio $\approx 0.25$), possibly explaining the disturbed isotopic ratios in CAIs Allende 3529-41 and Efremovka E40 (\S~\ref{sec:beryllium7}).

\subsubsection{The Solar Nebula as a Transition Disk?}
\citet{JacobsenEtal2011} found that as long as the disk was optically thin, so that SEPs were able to reach all the material in the disk, then irradiation of a solar composition reservoir yielded ${}^{36}{\rm Cl}/{}^{35}{\rm Cl} \approx 2 \times 10^{-5}$. 
They did not report the necessary SEP fluences.
\citet{DeschEtal2011} estimated ${}^{36}{\rm Cl}/{}^{35}{\rm Cl} \approx 2 \times 10^{-6}$ to $2 \times 10^{-5}$ for SEP fluences corresponding to irradiation of unshielded gas at 2.5 AU in 0.3 to 3 Myr. 
These findings indicate the validity of the irradiation model for explaining the abundances of $\clthirtysix$, provided the disk was characterized by surface densities $< 0.1 - 10 \, {\rm g} \, {\rm cm}^{-2}$, the stopping lengths of SEPs capable of inducing nuclear reactions.
This is far less than the typical surface density of the disk at a few AU, $> 10^3 \, {\rm g} \, {\rm cm}^{-2}$ \citep{Weidenschilling1977}, and is only possible in a disk that has substantially dissipated.
But this irradiation also must take place while some gas exists, so that ices and chondrites may form.
In this regard, the solar nebula must have resembled a {\it transition disk}\index{Transition disk}.

Infrared observations yield spectral energy distributions of protoplanetary disks show that 10-20\% are transition disks, with a central cavity largely cleared of dust \citep{WilliamsCieza2011}. 
This stage typically occurs after about 2 to 3 Myr of disk evolution, and the median cavity is typically several AU, as in the examples of TW Hya\index[obj]{TW Hya} and DM Tau\index[obj]{DM Tau}. 
As pointed out by \citet{DeschEtal2018}, there is evidence that this is precisely what happened in the solar nebula. 
While formation of carbonaceous chondrites\index{chondrites} continued in the outer disk beyond Jupiter\index{Jupiter} ($> 3$ AU) for $> 4$ Myr, there is no evidence of any chondrites from the NC reservoir inside Jupiter ($< 3$ AU) forming after about 2.5 Myr. 
Whether gas was lacking as well as dust is less clear. 
In one model, a giant planet may open a gap\index{Protoplanetary disk substructure!gaps} in a disk, and  photoevaporation by the central star may rapidly clear out the gas interior to the gap \citep{Clarke2007}.
On the other hand, observations suggest that gas may persist in the central cavity in many disks, suggesting a model in which giant planets form and open a gap that filters out solids but allows gas to flow through the disk \citep{ErcolanoPascucci2017}.
In any case, transition disks likely are formed by gaps opened by giant planets, and must eventually be characterized by lowered gas densities, suggesting a stage in which SEPs from the Sun could directly irradiate the inner edge of the disk outside Jupiter's gap\index{Jupiter}, creating $\clthirtysix$-rich ices that later were incorporated into carbonaceous chondrites. 

\subsubsection{Summary}

There are great challenges to producing sufficient quantities of any SLRs by SEP irradiation of solar nebula material, especially the significant loss of energy due to ionization of ${\rm H}_{2}$ gas by SEPs.
In addition, irradiation tends to produce $\beten$ more efficiently than any other SLR, so production of $\altwosix$ or other SLRs in the right quantities tends to overproduce $\beten$. The lack of variability of $\beratio$ ratios argues against widespread irradiation.
Nevertheless, the special case of $\clthirtysix$ does seem to demand SEP irradiation of materials late in the solar nebula, during a transition disk\index{Transition disk} stage. 
Few other SLRs would be produced, but spallogenic Li and B may have been incorporated into CAIs along with $\clthirtysix$. 

\subsection{The Sun's Birth Environment and Universality of the SLR Abundances}
\label{sec:usual_abundances}
 
Except for $\beten$, which was likely produced by GCR spallation of CNO nuclei in the molecular cloud, the SLRs all derive from stellar nucleosynthetic sources. 
As discussed in \S 3.3.2, a likely model, one which matches the more than a dozen SLR abundances with just two parameters, posits that molecular clouds in giant cloud complexes, and surrounding diffuse gas, were continuously ``self-enriched" by ongoing stellar nucleosynthetic sources, with ejecta from WR stars dominating over the ejecta from core collapse supernovae.
With these insights gained into the sources of the Sun's SLRs, we can ask the questions, how typical is the Sun's birth environment? And, are its SLR abundances somehow extraordinary?
We focus on the case of $\altwosix$, as radiogenic heating from its decay is especially important for the evolution of planetary materials, and it is a well-studied SLR that permits quantitative answers to these questions. 

The long-standing assumption \citep[e.g.,][]{CameronTruran1977} that the Solar System's $\alratio$ ratio was unusually high stems from perceptions of how long material from massive stars would take to enter forming solar systems, and from the perception that massive stars are rare. 
Even \citet{GaidosEtal2009}, who suggested WR stars as the source of the Solar System's $\altwosix$, concluded that this would be an unusual occurrence.
However, the observation that the solar initial $\alratio$ is similar to that in star-forming regions today \citep{Jura2013} suggests that the solar value was not unusual.  

%----------------------------
% FIGURE 9
\begin{figure*} [t!]
\begin{center}
 \includegraphics[width=0.85\textwidth]{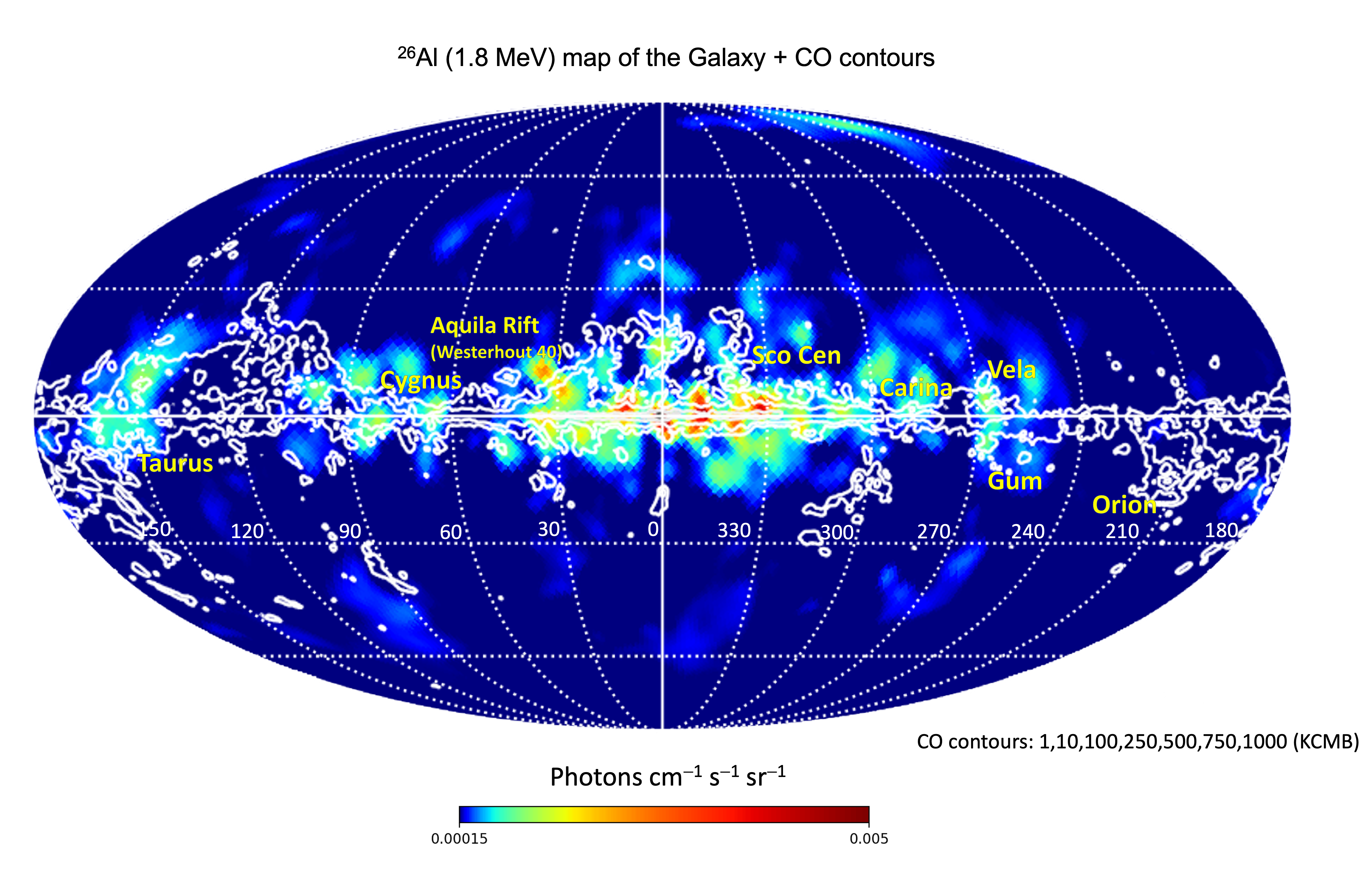}
 \caption{Mollweide projection showing the distribution of $\altwosix$ in the Galaxy\index{Radionuclides!Aluminum-26} as determined using the SPI spectrometer aboard the {\it INTEGRAL} observatory plotted by Alexandra Doyle (UCLA) using the data from \cite{Bouchet2015}.  Color contours indicate the intensity of $1.8$ MeV $\gamma-{\rm ray}$ emission.  Contours for CO as obtained from the Planck mission are overlain to serve as a tracer of molecular cloud material. Major star-forming regions in the Galaxy are labelled. }
 \label{fig:integral_data}
\end{center}
\end{figure*}

The fact that $\altwosix$ is concentrated in star-forming regions is suggested by its heterogeneous distribution across the Galaxy.  
The decay of $\altwosix$ is associated with emission of 1809 keV gamma rays.  
Figure~\ref{fig:integral_data} shows the distribution of $\altwosix$ across the Galaxy based on measured 1809 keV gamma rays as measured by the SPI spectrometer on board the {\it INTEGRAL} observatory.  
Overlain on the figure are contours for CO (from the NASA {\it Planck} data archive) to serve as tracers of molecular cloud material.  
Despite the limited spatial resolution, there is evidently at least a qualitative correlation between the distribution of $\altwosix$ and star-forming regions (e.g., Cygnus, Carina\index{Carina Nebula}, Scorpius-Centaurus\index{Scorpius-Centaurus}, and others).
% (Figure \ref{fig:integral_data}).   

WR stars and their rapidly rotating main-sequence immediate progenitors are prodigious sources of $\altwosix$ \cite[e.g.,][]{Arnould2006}. 
Numerous studies have concluded that WR stars were a likely source for $\altwosix$ in the early Solar System, and in general these remain an efficacious source of  $\altwosix$ today in the Galaxy \citep{Palacios2005, Gaidos2009, Gounelle2009, Young2014, Young2016}. 
For example, \cite{Voss2012} concluded that $\altwosix$ in the Carina region is largely attributable to WR winds.  
An inherent feature of the initial mass function (IMF) for star formation is that the most massive stars are the rarest.
As a result, in order to ensure that WR stars, with progenitor masses greater than about $25 M_\odot$, are present in a star-forming region, the stellar clusters must be composed of thousands of stars. 
It is expected, therefore, that in massive star-forming regions like Cygnus and Carina, as well as active regions like Scorpius-Centaurus, WR winds will enrich the star formation region; $\altwosix$ may well be a hallmark of massive star-forming regions \citep{Reiter2020}.
These winds may also be responsible for producing $\cafortyone$,  $\clthirtysix$, and much lesser amounts of $\fesixty$ and $\pdoneohseven$.   

The enrichment model requires WR winds to contribute on the order of $4000\times$ more $\altwosix$ to the Sun's molecular cloud than would be supplied by average ISM ejecta from, for example, core collapse supernovae.
The dominance of WR winds is an unexpected constraint on how the Sun acquired its SLRs. 
WR winds arise from progenitor stars with masses $> 25 \, M_{\odot}$, whereas core collapse supernovae are the explosions of stars with masses $> 8 \, M_{\odot}$. 
Both WR winds and supernovae ejecta produce several solar masses of material at 1000s of km/s, so would be injected into molecular clouds with comparable efficiency.
All else being equal, one would expect more material from supernovae, except that supernova ejecta from the most massive stars, e.g., $> 30 \, M_{\odot}$ \citep{Fryer1999,Smartt2017,EbingerEtal2020}, tend not to be fully ejected, instead experiencing ``fallback" onto the stellar core as it collapses to a black hole. 
Nevertheless, after integrating over the IMF, one would expect comparable masses of supernova ejecta as WR ejecta from more massive stars, or $\Lambda_{\rm W} \sim 1$. 
However, ejection from the lower-mass stars, especially those $< 20 \, M_{\odot}$, necessarily requires that they evolve for at least $\approx 15$ Myr, during which time they may be  tens of parsecs from the nearest molecular cloud material, due to their own velocities and/or due to dissipation of the clouds by photoevaporation due to the progenitor star \citep[e.g.,][]{HesterDesch2005}.
Therefore supernova ejecta might not contribute as strongly to newly forming stars as they do to the ISM, increasing $\Lambda_{\rm W}$, perhaps making it consistent with the continuous enrichment model. 
Although WR winds are less-visible end-stages of stellar evolution, they offer more opportunities to deliver the products of recent stellar nucleosynthesis into newly forming stars. 

All of these insights converge on the idea that the Sun formed 4.57 Gyr ago in a region of high SFR, associated with GMCs\index{Giant molecular clouds}, almost certainly in a spiral arm of the Galaxy.
The abundance of $\beten$ in the solar nebula suggests the SFR averaged over the nearest 1.3 kpc was a factor $\approx 4$ higher than the Galaxy-wide spatial average (\S 3.2), consistent with formation of the Sun in a region in a spiral arm with slightly higher than median SFR seen by stars. 
These regions are naturally associated with WR winds and high abundances of $\altwosix$, both in simulations (Figure 4) and in observations of the present-day Milky Way (Figure 8). 
All stars born in such regions should naturally acquire $\altwosix$ with $\alratio \approx 5 \times 10^{-5}$, as well as all the other SLRs.

Of course, many areas of the Galaxy are far from spiral arms and are characterized by low SFRs. 
In such regions, one can reasonably assume that the abundances of $\beten$ and $\altwosix$, and other SLRs, would be proportionally lower as well.
In a solar system with $\alratio < 5 \times 10^{-6}$ (equivalent to the solar nebula after $> 2.5$ Myr of decay of $\altwosix$), melting of asteroids\index{asteroids} by $\altwosix$ might not occur \citep{GrimmMcSween1993}, with dramatic effects on the evolution of planetary materials and possibly their volatile inventories \citep{LichtenbergEtal2019}. 
From a spatially averaged perspective, average SFR rates an order of magnitude lower than that experienced by the Sun might be found in 50\% of the Galaxy (Figures 4 and 5); however, only a few percent of stars actually form in these barren regions between spiral arms (Figure 5).
From the CDFs displayed in Figure 5, probably $> 80\%$ of newly formed stars see local SFRs and $\altwosix$ abundances within a factor of 2 of the initial solar concentration.   
In this respect, the Sun's SLR abundances appear to be fairly universal. 

\section{FUTURE DIRECTIONS}

The picture that has emerged is that the SLRs except $\clthirtysix$ appear to have been homogeneously distributed in the solar nebula at $t\!\!=\!\!0$. 
To better test this conclusion,  we suggest several meteoritic measurement campaigns. 
More coordinated studies of multiple systems (Pb-Pb, Al-Mg, Mn-Cr, etc.) in single objects are most useful for assessing discrepancies between chronometers and any heterogeneities between SLRs. 
The focus should be on samples that likely achieved isotopic closure simultaneously, including quenched angrites and other volcanic achondrites, or impact melts, of which the CB/CH chondrules may be an example \citep{DeschEtal2022b}.
We especially recommend coordinated studies of Cl-S and Al-Mg systematics in aqueously altered CAIs, to better understand the effect of alteration on the initial ${}^{36}{\rm Cl}/{}^{35}{\rm Cl}$ and $\alratio$ ratios in these objects, and to constrain the initial ${}^{36}{\rm Cl}/{}^{35}{\rm Cl}$ in the solar nebula.
We also recommend obtaining Pb-Pb ages, if possible, for fine-grained CAIs that are less likely to have suffered heating and subsolidus diffusion of Pb after their formation; these should provide a better estimate of the Pb-Pb age of $t\!\!=\!\!0$, just as fine-grained CAIs apparently record a higher $\hfratio$ than coarse-grained CAIs \citep{KruijerEtal2014}. 
We recommend more analyses to better constrain the initial ${}^{135}{\rm Cs}/{}^{133}{\rm Cs}$ and ${}^{107}{\rm Pd}/{}^{108}{\rm Pd}$ %(and ${}^{60}{\rm Fe}/{}^{56}{\rm Fe}$)
ratios in the solar nebula, which might provide stringent tests of the continuous self-enrichment model.
Improvements in diffusion coefficients (especially B diffusion in melilite)
%and Mg diffusion in CAI minerals) 
and SLR half-lives (especially those of low precision, like $\pdoneohseven$ and ${}^{53}{\rm Mn}$) are always useful.

We also recommend improvements to the astrophysical models and astronomical observations.
The models presented here constrain the residence time of material in clouds prior to incorporation into a star to be $\approx\!\!200$ Myr.
While consistent with a simple inference from comparing molecular masses to star formation rates, real-world confirmation of the relative efficiency, or inefficiency, of star formation could test these models.
Models should be updated to include such physical effects as fallback, to generate more-accurate predictions of the production and spatial distribution of ${}^{60}{\rm Fe}$ and ${}^{26}{\rm Al}$, connecting to both meteoritic data and $\gamma$-ray observations.
Models of the physical growth of the first Solar System solids, building on presolar grain data, could better test whether the observed variations of $\alratio$ in hibonite grains  can be explained as a chemical heterogeneity. 

There have been many advances since \citet{Urey1955} proposed that the forming Solar System acquired $\altwosix$ from dying stars in its birth environment, based on knowledge of how planetary materials had melted. 
Its existence has since been confirmed via internal isochrons constructed from isotopic analyses of meteoritic samples, and these analyses can be used to date events in the solar nebula.
Other SLRs have been discovered, and the concordancy between the isotopic systems has pointed strongly toward homogeneity of $\altwosix$, ${}^{53}{\rm Mn}$, ${}^{182}{\rm Hf}$, $\beten$, and other SLRs in the solar nebula.
While $\clthirtysix$ reveals events occurring as the protoplanetary disk dissipated, the other SLRs speak to their inheritance from the molecular cloud. 
We are beginning to understand how the Sun's molecular cloud acquired its SLRs and can place the Sun's formation in a Galactic context. 
Maps of ${}^{26}{\rm Al}$ $\gamma$-ray emission speak to the life-cycle of this and other SLRs as they were ejected from supernovae, neutron stars, low-mass giants, and especially WR stars, injected into nearby molecular clouds, including the Sun's, to be incorporated into meteoritic materials, completing the circle between protostars and planets.

%
%
%\bigskip
%\noindent
%\textbf{2.2 Extrasolar Planetary Systems}
%
%\begin{figure}[h]
% \epsscale{1.0}
% \plotone{fig1.eps}
% \caption{\small Figure Caption is here.}
% \label{fig0}
%\end{figure}
%

\bigskip

\noindent\textbf{Acknowledgments} 
SJD gratefully acknowledges that the work herein benefitted from collaborations and/or information exchange within NASA's Nexus for Exoplanetary System Science research coordination network sponsored by NASA's Space Mission Directorate (grant NNX15AD53G, PI Steve Desch).
ETD gratefully acknowledges support from a 51 Pegasi b Fellowship, grant \#2020-1829.
YF gratefully acknowledges computational resources of Oakforest-PACS provided by Multidisciplinary Cooperative Research Program in Center for Computational Sciences, University of Tsukuba (Project ID: xg21i003) and Oakbridge-CX provided by the Information Technology Center, The University of Tokyo through the HPCI System Research Project (Project ID: hp210015).

%\bibliography{pp7}
\bibliographystyle{pp7}
\bibliography{pp7.bib}

%\noindent\textbf{REFERENCES}
%\bigskip
%\parskip=0pt
%{\small
%\baselineskip=11pt
%%
%\refs Hayashi, C., 1961 {\it PASJ}, 13, 450
%%
%\refs Hayashi, C. \& Nakano, T., 1963 {\it Prog. Theor. Phys}, 30, 460
%%
%\refs Hayashi, C., et al., 1985, 
       %{\it Protostars and Planets II} (D. C. Black and M. S. Matthews, UAP), 1100
%%

%\bibliographystyle{pp7}
%\bibliographystyle{ppvi}
%\bibliographystyle{apj}
%\bibliographystyle{aasjournal}
%\bibliography{pp7.bib}

%\printindex
%\renewcommand{\indexname}{Object Index}
%\printindex[obj]

\end{document}